\documentclass[11pt]{article}
\usepackage[utf8]{inputenc}
\pdfoutput=1
\usepackage[T1]{fontenc}
\usepackage{amsmath}
\usepackage{amsfonts}
\usepackage{amssymb}
\usepackage[version=4]{mhchem}
\usepackage{stmaryrd}
\usepackage{mathtools}
\usepackage{graphicx}
\usepackage{lipsum}
\usepackage[export]{adjustbox}
\graphicspath{ {./} }
\usepackage{bbold}
\usepackage{hyperref}
\hypersetup{colorlinks=true, linkcolor=blue, filecolor=magenta, urlcolor=cyan,}
\title{Hadron masses from a Kaluza-Klein like Model }

\author{Manfred M. Vieten\\
University of Konstanz\\
E-mail: manfred.vieten@uni.kn}
\date{}

\begin{document}
\maketitle

\begin{abstract}
The purpose of this paper is to calculate the masses of the hadrons. More precisely the masses of the scalar and vector mesons as well as the baryons having $\frac{\hslash}{2}$ and $\frac{3 \hslash}{2}$ spin without orbital momentum are calculated. It is hypothized the Kaluza-Klein like model does deliver results in particle physics complementing those of the  standard model of particle physics. It is presumed the Standard Model and the Kaluza-Klein like model establish a dualism describing particle physics similar to wave-particle dualism. For this paper a Kaluza-Klein like model was developed based on the structure of the Standard Model. The model consists of 10 dimensions which are: one time, three usual macroscopic space and six compactified dimensions. Excitations - disturbances traveling with the speed of light on the 10D-spacetime - are introduced. An excitation on a compactified dimension can induce a mass in 4D-spacetime; it is accompanied by an integer, the excitation number, and has a well-defined spin. The model's parameters are determined using electron's g-factor and the measured masses of the charged leptons, the proton, the neutron, the mesons $\pi^{+}, \phi, \psi, \Upsilon$. The most important of the derived parameters are the compactification radius $\rho$, the weak coupling $\alpha_{w}$, the strong coupling $\alpha_{s}$ (for hadrons coupling strength) and the anti-neutrino to neutrino density ratio $\delta$. So far, exact  formulas for calculating hadron masses were not derived. The approximate formulas, however, are reported and applied to about one hundred composite particles, which are compiled within four separate tables. The comparison between the measured masses of 64 hadrons and the approximations shows relative errors below $0.05$ for 55\% and below $0.1$ in 81\% of all cases. Pearson's correlation coefficient of masses measured versus calculated is $r=0.998$ with a significance of $p<10^{-73}$. The results of another 36 hadron masses, not yet experimentally determined, are stated.\\
For a better assessment of the presented model's validity, and in addition to the paper's original purpose, the strong, electric, and weak forces were calculated. These forces together with the other parameters are discussed and compared with measurements or values from the standard model of particle physics.
\end{abstract}

Keywords: hadron masses, Kaluza-Klein theory, geodesic equation, Standard Model, antineutrino/neutrino density, forces (strong, electric, weak), neutrino mass

\section{Introduction }
The Standard Model of particle physics [1-7] is one of the most successful theories of physics. The strength of the model is its ability to explain the forces between particles as the exchange of bosons [8] whereby the photon in case of electromagnetism, the 8 gluons for the strong and the $\mathrm{W}^{+}, \mathrm{W}^{-}$and the $\mathrm{Z}$ for the weak interaction operate as mediators. One of the model's greatest successes is the calculation of the anomalous magnetic dipole moment of the electron through quantum mechanical corrections, as first introduced by Julian Schwinger in 1948 [9] with the first order correction on the predictions of the Dirac equation. Later, higher order corrections established the agreement between theoretical and experimental results of up to 12 significant digits [10].

There are however weaknesses such as the well-known failure to provide any information on dark matter and dark energy [11]; and the inability to describe gravity. Thus far, it appears that there is no room within the Standard Model for providing information with regards to dark matter and dark energy. For the missing gravitational interaction, some extensions in the form of quantum gravity [12] with the spin2 graviton that serves as an exchange boson, have been suggested. An additional shortcoming is the incomplete ability to calculate meson and baryon masses, the particles representing ordinary matter, and the unstable particles created in high-energy particle smashing experiments.

The Standard Model scenario involves the acting of various fields within the flat 4-dimensional spacetime. General relativity [13], on the other hand, uses curved spacetime to describe gravity on a larger scale - the main theory of cosmology [14]. But what if instead of looking for a graviton as the force creating boson of quantum gravity, the alternative could be a multidimensional spacetime to describe the strong, the electromagnetic and the weak interactions? The first attempt dates back to the year 1921, when Theodor Kaluza $[15,16]$ presented a 5-dimensional model to include electromagnetism. The quantum aspect was introduced in 1926 by Oskar Klein [17, 18]. A summary on the temporary development was published in 1987 by Appelquist, Chodos and Freund titled "Modern Kaluza-Klein Theories" [19]. In 1999 another summary [20] was published with a similar title by Paul Wesson. The update came in 2019 [21].

One of the focal points of many papers dealing with Kaluza-Klein theories concentrate on the inclusion of electromagnetism in 5-dimensional spacetime. These developments frequently involve the construction of a 5-dimensional line element and the composition of the respective Einstein tensor. Higher dimensional approaches come with string theory [22-24]. While string theory introduces a rich mathematical methodology, it is unable to deliver results that can be compared with data gathered from experiments. The Standard Model comes with the quantum structure of elementary particles which is capable of calculating the decay probabilities of particles [25]. However, a general method for calculating hadron masses is still lacking $[26,27]$. In addition, in order to describe the features and capabilities, a method should be found within the framework of a multi-dimensional spacetime in which to portray particle physics.

Is it reasonable to exchange the robust Standard Model for something that so far has been able to describe gravitation very well while at the same time be unable to contribute to the very small, particle physics? And is it possible that the Standard Model and a multidimensional relativity both be meaningful and able to produce results? More precisely, what if both approaches represent two sides of the same coin, such as in the sense of dualism? The famous double slit experiment [28] that displays wave and particle nature at the same time points towards such a possibility. Therefore, within this article a model is introduced which describes the outlines of the Standard Model in terms of a "Kaluza-Klein like theory": an approach for looking at the Standard Model from a different perspective, and not as a contradictory method. The following development consists of ten deformable dimensions: one time, three macroscopic space and another 6 compactified space dimensions. Within all dimensions, an excitation, a temporary deformation of this 10-dimensional canvas, always moves with the speed of light. In the usual 4D-space, an excitation is identified as a gravitational wave. In a compactified dimension, a stable excitation has a well-defined length, depending on the dimension's radius. The actual radii of the six compactified dimensions are able to change dynamically depending on the specific excitations, which are identified as particles. During the development of the model, the explicit spacetime structure is given. The next step is an explanation of the three interactions and the corresponding formulas for calculating particle masses. The particles are identified in terms of the 6 compactified dimensions. Afterwards, the model's constants are determined; the most important are the compactification radius $\rho$, the weak coupling $\alpha_{w}$, the strong coupling $\alpha_{s}$ and the neutrino $z_{v}$ to anti-neutrino $z_{\bar{v}}$ distribution ratio $\delta$. For this purpose, the measured masses of the charged leptons, the proton, the neutron, and the mesons $\pi^{+}, \phi, \psi, \Upsilon$ and the magnetic moment of the electron are used. From here onwards, there are no more free parameters available. Then, the calculation rules for the scalar and vector mesons; and for the spin $1 / 2$ and $3 / 2$ baryons without orbital angular momentum are stated. The masses of approximately 100 hadrons will be calculated and compared with the experimental data, if available. The relative error between the measured and the calculated mass is $<0.05$ for most of the comparable hadrons.\\
In Addition to the original purpose of calculating the hadron masses and for better evaluating the validity of the presented model, the strong, electric, and weak forces are calculated and stated as equations and graphs in the appendix.

\section{Model}
\subsection{Embedding and container space}
The start point is a 10-dimensional space consisting of 1 time and 9 space dimensions. The time and three space dimensions constitute the usual spacetime. The other six space dimensions are compact dimensions. For calculational reasons, this 10-dimensional space is embedded within a 20-dimensional container space. The connections between the physical 10D-spacetime $x^{i}$ and the container space $y^{j}$ is given as:

\begin{align}
&y^{0}=c t \\
&y^{1}=x^{1} \nonumber\\
&y^{2}=x^{2} \nonumber\\
&y^{3}=x^{3} \nonumber\\
&y^{i}=r^{i} \cdot \cos \left(\varphi_{i}\right)=r^{i} \cdot \cos \left(\frac{x^{i}}{r^{i}}\right) \nonumber\\
&y^{10}=0 \nonumber\\
&y^{11}=0 \nonumber\\
&y^{12}=0 \nonumber\\
&y^{13}=0 \nonumber\\
&y^{i+10}=r^{i} \cdot \sin \left(\varphi_{i}\right)=r^{i} \cdot \sin \left(\frac{x^{i}}{r^{i}}\right) \nonumber
\end{align}

As such, $r^{i}$ is the radius of the compact dimension with indices $i=4,5,6,7,8,9$. The container space has a simple flat metric of the form

\begin{align}
\eta_{a b}=\operatorname{diag}(1,9 \times (-1),1,9 \times (-1))
\end{align}

with $a$ and $b$ from 0 to 19.

The connection between the 10-dimensional physical space and the 20-dimensional container space is given as

\begin{align}
g_{\mu v}=\frac{\partial y^{a}}{\partial x^{\mu}} \frac{\partial y^{b}}{\partial x^{v}} \eta_{a b}
\end{align}

This results in a spacetime signature of $\operatorname{diag}(1 -1 -1 -1 -1 -1 -1 -1  -1 -1)$ on the undeformed 10D-spacetime.\\
The 6 additional compactified dimensions are subdivided into sections for the purpose of: the strong interaction - dimensions 4 to 6 -; the electric interaction - dimension 7 -; and the weak interaction - dimensions 8 and 9 .

\subsection{The angular momentum of a stable excitation on a compactified dimension}
A particle is a combination of associated excitations, which waves around some of the six compact cylinders forming one unit. The signal velocity around a cylinder is always $c$, the speed of light. A full revolution around a cylinder (double value) is

\begin{align}
4 \pi r^{i}=n^{i} \lambda^{i}
\end{align}

with $n^{i}$ being an integer. The wavelength is connected to the speed of light and the frequency $f$ in the usual way.

\begin{align}
c=f \cdot \lambda
\end{align}

The momentum of an excitation is

\begin{align}
p^{i}=\hbar k^{i}
\end{align}

whereby $k^{i}=2 \pi / \lambda^{i}$ is the wave vector. With these definitions, the spin (angular momentum) of an excitation on one cylinder is

\begin{align}
s^{i}=r^{i} \times p^{i}=r^{i} \times \frac{2 \pi \hbar}{\lambda^{i}}=r^{i} \times \frac{n^{i} \hbar}{2 r^{i}}=n^{i} \frac{\hbar}{2}
\end{align}

, independent of the cylinder radius.

\subsection{The electric interaction - dimension 7}
The start of the development is a point-particle current (as seen in the usual 4D-spacetime) of four-velocity $w^{\mu}$

\begin{align}
J_{e}^{\mu}=\frac{e}{3} \cdot o_{7} \cdot \delta^{3}(\vec{x}-\vec{y}) \cdot w^{\mu}
\end{align}

and the covariant Liénard-Wiechert potential [30] caused by a point-particle of four-velocity $v^{\mu}$.

\begin{align}
A_{e}^{\mu}=\frac{\frac{e}{3} \cdot n_{7}}{4 \pi \varepsilon_{0} c^{2}} \cdot \frac{v^{\mu}}{\left|\vec{y}-\vec{x}_{0}\right|}
\end{align}

The potential energy between these two particles is:

\begin{align}
&U_{n o}^{e}\left(\left|\vec{x}-\vec{x}_{0}\right|\right)=\iiint_{-\infty}^{\infty} J_{e}^{\mu} A_{\mu}^{e} d y_{1} d y_{2} d y_{3} \\
&=\frac{e^{2} n_{7} o_{7}}{8 \pi \varepsilon_{0} c^{2} 9} \cdot \frac{w^{\mu} v_{\mu}}{\left|\vec{x}-\vec{x}_{0}\right|}=\frac{\alpha_{e} c \hbar n_{7} o_{7} f_{n o}(z)}{9 \rho z} \nonumber
\end{align}

Here $\left|\vec{x}-\vec{x}_{0}\right|$ is abbreviated as $\rho \cdot z$, with a constant $\rho$ of dimension length, and a dimensionless parameter $z$. The invariant expression is condensed as:

\begin{align}
w^{\alpha} v_{\alpha} = g_{\alpha\beta} w^{\alpha} v^{\beta}= c^{2} f_{\mathit{no}} \! \left(z \right)
\end{align}

with

$$
w_{o}^{\mu}=\left[\begin{array}{cccccccccc}
\gamma_{o}  & \gamma_{o} \beta_{o}  & 0 & 0 & |{signum(o_{i})}| \end{array}\right] \cdot c
$$

and

$$
v_{n}^{\mu}=\left[\begin{array}{cccccccccc}
\gamma_{n}  & \gamma_{n} \beta_{n}  & 0 & 0 & |{signum(n_{i})}| \end{array}\right] \cdot c
$$

$i=4,5,6,7,8,9$ (see appendix 8.5 also). The $signum$ is defined with $signum(0)=0$. $f_{n o}(z)$ contains all z-depended parts of the expression. In particular, this includes the metric tensor $g_{\mu \nu}(z)$, and the normalized velocity $\beta=\frac{v}{c}$, as well as $\gamma=\frac{1}{\sqrt{1-\beta^{2}}}$. The force on such a particle is

\begin{align}
\vec{F}_{e}(z)=-\frac{1}{\rho} \cdot \frac{\partial}{\partial z} \left(\frac{\alpha_{e} c \hbar n_{7} o_{7} f_{n o}(z)}{9 \rho z}\right)=\frac{\alpha_{e} c \hbar n_{7} o_{7} f_{n o}(z)}{9 \rho^{2} z^{2}}-\frac{\alpha_{e} c \hbar n_{7} o_{7} \frac{\partial f_{n o}(z)}{\partial z}}{9 \rho^{2} z}
\end{align}

The rest and kinetic part of the electric energy of the excitation $o_{7}$ is anticipated as

\begin{align}
E_{o}^{e}(z)=\left(\hbar \omega\left(r_{o}\right)+\frac{A {| o_{7} |} \alpha_{e} c \hbar}{9 \rho}\right) \cdot \gamma_{o}(z)=\left(\frac{o_{7} c \hbar}{r_{o}(z)}+\frac{A {| o_{7} |} \alpha_{e} c \hbar}{9 \rho}\right) \cdot \gamma_{o}(z)
\end{align}

in which the second term represents a constant, multiplied by the $\gamma_{o}$-factor. The total electric energy is the sum of (13) and (10), which is conserved and hence it follows

\begin{align}
0&=\frac{\partial}{\partial z} (E_{o}^{e}(z)+ U_{n o}^{e}(z)) \\
0 &= 
-\frac{o_{7} \hslash  c \left(\frac{d}{d z}r \! \left(z \right)\right) \gamma_{o} \! \left(z \right)}{r \! \left(z \right)^{2}}+\left(\frac{o_{7} \hslash  c}{r \! \left(z \right)}+\frac{A {| o_{7} |} \alpha_{e} c \hslash}{9 \rho}\right) \left(\frac{d}{d z}\gamma_{o} \! \left(z \right)\right) \nonumber \\ 
&+\frac{\alpha_{e} c \hslash  n_{7} o_{7} \left(\frac{d}{d z}f_{\mathit{no}} \! \left(z \right)\right)}{9 \rho  z}-\frac{\alpha_{e} c \hslash  n_{7} o_{7} f_{\mathit{no}} \! \left(z \right)}{9 \rho  \,z^{2}}
\nonumber
\end{align}

This differential equation allows to calculate the cylinder radius of dimension 7 (with $o_{7}$ and $n_{7}$ being the excitation numbers).

\begin{align}
r_{o 7}(z)=\frac{9 \gamma_{o}(z) \rho z o_{7}}{-A {| o_{7} |} \alpha_{e} \gamma_{o}(z) z-\alpha_{e} n_{7} o_{7} f_{n o}(z)+9 C \rho z o_{7}}
\end{align}

The limit of the undisturbed radius at infinity is (see "The undisturbed cylinder radii" below):

\begin{align}
\lim _{z \rightarrow \infty} r_{o_{7}}=\frac{9 \rho o_{7} \gamma_{o}(\infty)}{-A {| o_{7} |} \alpha_{e} \gamma_{o}(\infty)+9 C \rho o_{7}}=\frac{9 \rho}{\alpha_{e} \operatorname{signum}\left(o_{7}\right)}
\end{align}

hence allowing to fix the constant $C$ as:

\begin{align}
C=\frac{\alpha_{e} \left|o_{7}\right| \left(1+A\right) \gamma_{o}(\infty)}{9 o_{7} \rho}
\end{align}

Here $\rho$ is identified as the compactification radius, which will be calculated later. The radius of equation (15) becomes

\begin{align}  r_{o_{7}}(z)=\frac{9 \gamma_{o}(z) \rho \  signum(o_{7})}{\alpha_{e} \cdot\left(\gamma_{o}(\infty) \cdot \left(1+A\right)-\frac{n_{7} \  signum(o_{7}) f_{n o}(z)}{z}-\gamma_{o}(z) A\right) }  \end{align}

The energy generated via the electric interaction of the particle is achieved by inserting (18) into equation (13).

\begin{align}
E_{o}^{e}(z)=\left(\frac{o c \hbar}{r_{o_{7}}(z)}+\frac{A \left|o_{7}\right| \alpha_{e} c \hbar}{9 \rho}\right) \cdot \gamma_{o}(z) \\
=\frac{\alpha_{e} c \hbar}{9 \rho} \cdot\left(\gamma_{o}(\infty) \left|o_{7}\right|  \left (1+A\right)-\frac{n_{7} o_{7} f_{n o}(z)}{z}\right) \nonumber
\end{align}

Equation (19) describes the electric rest and kinetic energy. To calculate the mass, the potential energy (10) must be added to (19) and following is the division by $c^{2}$. The electric mass becomes

\begin{align}
m_{o_{7}}=\frac{E_{o}^{e}(z)+U_{n o}^{e}(z)}{c^{2}}=\frac{\alpha_{e} \hbar \cdot \left|o_{7}\right| \left(1+A\right) \gamma_{o}(\infty)}{9 c \rho}
\end{align}

, which is not dependent on $z$.

\subsection{The strong interaction - dimensions 4, 5, and 6}
At this point the color representation is introduced as:

\begin{align}
c_{\text {red }}=\left(\begin{array}{c}
1 \\
-1 \\
0
\end{array}\right) \quad c_{\text {green }}=\left(\begin{array}{c}
0 \\
1 \\
-1
\end{array}\right) \quad c_{\text {blue }}=\left(\begin{array}{c}
-1 \\
0 \\
1
\end{array}\right)
\end{align}

For an anti-color each component is multiplied by $-1$. The calculations of the cylinder radii are done analog to the electric case. Again, starting with a point-particle current of four-velocity $w^{\mu}$ for each dimension the current is

\begin{align}
J_{s}^{\mu}=e_{s} \cdot \delta^{3}(\vec{x}-\vec{y}) \cdot w^{\mu} n_{i} && { with } && i=4,5,6
\end{align}

and the vector potential (simplified ansatz form [31-34]) depending linearly on the distance between the two color charges,

\begin{align}
A_{s}^{\mu}=\frac{e_{s}}{4 \pi \varepsilon_{0} c^{2}} \cdot \frac{\left|\vec{y}-\vec{x}_{0}\right|}{\rho^{2}} v^{\mu} o_{i}=\frac{e_{s}}{4 \pi \varepsilon_{0} c^{2}} \cdot \frac{z}{\rho} v^{\mu} o_{i}
\end{align}

the potential energy between these two particles is calculated in the usual way.

\begin{align}
&U_{n o}^{s}(z)=\iiint_{-\infty}^{\infty} J_{s}^{\mu} A_{\mu}^{s} d y_{1} d y_{2} d y_{3} \\
&=\frac{e_{s}^{2} n o}{4 \pi \varepsilon_{0} c^{2}} \cdot w^{\mu} v_{\mu} \cdot n_{i} o_{i} \cdot \frac{z}{\rho} \nonumber\\
&=\frac{\alpha_{s} c \hbar n_{i} o_{i}}{\rho} \cdot f_{n o}(z) \cdot z \nonumber
\end{align}

The force becomes:

\begin{align}
&\vec{F}_{s}=-\frac{1}{\rho} \cdot \frac{\partial}{\partial z} U_{n o}^{s}(z) \\
&=-\frac{\alpha_{s} c \hbar n_{i} o_{i}}{\rho^{2}} \cdot\left(\frac{d f_{n o}(z)}{d z} \cdot z+f_{n o}(z)\right) \nonumber
\end{align}

The rest and kinetic color energy of a cylinder's excitation is assumed to be of the same format as in the electric case:

\begin{align}
E_{o}^{s}(z)=\left(\frac{o_{i} c \hbar}{r_{o_{7}}(z)}+\frac{\alpha_{s} c \hbar A \left|o_{i}\right|}{\rho}\right) \cdot \gamma_{o}(z)
\end{align}

and as before it follows:

\begin{align}
0=&\frac{\partial}{\partial z} (E_{o}^{s}(z)+U_{n o}^{s}(z)) \\
=&\left(\frac{o_{i} c \hbar}{r_{o}(z)}+\frac{\alpha_{s} c \hbar A \left|o_{i}\right|}{\rho}\right) \cdot \frac{\partial \gamma_{o}(z)}{\partial z}-\frac{\left.o_{i} c \hbar \gamma_{o}(z)\right)_{o}}{r_{o}^{2}(z)} \cdot \frac{\partial r_{o}}{\partial z} \nonumber \\ &+\frac{\alpha_{s} c \hbar n_{i} o_{i}}{\rho} \cdot\left(\frac{d f_{n o}(z)}{d z} \cdot z+f_{n o}(z)\right) \nonumber
\end{align}

The radii of the color cylinders $(i=4,5,6)$ become:

\begin{align}
r_{o_{i}}(z)=\frac{\rho o_{i} \gamma_{o}(z)}{-\alpha_{s} A \gamma_{o}(z)-\alpha_{s} n_{i} o_{i} f_{n o} z+C \rho o_{i}}
\end{align}

Fixing the constant is archived:

\begin{align}
\lim _{z \rightarrow 0} r_{o_{i}}=\frac{o_{i} \rho \gamma_{o}(0)}{-\alpha_{s} A \left|o_{i}\right| \gamma_{o}(0)+o_{i} \rho C}=\frac{\rho}{\alpha_{s} \operatorname{signum}\left(o_{i}\right)}
\end{align}

with the result:

\begin{align}
C=\frac{\alpha_{s} \left|o_{i}\right| \left(1+A\right) \gamma_{o}(0)}{o_{i} \rho}
\end{align}

The cylinder radius becomes:

\begin{align}
r_{o_{i}}(z)=\frac{\rho \ signum(o_{i}) \gamma_{o}(z)}{\alpha_{s}\left(\gamma_{o}(0) \cdot \left(1+A\right)-signum(o_{i}) \ n_{i} f_{n o)}(z) \cdot z-\gamma_{o}(z) A \right)}
\end{align}

At this point, without the potential energy, the color energy becomes

\begin{align}
E_{o}^{S}(z)=\frac{\alpha_{s} c \hbar}{\rho}\left( \left|o_{i}\right| \left(1+A\right) \gamma_{o}(0)-n_{i} \ signum(o_{i}) \cdot f_{n o} \cdot z\right)
\end{align}

Finally, the color mass is

\begin{align}
m_{o_{i}}=\frac{E_{o}^{S}(z)+U_{n o}^{s}(z)}{c^{2}}=\frac{\alpha_{s} \hbar}{c \rho} \cdot \left|o_{i}\right| \, \left(1+A\right) \gamma_{o}(0)
\end{align}

\subsection{The weak interaction - dimensions 8 and 9}
The point-particle currents of the weak interaction are

\begin{align}
J_{w}^{\mu}=e_{w} \cdot \delta^{3}(\vec{x}-\vec{y}) \cdot w^{\mu} n_{i}  && { with } && i=8,9
\end{align}

In the electric case, the vector potential shows a $1 / z$ dependence. For the strong interaction the vector potential has a $z$ dependency. The electric interaction needs one dimension while the strong interaction occupies three dimensions. The weak interaction uses two dimensions. Therefore, the conjecture for the weak potential is a logarithmic dependence since this appears to be the plausible functional behavior for this number of dimensions.

\begin{align}
A_{w}^{\mu}=\frac{e_{w}}{4 \pi \varepsilon_{0} c^{2}} \cdot \frac{v^{\mu}}{\rho} o_{i} n_{i} \cdot \ln (z)
\end{align}

The associated potential energy as previously demonstrated

\begin{align}
&U_{n o}^{w}(z)=\iiint_{-\infty}^{\infty} J_{w}^{\mu} A_{\mu}^{w} d y_{1} d y_{2} d y_{3} \\
&=\frac{e_{w}^{2}}{4 \pi \varepsilon_{0} c^{2}} \cdot \frac{w^{\mu} v_{\mu}}{\rho} o_{i} n_{i} \cdot \ln (z) \nonumber\\
&=\alpha_{w} \cdot \frac{c \hbar o_{i} \cdot n_{i} \cdot \ln (z) f_{n o}(z)}{\rho} \nonumber
\end{align}

with the force

\begin{align}
&\vec{F}_{w}=-\frac{1}{\rho} \cdot \frac{\partial}{\partial z} U_{n o}^{w}(z) \\
&=-\frac{\alpha_{w} c \hbar o_{i} n_{i}}{\rho^{2}} \cdot\left(\ln (z) \cdot \frac{\partial f_{n o}(z)}{\partial z}+\frac{f_{n o}(z)}{z}\right) \nonumber
\end{align}

With the same consideration as above the rest and kinetic energy is written as

\begin{align}
E_{o}^{w}(z)=\left(\frac{o_{i} c \hbar}{r_{o_{i}}(z)}+\frac{\alpha_{w} c \hbar A \left|o_{i}\right|}{\rho}\right) \cdot \gamma_{o}(z)
\end{align}

As before the total energy, the sum of (38) and (36) is conserved, which results in:

\begin{align}
0=&\frac{\partial}{\partial z} (E_{o}^{s}(z)+ U_{n o}^{s}(z)) \\
0=&\left(\frac{o c \hbar}{r_{o}(z)}+\frac{\alpha_{w} c \hbar A \left|o_{i}\right|}{\rho}\right) \cdot \frac{\partial \gamma_{o}(z)}{\partial z}-\frac{o c \hbar \gamma_{o}(z)}{r_{o}^{2}(z)} \cdot \frac{\partial r_{o}}{\partial z} \nonumber \\ &+\frac{\alpha_{w} c \hbar o_{i} n_{i}}{\rho} \cdot\left(\ln (z) \cdot \frac{\partial f_{n o}(z)}{\partial z}+\frac{f_{n o}(z)}{z}\right) \nonumber
\end{align}

Once again, (39) allows for calculating the cylinders' radii with the result

\begin{align}
r_{o_{i}}(z)=\frac{o_{i} \rho \gamma_{o}(z)}{-A \left|o_{i}\right| \alpha_{w} \gamma_{o}(z)-\alpha_{w} o_{i} n_{i} \ln (z) f_{n o}(z)+C \rho o_{i}}
\end{align}

for $i=8,9$.

Not mentioned so far, is a screening effect playing a role here, too. Every particle reacting to the weak interaction is also affected by the nearby neutrinos and anti-neutrinos. This means that besides the interaction with a specific second particle, there is an additional contribution of nearby neutrinos and anti-neutrinos. Be $\delta$ the average distance $z_{v}$ of the nearest neutrino compared to the average distance $z_{\bar{v}}$ of the nearest anti-neutrino. $\delta$ is also equal to the cubic root of the anti-neutrino density $\rho_{\bar{v}}$ divided by the neutrino density $\rho_{v}$. For a homogeneous density, this connection is for any given volume $V=n_{v} z_{v}^{3} \rho^{3}=n_{\bar{v}} z_{\bar{v}}^{3} \rho^{3}$

\begin{align}
\frac{\rho_{\bar{v}}}{\rho_{v}}=\frac{n_{\bar{v}}}{V} \cdot \frac{V}{n_{v}}=\frac{n_{\bar{v}} n_{v} z_{v}^{3} \rho^{3}}{n_{v} n_{\bar{v}} z_{\bar{v}}^{3} \rho^{3}}=\frac{z_{v}^{3}}{z_{\bar{v}}^{3}}=\delta^{3}
\end{align}

This establishes the neutrino $z_{v}$ to anti-neutrino $z_{\bar{v}}$ rate

\begin{align}
&\qquad \delta=\frac{z_{v}}{z_{\bar{v}}}=\left(\frac{\rho_{\bar{v}}}{\rho_{v}}\right)^{\frac{1}{3}}
\end{align}

$\delta$ alters the undisturbed cylinder radii of the weak interaction. Therefore, the fixing of the free constant becomes

\begin{align}
\lim _{z \rightarrow 1} r_{o_{i}}=\frac{o_{i} \rho \gamma_{o}(1)}{C \rho o_{i}-\alpha_{w} A \left|o_{i}\right| \gamma_{o}(1)}=\frac{\rho}{\alpha_{w} \cdot\left(\operatorname{signum}\left(o_{i}\right)-\ln (\delta)\right)}
\end{align}

, which results in

\begin{align}
r_{o_{i}}(z)= \\ \frac{\rho \operatorname{signum}\left(o_{i}\right) \gamma_{o}(z)}{Numerator} \nonumber \\
Numerator=\alpha_{w}\left(\left[\left(1-\operatorname{signum}\left(o_{i}\right) \ln (\delta)\right) +A \right] \cdot \gamma_{o}(1) \right. \nonumber \\
\left. -f_{n o}(z) \cdot \ln (z) \cdot n_{i} \operatorname{signum}\left(o_{i}\right)-A \gamma_{o}(z)\right) \nonumber
\end{align}

The overall distribution of neutrinos and anti-neutrinos expressed though $\delta$, provides a very small contribution to a particle's energy

\begin{align}
E_{o} \! \left(z \right) = 
\frac{\alpha_{w} \hslash  c \left(\left(\left(A +1\right) {| o_{i} |}-\ln \! \left(\delta \right) o_{i} \right) \gamma_{o} \! \left(1\right)-\ln \! \left(z \right) f_{\mathit{no}} \! \left(z \right) n_{i} o_{i} \right)}{\rho}
\end{align}

Once again, the weak mass becomes

\begin{align}
m_{o_{i}}=\frac{E_{o}(z)+U_{n o}^{w}(z)}{c^{2}}=\frac{\alpha_{w} \gamma_{o} \! \left(1\right) \hslash  {| o_{i} |}}{\rho  c} \cdot\left(A +1-\ln \! \left(\delta \right) \, \mathrm{signum}\! \left(o_{i} \right)\right) 
\end{align}

\subsection{The undisturbed cylinder radii}
The undisturbed radius of a cylinder is the radius not altered by any potential energy depending on $z$. The remaining energy is

\begin{align}
E_{\text {undisturbed }}=\frac{c \hbar o \gamma_{o}}{r(z=\infty, 0,1)}=\frac{\alpha c \hbar o \gamma_{o}}{\rho} \cdot F(o, n, \ldots)
\end{align}

, and must be non-zero, not to violate the energy conservation. The solution for $r$ yields

\begin{align}
r=\frac{\rho}{\alpha \cdot F(o, n, \ldots)}
\end{align}

The ansatz for the interactions is

\begin{align}
&F\left(o_{7}, n_{7}\right)=\operatorname{signum}\left(o_{7}\right) \cdot \operatorname{signum}\left(\left|n_{7}\right|\right) \quad \Rightarrow r=\frac{\rho}{\alpha_{e} \cdot\left(\operatorname{signum}\left(o_{7}\right)\right)} \\
&F\left(o_{4,5,6}, n_{4,5,6}\right)=\operatorname{signum}\left(o_{4,5,6}\right) \cdot \operatorname{signum}\left(\left|n_{4,5,6}\right|\right) \Rightarrow r=\frac{\rho}{\alpha_{s} \cdot\left(\operatorname{signum}\left(o_{4,5,6}\right)\right)} \nonumber\\
&F\left(o_{8,9}, n_{8,9}\right)=\operatorname{signum}\left(o_{8,9}\right) \cdot \operatorname{signum}\left(\left|n_{8,9}\right|\right)-\ln (\delta) \nonumber\\
&\Rightarrow\left\{\begin{array}{c}
r=\frac{\rho}{\alpha_{w} \cdot\left(\operatorname{signum}\left(o_{8,9}\right)-\ln (\delta)\right)} \text { if entangled with another particle } \nonumber\\
r=\frac{-\rho}{\alpha_{w} \cdot \ln (\delta)} \text { if not entangled }
\end{array}\right.
\end{align}

The additional term $\ln (\delta)$ is a consequence of the formula of the potential energy (36) of the weak interaction in which the mean distances to the neutrinos and anti-neutrinos play a role.

\begin{align}
U^{w}=\frac{\alpha_{w} c \hbar o_{i} \gamma_{0}}{\rho} \cdot\left(-\ln \left(z_{v_{e}}\right)+\ln \left(z_{\bar{v}_{e}}\right)\right)=-\frac{\alpha_{w} c \hbar o_{i} \cdot \gamma_{o}}{\rho} \cdot \ln (\delta)
\end{align}

This "potential" is independent of $z$ and does not give rise to any force.

\subsection{The particle mass of strong, electric, and weak interaction}
The Lorentz factors for all three interactions at their specific fix points (see appendix 8.9) at which their radii are equal to the undisturbed radii are

\begin{align}
\quad \gamma_{os}(0) = 1\:\:\:\:\: \gamma_{oe}(\infty) = 1\:\:\:\:\: \gamma_{ow}(1) =1
\end{align}

For all the additional dimensions - the three interactions strong, electric, and weak - the total mass $m_{o}$, without the contribution of "magnetism", of a particle is

\begin{align}
&m_{o}=\frac{\hbar}{c \rho} \cdot\left[ \left[ \alpha_{s} \left(\left|o_{4}\right|+\left|o_{5}\right|+\left|o_{6}\right|\right)+\frac{\alpha_{e}\left|o_{7}\right|}{9}\right] \left(1+A\right) \right. \\
&\left.+\frac{\alpha_{w}}{2} \sum_{i=8}^{9} \left|o_{i}\right| \left(\left[1+A-\operatorname{signum}\left(o_{i}\right) \ln (\delta)\right] \right)\right] 
 \nonumber
\end{align}

\subsection{Masses depending solely on the weak sector}
The situation differs somewhat for particles that are affected exclusively by the weak interaction. Only nearby neutrinos and anti-neutrinos effectively alter the value of the weak cylinders' radii. Equation (36) reduces to\\
\begin{align}
U_{o}^{w}(z)=0
\end{align}

Consequently, there is no force

\begin{align}
\vec{F}_{w}=-\frac{1}{\rho} \cdot \frac{\partial}{\partial z} U_{n o}^{w}(z)=0
\end{align}

The energy of the excitation in the rest frame has the identical structure as for the general weak interaction (38) but without the dependency on $z$ :

\begin{align}
E_{o}^{w}=\frac{o_{i} c \hbar}{r_{o_{i}}}+\frac{\alpha_{w} c \hbar A \left|o_{i}\right| }{\rho}
\end{align}

The non-varying cylinder radius is

\begin{align}
r_{o_{i}}=\frac{-\rho}{\alpha_{w} \ln (\delta)}
\end{align}

, which then results in the excitation energy

\begin{align}
E_{o}(z)=\frac{\alpha_{w} c \hbar \left|o_{i}\right|}{\rho} \cdot\left(A-signum(o_{i} )\ln (\delta)\right)\end{align}

The mass becomes $(i=8,9)$

\begin{align}
m_{o_{i}}=\frac{E_{o}(z)+U_{o}^{w}(z)}{c^{2}}=\frac{\alpha_{w} \hbar \left|o_{i}\right|}{c \rho}\left(A-signum(o_{i} )\ln (\delta)\right)
\end{align}

\subsection{Calculation of particle velocity, acceleration, and force}
In the usual four-dimensional spacetime with $u^{\alpha}=\gamma \beta^{\alpha} c$, the expression

\begin{align}
u^{\alpha} u_{\alpha}=c^{2} \text { with } \alpha=0,1,2,3
\end{align}

is conserved. The 10-dimensional version, conserved as well, changes with the number of compact active dimensions - $n_{cd}$. It has the form of

\begin{align}
u^{\mu} u_{\mu}= (1-n_{cd})c^{2} 
\end{align}

The additional dimensions are indexed 4 to 9 . For a coordinate system, with the two particles located on the $z=z^{1}$-axis and $z^{2}=z^{3}=0$, the velocity can be calculated as a function of the distance between the particles. By means of the metric tensor of equation (3), the calculation of the Christoffel symbols can be performed in the usual way with repeated indices summed form 0 to 9.

\begin{align}
\Gamma_{v \xi}^{\mu}=\frac{1}{2} g^{\mu \kappa}\left(\partial_{v} g_{k \xi}+\partial_{\xi} g_{k v}-\partial_{\kappa} g_{v \xi}\right)
\end{align}

The accelerations are calculated by means of the geodesic equation.

\begin{align}
a_{\texttt{s,e,w}}^{\mu}=-\Gamma_{v \xi}^{\mu} u^{v} u^{\xi}
\end{align}

The forces are established as:

\begin{align}
F_{\texttt{s,e,w}}^{\mu}(z)=m_{\texttt{s,e,w}} \cdot a_{\texttt{s,e,w}}^{\mu}
\end{align} For the strong force the off-diagonal elements $g_{\texttt{1i}} = g_{\mathit{i1}} , g_{\texttt{1j}} = g_{\mathit{j1}}$ of the metric tensor with $i,j = 4,5,6$ and $i \neq j$ are non-zero. The electric force comes with the non-zero off-diagonal tensor elements $g_{17} = g_{71}$ and the weak force has the non-zero elements $g_{18} = g_{81} , g_{19} = g_{91}$.

\subsection{Coulomb force between two slow moving charges}
The electric force between two charges $q_{1}=\frac{e}{3} \cdot o_{7}$ and $q_{2}=\frac{e}{3} \cdot n_{7}$ of first-generation particles $-A=0$, which will be shown below - is calculated for slow movement $\beta \cong 0$ and $f_{n o}=-1$ (see appendix 8.7). The radius (18) reduces to:

\begin{align}
r_{o_{7}}(z)=\frac{9 \rho \operatorname{signum}\left(o_{7}\right)}{\alpha_{e} \cdot\left(1+\frac{n_{7} \operatorname{signum}\left(o_{7}\right)}{z} \right)}
\end{align}

The mass $m_{o_{7}}$ associated with equation (20) is used. While the full solution of $(63)$ is rather long, the first terms of the series in $\rho$ are manageable.

\begin{align}
&F^{1}(R) =-m_{o_{7}} \cdot \Gamma^1 _{\mu\nu} u^{\mu} u^{\nu}=-m_{o_{7}} \cdot \Gamma_{77}^{1} \cdot c^{2}+\cdots \\ &=\frac{\alpha_{e} \hslash c n_{7} o_{7}}{9 R^{2}}-\frac{{\alpha_{e}} \hslash c n_{7}^{2} | o_{7}| \rho}{9 R^{3}}+O\! \left(\rho^2 \right) = \frac{q_{o} q_{n}}{4 \pi  \epsilon_{0} R^{2}} - ... \nonumber
\end{align}

$R=\rho \cdot z$ is the distance in meters. $\rho$ is of the order of femtometer as shown below. For distances above $10^{-13} \mathrm{~m}$, the first term of the second line of (65) is dominant, and represents the well-known formula of the Coulomb force of electrostatics with $q_{o}$ and $q_{n}$ the two charges and $\epsilon_{0}$ the vacuum permittivity.

\section{Calculation of the model's constants}
\subsection{The compactification radius $\rho$}
The time for an electron excitation (of length $\lambda=4 \pi |r| / 3$ and velocity $c$) to revolve around the cylinder is:

\begin{align}
\Delta T=\frac{3 \lambda}{c}=\frac{4 \pi |r|}{c}
\end{align}

With this, the electric current calculates as:

\begin{align}
I=-\frac{e}{\Delta T}=-\frac{e c}{4 \pi |r|}
\end{align}

The classic magnetic dipole moment is defined ( $\mathrm{n}=2$ being the winding number, $\mathrm{A}$ is the area the current flows around) as:

\begin{align}
\mu=n I A=-2 \cdot \frac{e c}{4 \pi |r|} \cdot \pi \cdot |r|^2=-\frac{e c |r|}{2}
\end{align}

The magnetic dipole moment of a single electron within the compactified dimension with $|r|= \frac{9 \rho}{\alpha_{e}}$, the Bohr magneton $\mu_{B}=\frac{e \hbar}{2 m_{e}}$, and the z-component of the spin $S_{e}= 3 \hbar / 2$ - the spin of the electric dimension - is:

\begin{align}
\mu_{S}=-\frac{g_{e} \mu_{B} S_{e}}{\hbar}=-\frac{3 g_{e} e \hbar}{4 m_{e}}
\end{align}

Comparing these two expressions gives the value of the compactification radius $\rho$ by using the measured magnetic moment $g_{e}=2.00231930436182(52)$ as:

\begin{align}
\rho=\frac{g_{e} \hbar \alpha_{e}}{6 m_{e} c}=9.4040252(14) \cdot 10^{-16} \mathrm{~m}
\end{align}

\subsection{The weak coupling $\alpha_{w}$ and the neutrino distribution ratio $\delta$}
The mass of the three leptons $e, \mu, \tau$ are calculated according to equation (52):

\begin{align}
m_{e, \mu, \tau}=\frac{\hbar}{c \rho} \cdot\left[\frac{\alpha_{e}}{3}\cdot{(1+ A_{e, \mu, \tau})}+2 \alpha_{w}\cdot(1+\ln (\delta)+ A_{e, \mu, \tau})\right]
\end{align}

The coupling for the  weak interaction $\alpha_{w}$, and the constant $\delta$ are at this point not yet known. But solving the equation (71) for the electron with $A_{e}=0$ (72) and the equation for a positron (73)

\begin{align}
m_{e^{-}}=\frac{\hbar}{c r_{e}}+\frac{\hbar}{c r_{w}}=\frac{\hbar}{c \rho}\left[\frac{\alpha_{e}}{3}+2 \alpha_{w} \cdot(1+\ln (\delta))\right]
\end{align}\\
\begin{align}
m_{e^{+}}=\frac{\hbar}{c r_{e}}+\frac{\hbar}{c r_{w}}=\frac{\hbar}{c \rho}\left[\frac{\alpha_{e}}{3}+2 \alpha_{w} \cdot(1-\ln (\delta))\right]
\end{align}\\
allows determining $\alpha_{w}$ and estimate $\delta$. This is possible using masses of positron and electron and their upper mass difference limit ${\frac { \left| m_{{e^+}}-m_{{e^-}} \right| }{m_{{\it average}}}}
<8 \cdot 10^{-9}$ $[29,37]$.

\begin{gather}
\alpha_{w}=\frac{3 c \rho m_{\text {average }}-\hbar \alpha_{e}}{6 \hbar}=1.41040(26) \cdot 10^{-6} \\
\delta \gtrless \exp \left(\mp \frac{c \rho m_{\text {average }} \cdot 2 \cdot 10^{-9}}{\hbar \alpha_{e}}\right)=\left\{\begin{array}{l}
0.9999965467(64)  \text { if } m_{e^{+}}>m_{e^{-}} \\
1.0000034530(64) \text { if } m_{e^{+}}<m_{e^{-}} 
\end{array}\right.
\end{gather}

It is known that $\alpha_{s}$ is in the order of one and $\alpha_{w}$ is about $10^{-6}[36]$, which is in good agreement with the result of (74), which gives credits to the assumption of $A_{e}=0$. In the wake of it all, $A^{\prime} s$ of the first generation are presumed to be zero. There is a second argument supporting the accuracy. Later in this article the higher dimensional structure of the photon will be deduced. For the photon to be massless $A_{\text {first generation }}=0$ must hold true. From here onward the calculations will be done accordingly: $A_{e}=A_{v_{e}}=A_{d}=A_{u}=0$. To complete the list of constants in terms of the leptons with electric charge, the measured masses of the muon and the tau plus equation (71) are used to calculate the lepton constants (Table 1).

Table 1: Values of the lepton constants

\begin{center}
\begin{tabular}{|c|c|c|}
\hline
$A_{e}=0$ & $A_{\mu}=205.768283(44)$ & $A_{\tau}=3476.22830(74)$ \\
\hline
$A_{v_{e}}=0$ & $A_{v_{\mu}}=$ unknown & $A_{v_{\tau}}=$ unknown \\
\hline
\end{tabular}
\end{center}

The electron's magnetic field energy was neglectable because it is smaller than $2.5 \cdot 10^{-6}$ times the u-quark's field energy (see appendix 8.4).

\subsection{The hadron mass formulas}
By means of equation (52), the strong, electric, and weak mass ratios of the quarks, can be stated.

\begin{align}
& m_{d, s, b}=\frac{\hbar}{c \rho}\left[2 \alpha_{s}+\frac{1}{9} \alpha_{e}\right]\cdot{(1+ A_{d, s, b})+2 \alpha_{w} \cdot(1+\ln (\delta)+ A_{d, s, b})} \\
& m_{\overline{d}, \overline{s}, \overline{b}}=\frac{\hbar}{c \rho}\left[2 \alpha_{s}+\frac{1}{9} \alpha_{e}\right]\cdot{(1+ A_{d, s, b})+2 \alpha_{w} \cdot(1-\ln (\delta)+ A_{d, s, b})} \\
& m_{u, c, t}=\frac{\hbar}{c \rho}\left[2 \alpha_{s}+\frac{2}{9} \alpha_{e}\right]\cdot{(1+ A_{u, c, t})+ \alpha_{w} \cdot(1-\ln (\delta)+ A_{u, c, t})}  \\
& m_{\overline{u}, \overline{c}, \overline{t}}=\frac{\hbar}{c \rho}\left[2 \alpha_{s}+\frac{2}{9} \alpha_{e}\right]\cdot{(1+ A_{u, c, t})+ \alpha_{w} \cdot(1+\ln (\delta)+ A_{u, c, t})}
\end{align}

The differences of the quark masses are not caused exclusively by the strength of the interaction itself, rather it is strongly induced by the constants $A_{d,s,b,u,c,t}$. As previously mentioned, all constants of the first generation are treated as zero $A_{e}=A_{v_{e}}=A_{d}=A_{u}=0$. Not including the magnetic fields, the first-generation meson masses $\widetilde{m}$ become:

\begin{align}
&\widetilde{m}_{\pi^{0}}=\frac{\widetilde{m}_{d \bar{d}}+\widetilde{m}_{u \bar{u}}}{2}=\frac{m_{d}+m_{\bar{d}}}{2}+\frac{m_{u}+m_{\bar{u}}}{2}=\widetilde{m}_{\rho^{0}} \\
&\widetilde{m}_{\pi^{+}}=\widetilde{m}_{u \bar{d}}=m_{u}+m_{\bar{d}}=\widetilde{m}_{\rho^{+}} \nonumber
\end{align}

Each of the quarks and anti-quarks represents a dipole and possesses fields for each excited compactified dimension as well. The energy of a dipole $\vec{\mu}_{o{_{_i}}}$ within the field $\vec{B}_{\mathit{n{_{_i}}}}$ with $i = 4,5,6,7,8,9$ is written as

\begin{align}U_{i}^{\mathit{dipole}}
=-\vec{\mu}_{o_{i}} \cdot \vec{B}_{n_{i}}=\frac{\ c^2 \  |n_{i}| \cdot M }{\alpha_{s},\alpha_{e}/9,\alpha_{w}} 
\end{align}

The energy density of one quark's cylinder excitation $i$ within a given volume is $\frac{B_{i}^2}{2\mu_{0}}$ and following is the energy

\begin{align}
U_{s,e,w}^B=\int \frac{B_{s,e,w}^{2}}{2 \mu_{0}}dV =  
\frac{c^2 |o_{i}| \cdot N }{\alpha_{s},\alpha_{e}/9,\alpha_{w}} 
\end{align}

The details can be found in appendix 8.4. In combination with the other quarks, the fields must first be combined in order to calculate the energy of a meson or baryon. The meaning of the symbols is $\vec{B_{i}}$: fields, $o_{i}$; excitation numbers of quark one; $n_{i}$; excitation numbers of quark two. The mass contributions for mesons and baryons differ.

\subsubsection{The mesons' mass formulas}
For mesons, it can be stated as ($m = \frac{U}{c^2}$):

\begin{align}
m_{q \bar{q}} = 
m_{q}+m_{\bar{q}}+m_{q \bar{q}}^{B}+m_{q\bar{q}}^{\text {dipole }}
\end{align}

The dipole masses for scalar mesons are all negative, which reduces their mass. For vector mesons dipole masses are positive with a mass increasing effect. In addition, there are differences in calculating field contributions for scalar or vector mesons. The calculation for mesons is - starting with parallel spin orientation - a spin flip causes a quark or anti-quark excitation number to be multiplied by $-1$. Hence, the calculation rule for a scalar meson is:

\begin{align}
m_{q \bar{q}}^{B}  -> (99)  \quad m_{q\bar{q}}^{\text {dipole }}->(100)
\end{align}\\
where quark's spin orientation needs to be considered as well as the difference caused when anti-quarks are involved.

Now it is possible to state the mass formulas for the first-generation mesons.

\begin{align}
&m_{\pi^{0}}=
\frac{\left(m_{d} +m_{u} +m_{\bar{d}}+m_{\bar{u}}\right)}{2} +\frac{ 90 N}{\alpha_{e}}+\frac{ 27 M}{\alpha_{e}}\\
&m_{\pi^{+}}=m_{u} +m_{\bar{d}} +\frac{81 N}{\alpha_{e}}+\frac{27 M}{\alpha_{e}} \nonumber\\
&m_{\rho^{0}}=\frac{\left(m_{d} +m_{u} +m_{\bar{d}}+m_{\bar{u}}\right)}{2} -\frac{ 27 M}{\alpha_{e}} \nonumber\\
&m_{\rho^{+}}=m_{u} +m_{\bar{d}} + \frac{9 N}{\alpha_{e}}-\frac{27 M}{\alpha_{e}}\nonumber
\end{align}

By using three mass formulas of (85) together with the respective measured masses, it is possible to calculate the numerical values of the magnetic multipliers and the strong coupling. However, to get the most accurate results it is appropriate to select the particles with most precise mass measurements. Those are the charged pion, the proton, and the neutron. As a consequence first the mass formulas for hadrons are derived.

\subsubsection{The baryons' mass formulas}
Baryons consist of 3 quarks. Those without orbital momentum come as $\hbar / 2$ and $3 \hbar / 2$ spin versions. As with the mesons, the calculations of the field masses differ depending on the spin status. The general mass statement is identical for both versions, but the specific calculation of the various contributions differs. The baryon mass is

\begin{align}
m_{q_{o} q_{n} q_{p}}=m_{q_{o}}+m_{q_{n}}+m_{q_{p}}+m_{q_{o} q_{n} q_{p}}^{B}+m_{q_{o} q_{n} q_{p}}^{\text {dipole }}
\end{align}

The spin $\hbar / 2$ baryons' magnetic field and dipole masses are

\begin{align}
&m_{q_{o} q_{n} q_{p}}^{B}->(103) \quad m_{q_{o} q_{n} q_{p}}^{\text {dipole }}->(104) \end{align}

Here, one of the quarks, the $p$-quark, performed a spinflip, therefore breaking the symmetry. Consequently, different configurations can result in different masses, while the quark content remains the same. An example is the different masses of the $\Lambda^{0}(1116\, MeV)$ and the $\Sigma^{0}(1193\, MeV)$ baryons, both of which consist of one $d$, one $u$ and one $s$ quark. This symmetry breaking together with the selection rules (details see appendix 8.4.2) then causes changes with the calculation of the additional mass terms.\\
For the quark content $duu$, the proton, with one $u$-quark flipped does exist one configuration only. The complete proton mass is

\begin{align}
m_{\mathit{Proton}} = m_{d} +2 m_{u} -\frac{117 N}{\alpha_{e}}-\left(\frac{2}{\alpha_{s}}+\frac{27}{\alpha_{e}}\right) M
\end{align}

, while the neutron, quark content $ddu$ with one $d$-quark flipped, becomes

\begin{align}
m_{\mathit{Neutron}} = 2 m_{d}+m_{u}-\frac{90 N}{\alpha_{e}}-\left(\frac{2}{\alpha_{s}}+\frac{27}{\alpha_{e}}\right) M
\end{align}

Now the three most precise mass measurements of hadrons, those of the charged pion (85b), the proton (88), and the neutron (89) can be compared to calculate the constants  $N$ and $M$ as well as the strong coupling $\alpha_{s}$ (for the hadrons' binding energy).

Table 2: The field multiplier values and the strong coupling\\
\[
\begin{array}{|c|}
\hline M=-1.42892671(58) \cdot 10^{-31}\ {\mathrm{kg}} \\
\hline \mathrm{N}=7.050561(40) \cdot 10^{-34}\ {\mathrm{kg}} \\
\hline \alpha_s=0.51379966(12) \\
\hline
\end{array}
\]

For the spin $3 \hbar / 2$ baryons' field induced masses are found in the appendix as equations

\begin{align}
m_{q_{o} q_{n} q_{p}}^{B}->(105) 
\quad m_{q_{o} q_{n} q_{p}}^{\text {dipole }}->(104) 
\end{align}

The spin and quark settings for $3 \hbar / 2$ baryons must be applied. Each quark combines with their two neighboring quarks; therefore, the order of the quarks is of no concern, and consequently each quark combination has one mass value only.

\section{Particle identification}
The identification of the fermions is straight forward. Color and electric charge are easily set. A quark comes with one of the colors: red, green, or blue, and the respective excitation numbers. The electric excitation number is minus one for a down quark, two for an up quark and minus three for an electron. Each excitation number adds up a spin $\hbar / 2$ times excitation number. To get an overall spin $\hbar / 2$, the weak excitations adjust respectively. A neutrino has weak excitations, only. This particle comes as a superposition of the weak excitations. Hence, the excitation number formally written as $(-1)$, is the same as for the weak contribution of a u-quark (Table 3).

Table 3: Fermions of the first generation\\
\includegraphics[max width=\textwidth, center]{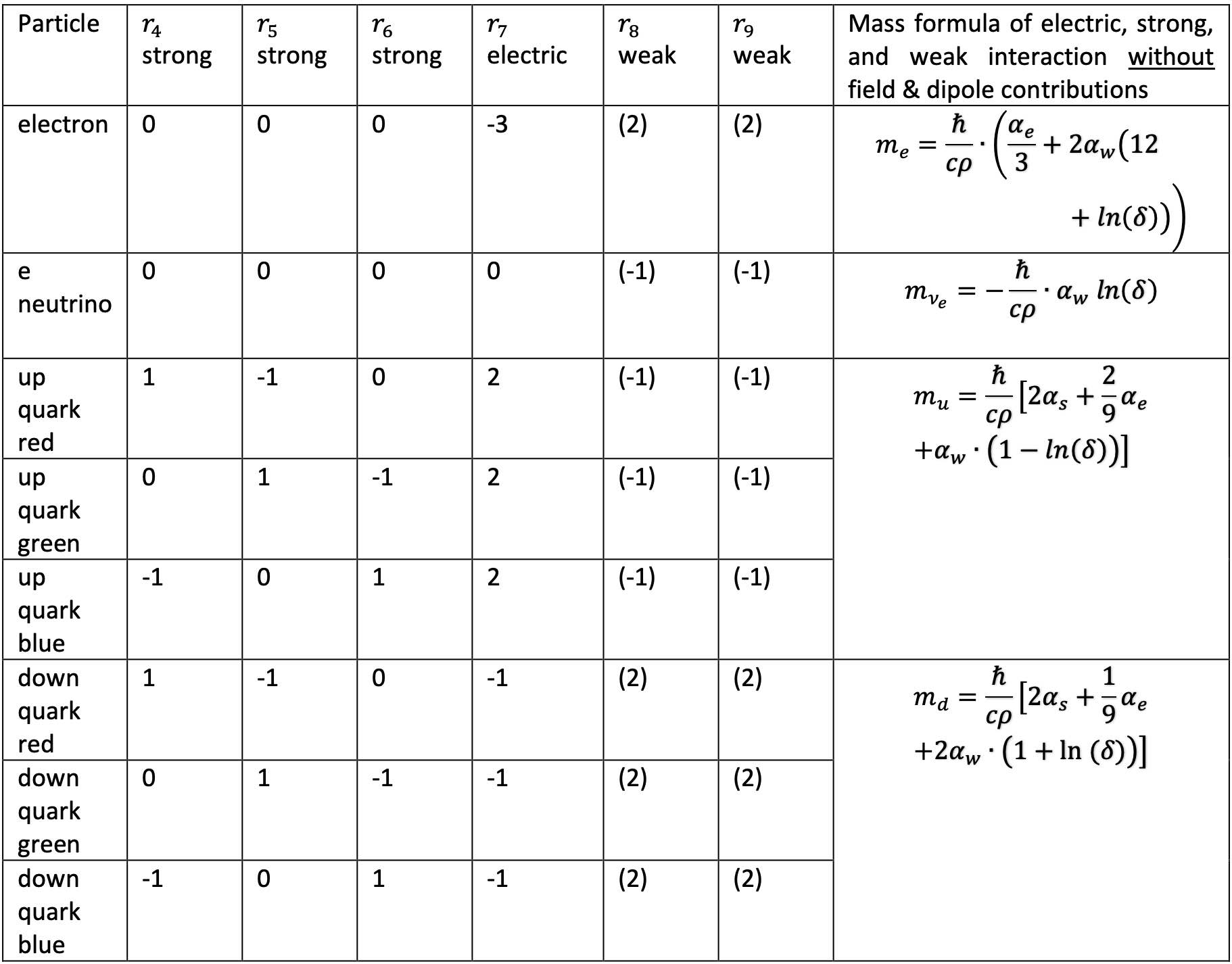}

For the higher generation particles, the excitation settings are identical to the one from the first generation. The difference comes solely with the flavour constants. The respective anti-particles get all the excitation numbers multiplied by $-1$. The Standard Model gauge bosons' excitation numbers can be deduced from known decay processes. In the process of electron-positron annihilation [38] two photons are produced (Figure 1).

$$
e^{+}+e^{-} \rightarrow 2 \gamma \nonumber
$$

$$
\left(\begin{array}{c}
0 \\
0 \\
0 \\
3 \\
(-2) \\
(-2)
\end{array}\right)+\left(\begin{array}{c}
0 \\
0 \\
0 \\
-3 \\
(2) \\
(2)
\end{array}\right) \rightarrow\left(\begin{array}{c}
0 \\
0 \\
0 \\
0 \\
1 \\
-1
\end{array}\right)+\left(\begin{array}{c}
0 \\
0 \\
0 \\
0 \\
-1 \\
1
\end{array}\right)
$$

Figure 1: Electron-positron annihilation into two photons

Two real measurable particles are created. The photons, as can be seen using equation (58), are massless. However, how can the other gauge bosons be accounted for? Do they exist as autonomous particles similar to the photon? The $\eta_{b}$ meson has one decay mode evolving into a three-gluon event (Figure 2).

$$
b \bar{b} \rightarrow 3 g
$$

$$
\left(\begin{array}{c}
1 \\
-1 \\
0 \\
-1 \\
(2) \\
(2)
\end{array}\right)\left(\begin{array}{c}
-1 \\
1 \\
0 \\
1 \\
(-2) \\
(-2)
\end{array}\right) \rightarrow\left(\begin{array}{c}
1 \\
-2 \\
1 \\
0 \\
0 \\
0
\end{array}\right)+\left(\begin{array}{c}
1 \\
1 \\
-2 \\
0 \\
0 \\
0
\end{array}\right)+\left(\begin{array}{c}
-2 \\
1 \\
1 \\
0 \\
0 \\
0
\end{array}\right)
$$

Figure $2: \boldsymbol{\eta}_{b}$ decay into three gluons

The photon and the gluon are presented in Table 4 .

Table 4: Intermediate bosons\\
\includegraphics[max width=\textwidth, center]{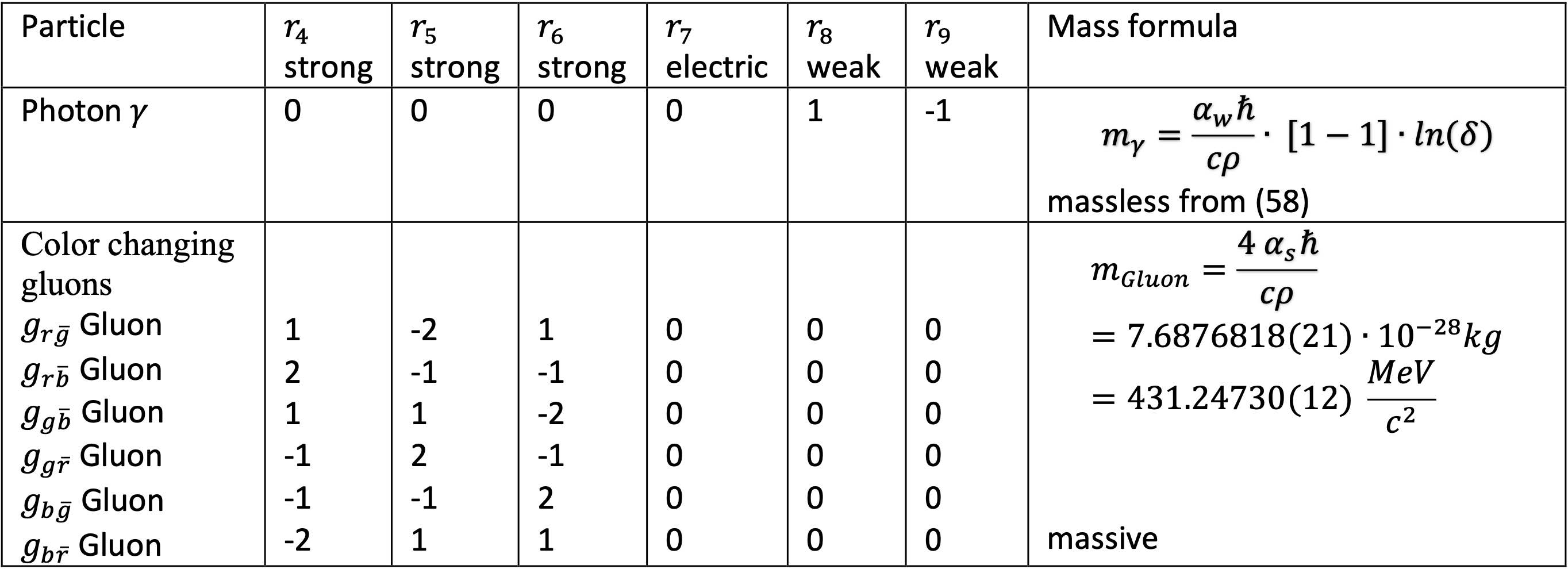}

Masses are created as excitations of the compactified dimensions. An excitation always travels with the speed of light, while manifesting as a mass in ordinary spacetime when induced through the higher dimensions.

\section{Hadron masses}
The field multipliers as well as the couplings are not dependent on particle generation. The quark flavor constants $A_{s,c,b}$ are easily calculated using the formulas and measured masses of the mesons $\phi, \psi, \Upsilon$. That allows to compute the hadron masses. First the meson masses scalar and vector - and afterwards the baryon masses - spin $\hbar / 2$ and $3\hbar / 2$ - are calculated. The search for lists of particles ended by using those of Wikipedia $[39,40]$, where the lowest mass values of the quark combinations are stated. It is not clear, if these mass values, specifically for baryons are the right choice for comparison. However, the measured particle masses shown are used in comparison to the calculated values and to estimate the relative error caused by the uncertainties of the experimentally determined values. All other calculations, however, rest on data of the latest PDG publications [29]. All lists include the particle name, the symbol, quark content, measured and calculated masses in $\mathrm{MeV} / \mathrm{c}^{2}$ and $\mathrm{kg}$, together with the standard deviations, the relative error, and the mass formula. The measured masses in $\mathrm{kg}$ are not stated with standard deviation, since these numbers are no longer used in any further calculation.

\subsection{Meson masses}
The spin of scalar mesons is anti-parallel, hence resulting in a net spin of $S=0$ (Table 5).

Table 5: Masses of scalar mesons\\
\includegraphics[max width=\textwidth, center]{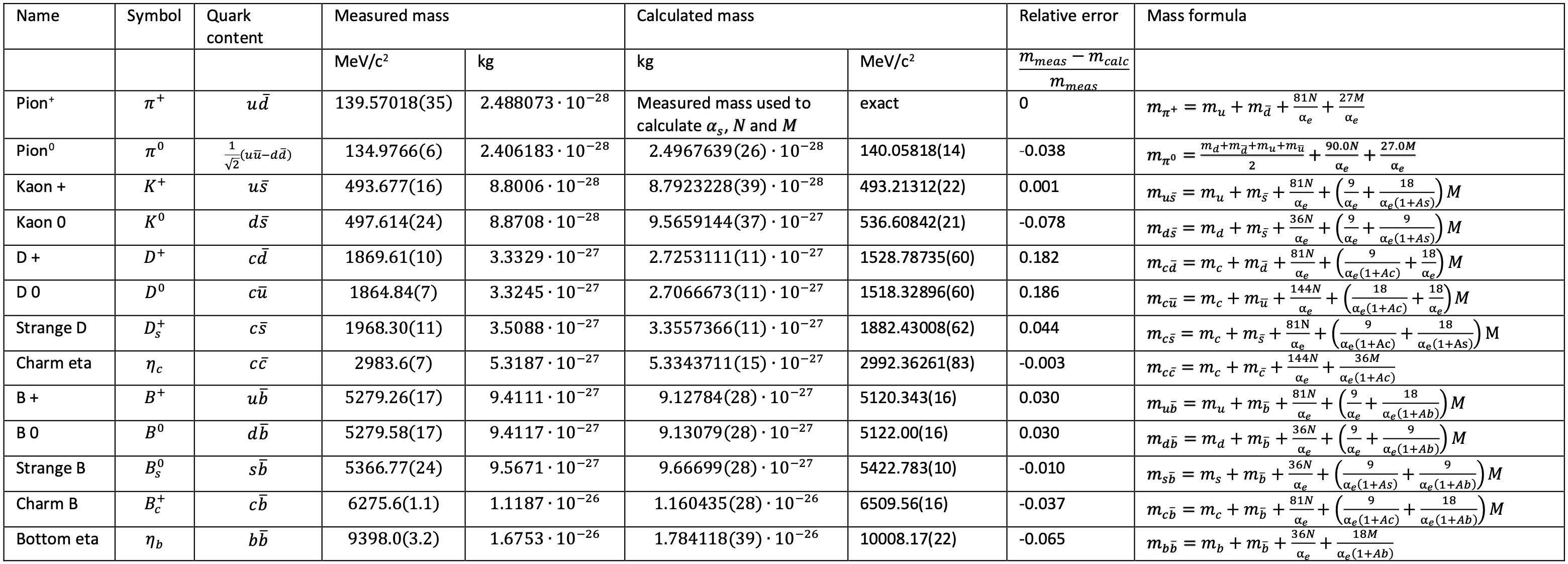}

The vector mesons have a net spin of $S=\hbar$. The quark and the anti-quark spins point in the same direction (Table 6).

Table 6: Masses of vector mesons\\
\includegraphics[max width=\textwidth, center]{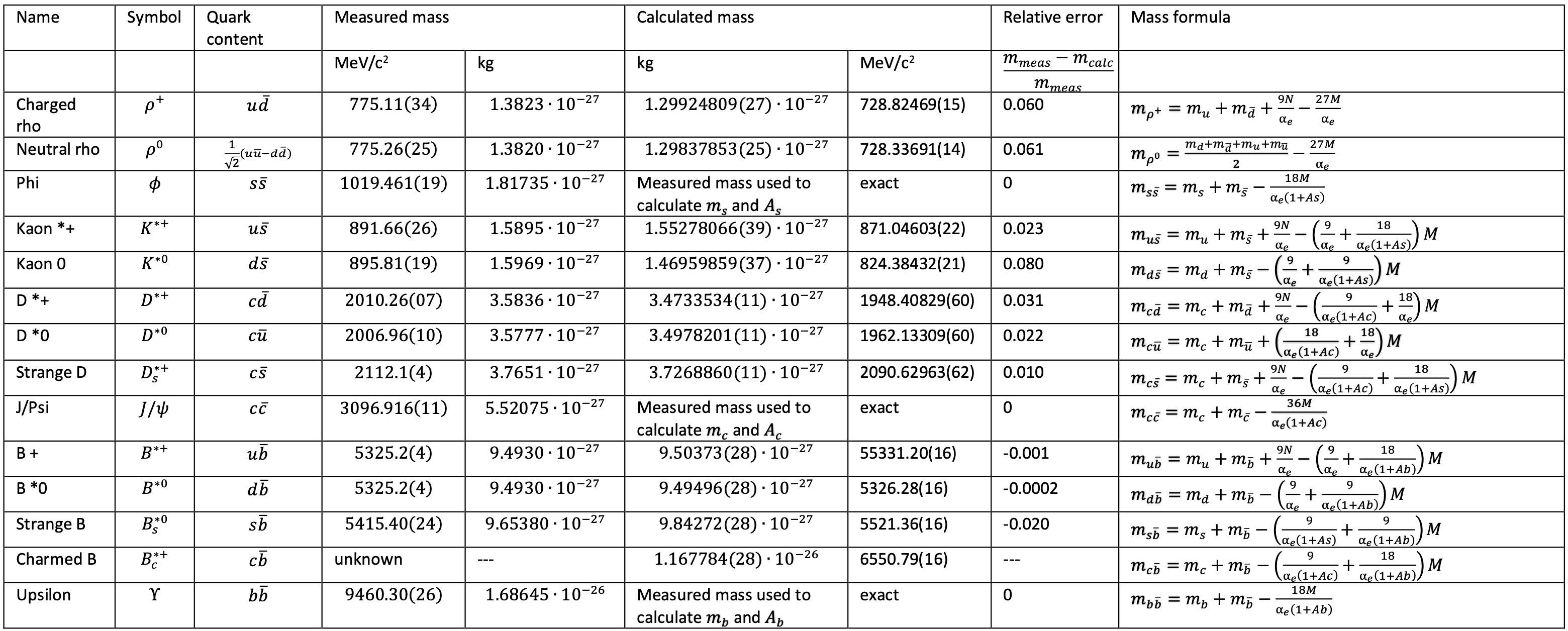}

\subsection{Baryon masses}
Three quarks, each of spin $\hbar / 2$, with one undergone a spin flip resulting in baryons of  spin $\hbar / 2$ (Table 7). The masses are calculated using question (86) with the setting of (87).\\
Table 7: Masses of Baryons with spin $\hbar / 2$\\
\includegraphics[max width=\textwidth]{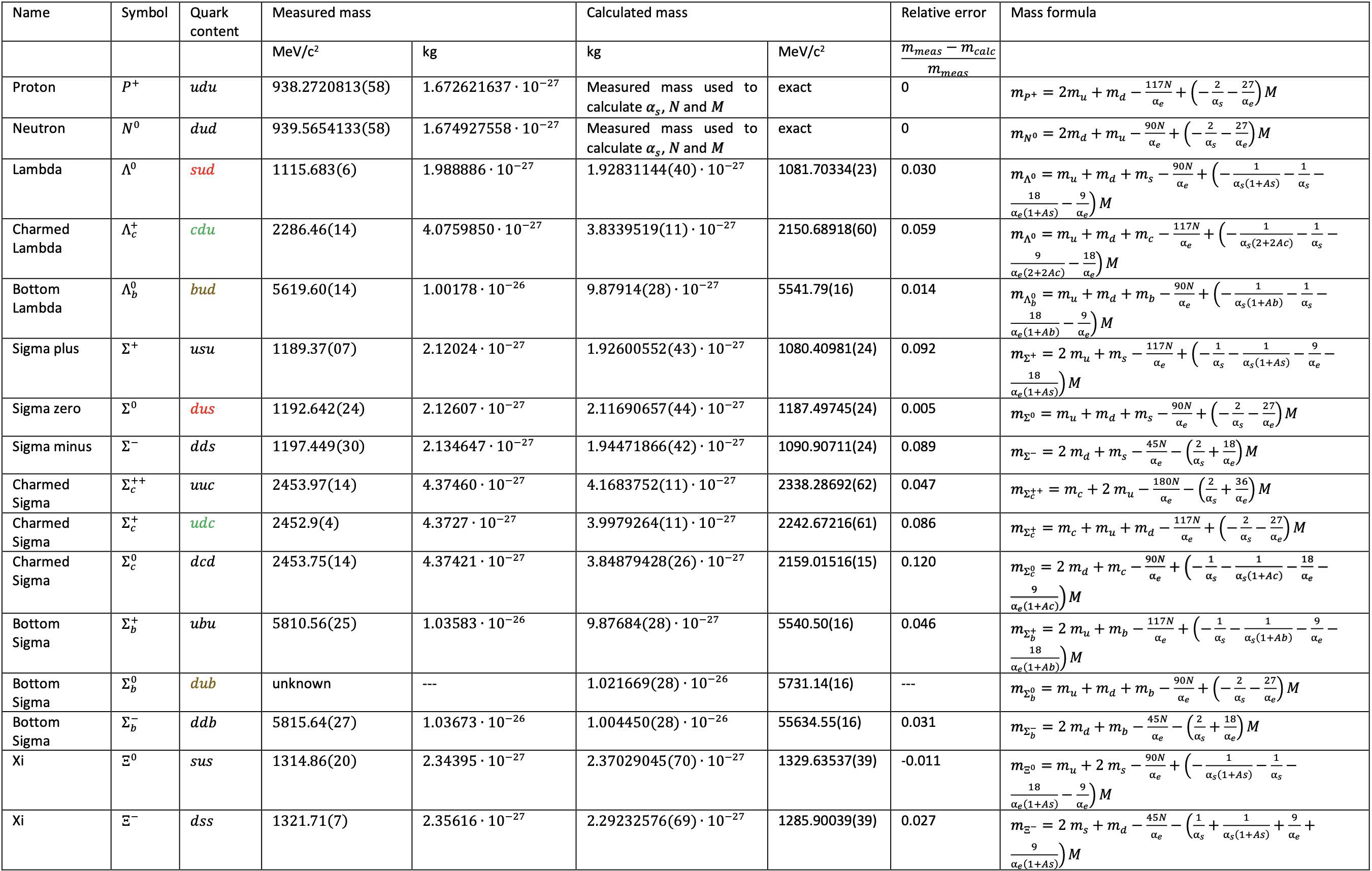}\\
\includegraphics[max width=\textwidth]{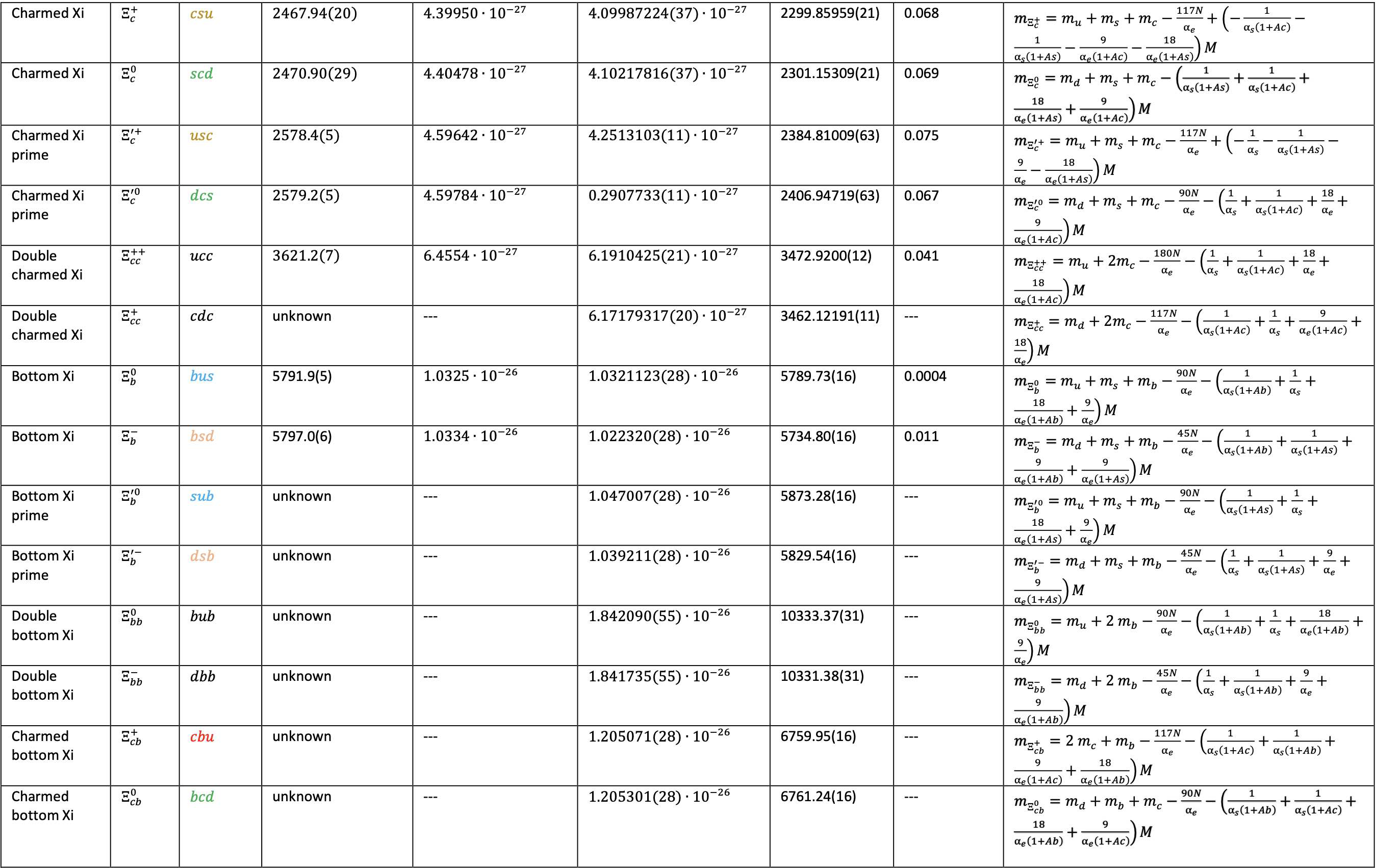}\\
\includegraphics[max width=\textwidth, center]{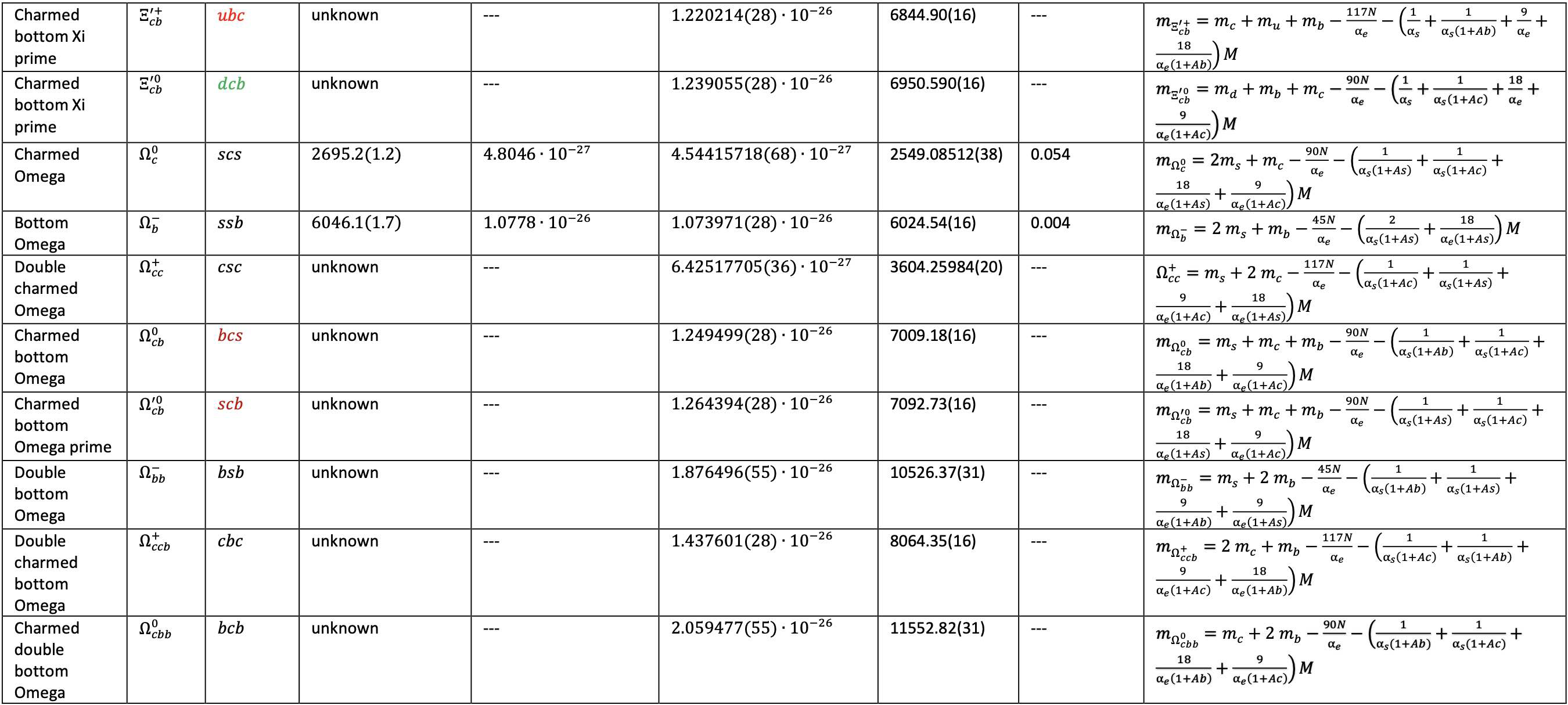}

Three particles, each of spin $\hbar / 2$ all pointing in one direction resulting in baryons of  spin $3 \hbar / 2$ (Table 8). Calculations are done according to equation (86) with setting (90).

Table 8: Masses of Baryons with spin $3 \hbar / 2$\\
\includegraphics[max width=\textwidth]{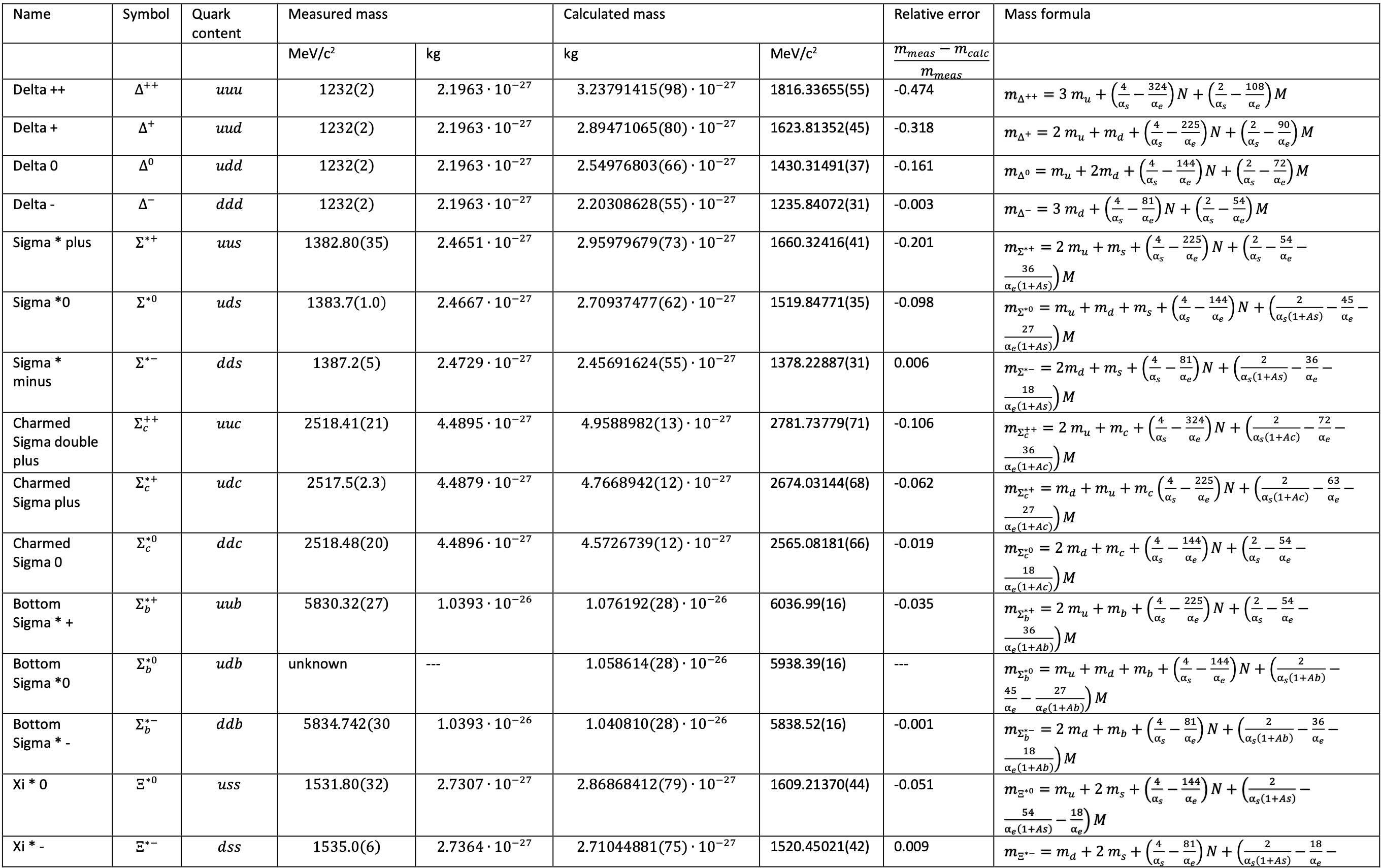}\\
\includegraphics[max width=\textwidth]{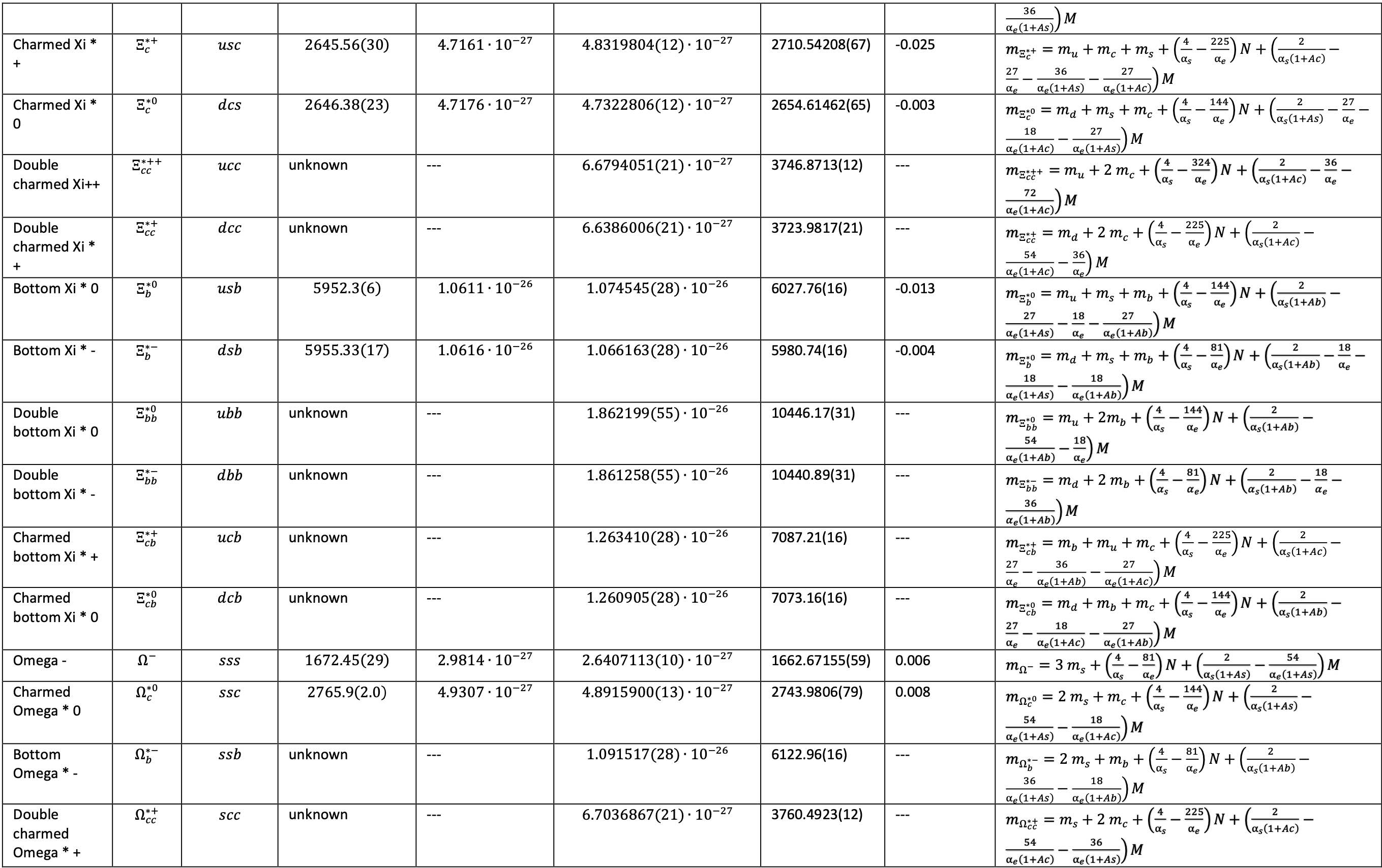}\\
\includegraphics[max width=\textwidth, center]{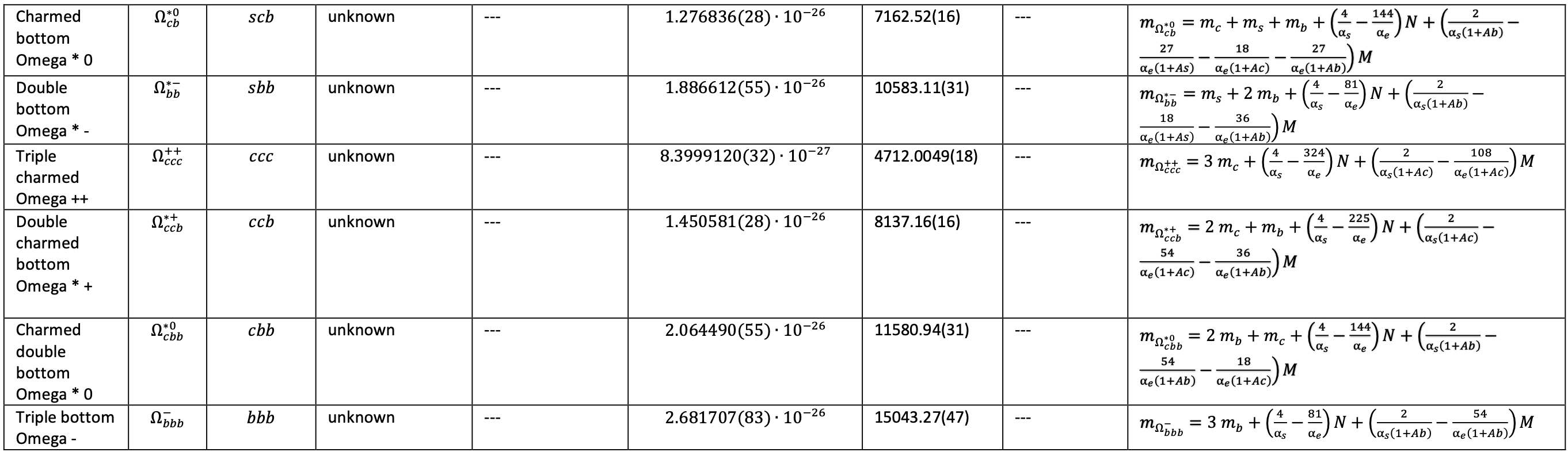}

\subsection{Correlation between measured and calculated hadron masses}
The correlations between measured and calculated hadron masses displays the suitability of the mass calculation (Table 9 and Figure 3).\\
Table 9: Pearson's correlation between measured and calculated hadron masses and the associated significance $p$. Proton, neutron, the charged pion, phi, J/psi, and upsilon used for determining the model's constants are not included within the correlation and not displayed in Figure 3.\\
\[
\begin{array}{|l|l|l|c|}
\hline \text { Number of hadrons } & \text { Particle group } & \text { Pearson's } r & p \\
\hline 12 & \text { Scalar Mesons } & 0.998 & 4.8 \cdot 10^{-13} \\
\hline 10 & \text { Vector Mesons } & 1 & 1.1 \cdot 10^{-15} \\
\hline 22 & \text { Spin 1/2 Baryons } & 0.999 & 4.5 \cdot 10^{-28} \\
\hline 20 & \text { Spin 3/2 Baryons } & 0.996 & 2.6 \cdot 10^{-20} \\
\hline
\end{array}
\]

\includegraphics[max width=\textwidth]{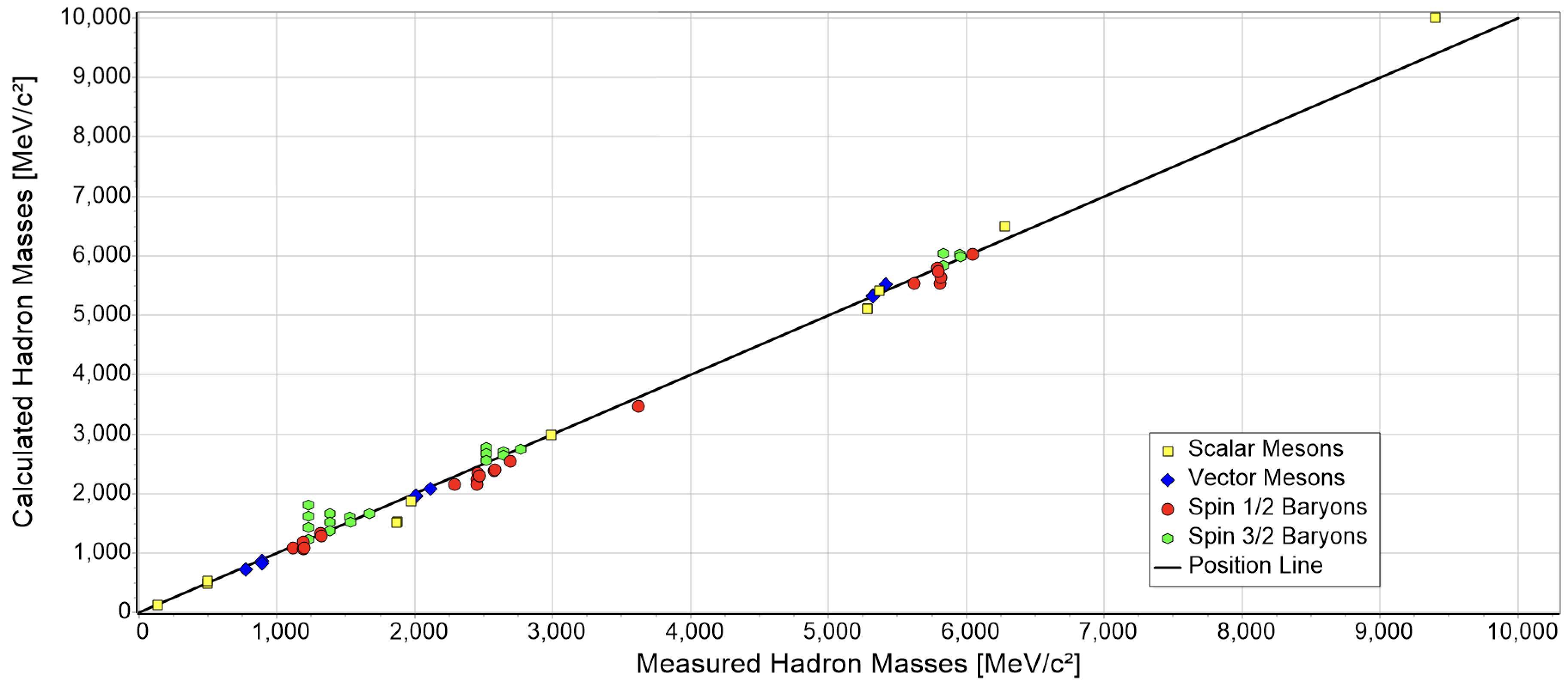} Figure 3: Correlation between measured and calculated masses. The position line marks the location of complete match between measured and calculated masses.

\section{Discussion}
This Kaluza-Klein like model enables the approximation of meson and baryon masses. In 55\% of all comparisons the relative error between measured and calculated masses is smaller than 0.05 and 81\% show deviations below 0.1. The model is based on the structure of the standard model of particle physics with the three interactions strong, electromagnetic, and weak but it does use the mathematical methods of general relativity. Consequently, gravity is an integral part of this 10-dimensional construct. Beside the usual 4-dimensional spacetime there are 6 compactified dimensions responsible for the strong (dimensions 4 to 6), the electric (dimension 7), and weak (dimensions 8 \& 9) interactions. Excitations in 10 dimensions always travel with the speed of light. Stable waves around the cylinders evolve naturally within the framework of the 10-dimensional spacetime and are responsible for the main fraction of mass. A substantial contribution to the particle mass, at least for the light quarks, results from the fields of the 6 compactified dimensions. E.g. the mass contribution from the magnetic fields and the dipoles are: proton ($udu$) 31\%, neutron ($dud$) 31\%, Lambda ($sud$) 25\%, Bottom Xi ($bus$) 18\%, charged rho ($u \overline{d}$) 41\%, upsilon ($b \overline{b}$) 0.09\%. However, there appear to be subtle mass contributing effects, which are not rendered completely. For example, the radii as given in (18), (31), and (44) are expected to be identical and valid for the second and third generation hadrons as well. Within the above stated equations the flavor constants for the second and third generation might not completely cancel for quarks moving relative to the hadron's center of gravity, resulting in slightly deviant radii. The calculation of the field constants $N$,$M$ and the strong coupling $\alpha_{s}$ are done using the most precise measured hadrons masses, those of the pion plus, proton, and neutron. However, other first generation meson masses do display small deviations between measured and calculated masses. Nevertheless, calculating the masses of the composite particles mesons and baryons (Tables 5 to 8), even using the approximated formula, produces sensible results. Differences between measured and calculated masses are small. However, there is the major exception of the abnormally large difference found with the $\Delta^{++}$ baryon, having spin $3 \hbar / 2$. It has a calculated mass of approximately $47 \%$ higher than the measured value. This could be caused by the mass measurement, which does not discriminate between the delta mesons with different electric charge. Nonetheless, it seems that the extremely high magnetic field strength of this triple up quark arrangement combines differently than other baryons with a lower field strength. Something similar can be observed with the charmed sigma double plus $\Sigma_{c}^{++}$, which has an identical field strength. The quark content is uuc, hence the relative influence of its magnetic field on the mass is smaller and has a deviation of $10 \%$. It can be expected to find the mass calculation of the so far not yet measured baryons $ucc$ and $ccc$ being overstated as well. Spin $\hbar / 2$ baryons consisting of three different quark flavors do exist with 2 slightly different masses. This model is capable calculating the different masses. For example, when comparing the $\Lambda^{0}$ to $\Sigma^{0}$, both have the same quark content of $d s u$, but with different mass. During the developmental process of the model, it was not only possible but also necessary to calculate some important constants: the compactification radius $\rho$, the weak $\alpha_{w}$ and the strong $\alpha_{s}$ coupling and $\delta$ the neutrino distribution ratio. $\delta$ evolves from the two compactified dimensions responsible for the weak interaction with a logarithmic dependency on the mean distance to the neutrino/anti-neutrino. In this context $\delta$ is responsible for the extremely low neutrino mass. Therefore, the estimate for the neutrino's upper mass border from the maximal possible mass difference between positron and electron, is $m_{v_{e}}=\frac{2 \hbar}{c \rho} \cdot \alpha_{w} \ln (\delta) \leq$ $0.001 \frac{e V}{c^{2}}$, for the lightest neutrino. It is not clear what the mass of the second or third generation neutrino might be. By taking into account the number received for $m_{v_{e}}$ is the upper bound, the value appears to be in agreement with the upper limits of $1.1 \frac{e V}{c^{2}}$ reported by the KATRINA collaboration [41], or the $0.33 \frac{\mathrm{eV}}{\mathrm{c}^{2}}$ of the astrophysical Chandra observations [42]. However, additional theoretical and experimental work is needed. Could there be areas in space with a different neutrino to anti-neutrino distribution? And can this difference be measured? The particle mass, e.g., of the electron, would change slightly, as would the magnetic moment. In the calculation above, using the mass data from PDG [29] for $e^{-}$and $e^{+}$, the bigger $\delta$ value was used, which means the number of anti-neutrinos is bigger than the one of neutrinos. As a consequence, the neutrino mass is positive and the anti-neutrino mass negative. The overall combined mass of neutrinos and anti-neutrinos is negative. In the event of a smaller number of anti-neutrinos compared to neutrinos, the neutrino mass would be negative and the mass of anti-neutrinos positive, yielding a negative total mass as well.\\
In the context of this paper, the couplings play the roles of constants, since mesons and baryons are described in its 'stable' situation and not in the process of decay. This does not mean that quarks and anti-quarks exist in a static configuration. The hadrons' masses arise as "solid" due to the combination of kinetic, potential, and field energies connected to strong, electric, and weak interaction and the contribution from the magnetic fields and the dipoles.\\
The tensor calculus of general relativity is used for each force separate. This is necessary because masses induced by strong, electric, or weak force are effected by the respective strong, electric, or weak metric tensor only. However, quarks simultaneously are effected by all three forces. As a consequence equation (60) cannot be applied for all three forces within one single metric tensor. But three calculations using equation (63), one for each type of interaction lead to the forces of the strong, electric, and weak interactions, which are given in detail - equations and graphics - in appendix 8.8. The force calculations are possible because invariant function $f_{\mathit{no}} \! \left(z \right) = \frac{w^{\mu} v_{\mu}}{c^{2}}$ was found to be constant but different for leptons, baryons, and mesons with $f_{no}^{leptons}=-1$, $f_{no}^{baryons}=-2$ and $f_{no}^{mesons}=-3$ (see appendix 8.7). As a consequence the particles' radii can be approximated, and via equation (3) the metric tensors too.\\
All forces are calculated using (63), which dependent on the geodesic equation (62) of the 10-dimensional spacetime. In such sense all forces within this model are exclusively caused by the curvature of 10D-spacetime.

\section{Conclusion}
The presented Kaluza-Klein like model uses the simplest version of a Calabi-Yau space, namely a 10-dimensional space with one time, three ordinary space, and 6 compactified dimensions in the form of cylinders. All of them are perpendicular to each other. This 10D-spacetime is a canvas representing all the structure available. No strings or other entities do populate it. But there are disturbances all moving with the speed of light deforming the original topology and violate Ricci flatness. Most of the disturbances are fleeting ripples of the 10D-spacetime but some are stable excitations on the compactified dimensions, still moving at light speed on 10D and at the same time perceived as particles in 4D. A stabile excitation on a compactified dimension of radius $r$ is a sinus wave of wavelength $\lambda$ with $4 \pi r = n \lambda$ and $n$ being an integer. The excitation’s energy is, using de Broglie’s formula [43], $E=\hbar \cdot \omega=\frac{c \hbar n}{r}$. The full energy of a particle is a combination of associated excitations, potential energy plus contributions from generalized magnetic fields and dipoles. Handedness is not considered within this work. The purpose of this paper was to calculate the hadron masses. So far, it was not possible to come up with an exact solution for the masses. However, the developed approximate mass formula yields sensible results. Relative errors of the 64 mass comparisons - measured versus calculated - are mostly below 5 \%. But calculated masses are mainly outside the margin of error of the measured masses. So, what is the conclusion regarding the validity of the model? Are this numbers purely accidental? If so, how big is the probability to get these 64 mass calculations close to the measured masses purely by chance? There are two other means to look at the model’s validity.\\
a. How much tuning does the model need to produce the mass results?\\
b. What other outcomes does the model deliver, which are comparable to measurements or findings of the standard model?\\
a. Input parameters are 9 particle masses (Table 12) and 4 constants (Table 13). There are no free parameters to fix the model’s output.\\
b. Beside the particles’ mass approximation, the model produces the values of the weak $\alpha_{w} \approx  1.410 \cdot 10^{-6}$ and the strong coupling (inside hadrons) $\alpha_{s} \approx  0.5138$, which are in accordance with the expected numbers. The standard model describes the dominating force between the quarks within hadrons via gluon exchange. The gluon mass in compliance with this Kaluza-Klein like model presented is about $431 \frac{MeV}{c^2}$. Using the evolution equation [3] to scale-up to the $Z_{0}$-boson energy yields

\begin{align}
\alpha_s\left(Z_0\right)=\frac{ \alpha_s }{1 + \frac{11 N_c-2 N_f}{12 \pi} \cdot \alpha_s\cdot \ln \left(\frac{m_{Z_0}^2} {m_{g l u o n}^2}\right)}=0.1179
\end{align}

with three colors $N_{c}=3$ and five flavors $N_{f}=5$. The particle data group lists the strong coupling for the $Z_{0}$-boson exchange as $0.1180(9)$, which is in good agreement with the result above. After calculating the invariant expression $f_{no}(z)$ - equation (11) and appendix 8.7 -, which was found to be constant but with different values for leptons, baryons, and mesons the calculations of the strong, electric, and weak forces were done. All equations of the forces are collected in appendix 8.8 as functions of $z$ and $\beta_{o}$. In addition, graphs of the forces are displayed. The strong force is, as expected, the dominating force within hadrons with maximal attractive force values of about $10\ kN$ for baryons and $44\ kN$ for mesons and small force values (asymptotic freedom) for $z < 0.8$ for baryons and $z < 0.6$ for mesons. Electric forces within hadrons are in the range of $10^{-3} N$ for distances above $z > 0.1$. At $z = 1.72 \cdot 10^{-20}$, the equivalence of the Planck-length, the force increases to $> 10^{15} N$ and is repulsive for all quark/quark and quark/anti-quark combinations independent of the signs of the loads! Leptons, however, act in the expected way of $\lim _{z \rightarrow 0} F_{e}\left(z, e^{-} e^{+}\right)=-\infty$ and $\lim _{z \rightarrow 0} F_{e}\left(z, e^{-}e^{-}/e^{+}e^{+}\right)=+\infty$. They do, however, show repulsion independent of the load sign combinations in the range $8 < z < 90$. For $200 < z$ the force follows the Coulomb law. The weak force for $\beta_{o} = 0$ has a strength of about $10^{-12} N$ within hadrons and is zero at $z=0$. But, things get more complicated for non-vanishing $\beta_{o}$! Here the weak force approaches +infinity again for all quark/quark and quark/anti-quark combinations within hadrons as well as lepton combinations. Another interesting aspect of the model is the force between $ud$-quarks within baryons. While the strong force, keeping the quarks combined, acting on each quark with the same strength, the electric and the weak force discriminate between the two flavours. A net force of about $0.5 \ mN$ caused by each $ud$-quark combination acts on a baryon. As a consequence a proton or a neutron, not locked into position by external forces, would perform a "random walk". Similar, a net force is active in $u\overline{d}$ combinations of mesons as well.

The explanation of the various parts of the model is partly satisfying. The masses of the excitations in terms of the waves around the compactified dimensions and the explanation of the associated mass via the de Broglie interpretation seams appropriate and all forces are caused by the curvature of spacetime. The flavour constants describe the masses of the second and third generation quark families. These numbers are no integers, which makes its interpretation difficult. If these constants came as even integers, an interpretation as waves with excitation numbers half positive and half negative were possible. So, there would be equal numbers $n_{p}$ and $n_{n}$ not changing the original spin. Also, combinations with zero net excitation numbers would be possible, resulting in particles with purely gravitational interaction. The understanding of the generalized magnetic fields and the affected dipoles need a construal in terms of the 10D-spacetime, where frame-dragging might be a way to give an explanation in terms of topology.\\
So far, this model was used to calculate hadron masses. Decay processes were not considered. But, the model might have the potential for looking at the creation and decay processes of particles without the use of Feynman diagrams [44].

\section{Appendix}
\subsection{Statistical error analysis}
The newly established system constants are $\rho$ : compactification radius, $\alpha_{w}$ : weak coupling, $\alpha_{s}$ : strong coupling, $\delta$ : neutrino distribution ratio, $M \& N$ : magnetic constants, $A_{d, u, s, c, b, t}$: flavor constants, $A_{e, \mu, \tau, v_{e}, \nu_{\mu}, v_{\tau}}$: lepton constants. To calculate these constants the measured masses (Table 11) of the charged leptons, the proton, the neutron and the mesons $\pi^{+}, \phi, \psi, \Upsilon$ are used as well as the anomalous magnetic dipole moment $g_{e}$ of the electron (Table 13). The statistical deviations of the calculated constants, quark and hadron masses were obtained using Gauss error propagation (92).

\begin{align}
s_{Z}=\sqrt{\left(\frac{\partial Z}{\partial a}\right)^{2} \cdot s_{a}^{2}+\left(\frac{\partial Z}{\partial b}\right)^{2} \cdot s_{b}^{2}+\cdots}
\end{align}

Equation (92) is used to compute the standard deviation of a quantity $Z$, which is calculated from parameters given as the mean values $a, b, \ldots$ and standard deviations $s_{a}, s_{b}, \ldots$ for the numerical result of equations (70), (74), (75), Tables 1, 2, 5, 6, 7, 8, 10, and 11. Error statements are exclusively of statistical nature. No other sources of deviations are included.

\subsection{Parameters calculated within the model}
Table 10: List of derived system constants\\
\includegraphics[max width=\textwidth, center]{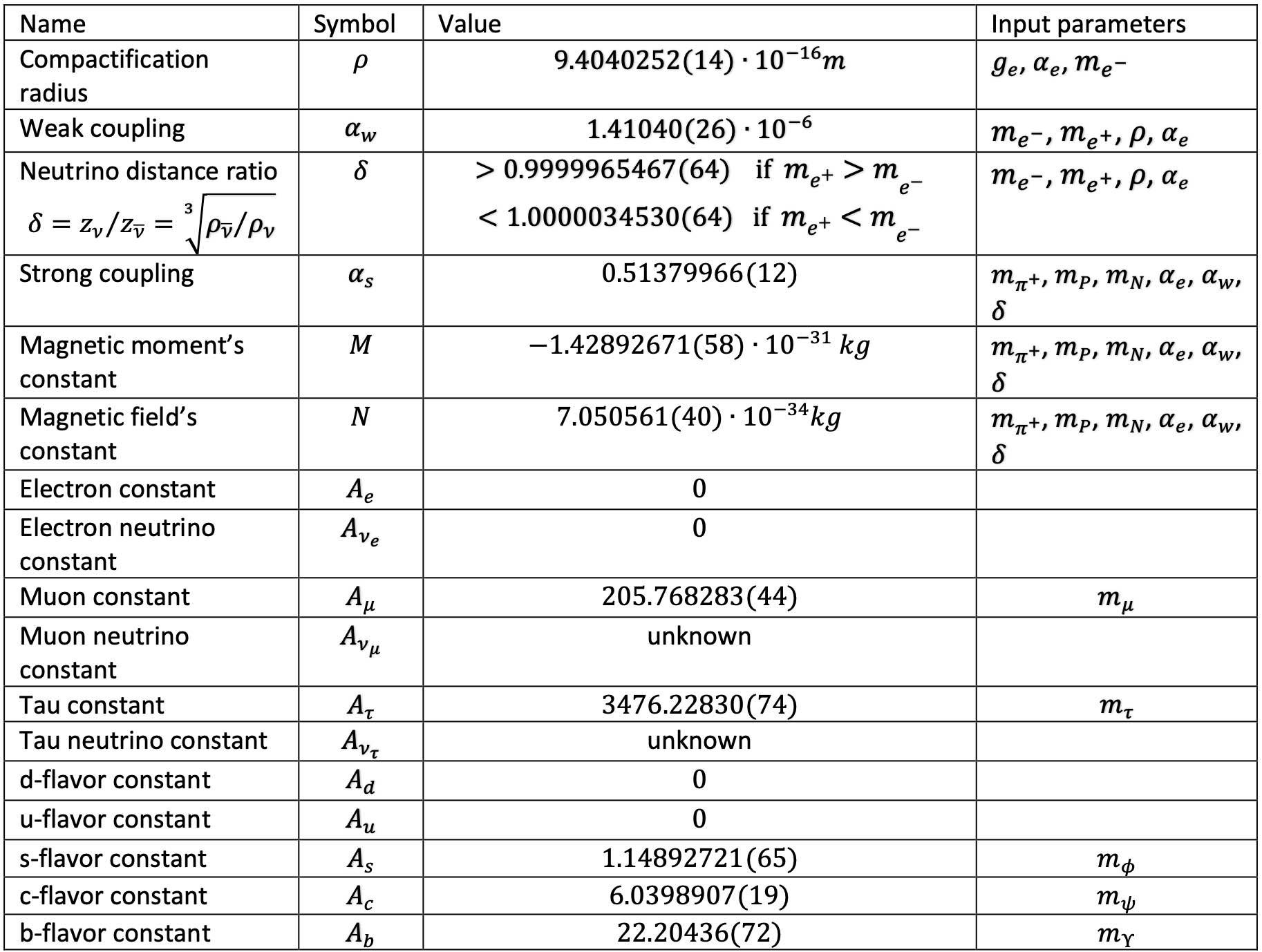}

Table 11: List of quark masses (without contributions of magnetism)\\
\includegraphics[max width=\textwidth, center]{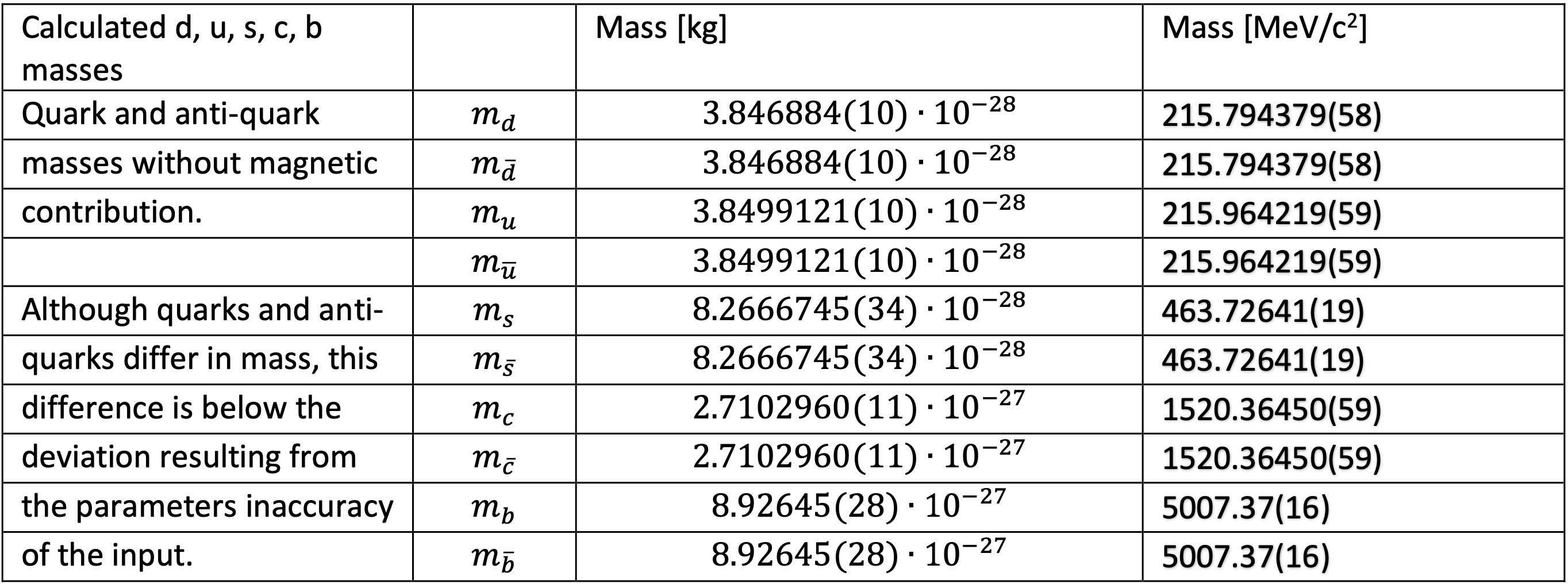}

\subsection{Parameters from literature used in the calculations}
Table 12: List of measured particle masses used in the calculations [29]

\begin{equation}
\begin{array}{|l|c|c|l|}
\hline \begin{array}{l}
\text { Measured particle masses } \\
\text { used within the calculations }
\end{array} & & \text { Mass [kg] } & \mathrm{Mass}\left[\mathrm{MeV} / \mathrm{c}^{2}\right] \\
\hline \text { Electron/Positron } & e & 9.109384033(55) \cdot 10^{-31} & 0.5109989461(31) \\
\hline \text { Muon } & \mu & 1.883531692(43) \cdot 10^{-28} & 105.6583745(24) \\
\hline \text { Tau } & \tau & 3.16754(21) \cdot 10^{-27} & 1776.86(12) \\
\hline \text { Pion }^{+} & \pi^{+} & 2.4880683(32) \cdot 10^{-28} & 139.57039(18)\nonumber \\
\hline \text { Proton} & P & 1.672621637(10) \cdot 10^{-27} & 938.2720813(58) \\
\hline \text { Neutron } & N & 1.674927558(10) \cdot 10^{-27} & 939.5654133(58) \\
\hline \text { Phi } & \phi & 1.817354(29) \cdot 10^{-27} & 1019.461(16) \\
\hline \text { J/Psi } & J / \psi & 5.520726(11) \cdot 10^{-27} & 3096.900(6) \\
\hline \text { Upsilon } & \Upsilon & 1.786808(55) \cdot 10^{-26} & 10023.26(31) \\
\hline
\end{array}
\end{equation}

Table 13: Constants from literature used in the calculations [29]

\begin{center}
\begin{tabular}{|l|c|c|}
\hline
Name & Symbol & Value \\
\hline
Speed of light & $c$ & $299792458 \mathrm{~m} / \mathrm{s}$ \\
\hline
Reduced Planck constant & $\hbar=\frac{h}{2 \pi}$ & $1.054571817 \cdot 10^{-34} \mathrm{Js}$ \\
\hline
Electron's spin g-factor & $g_{e}$ & $2.00231930436182(52)$ \\
\hline
Electric coupling & $\alpha_{e}$ & $7.2973525693(11) \cdot 10^{-3}$ \\
\hline
\end{tabular}
\end{center}

\subsection{Strong, electric, and weak  "magnetic" mass contributions}
The currents,\\
\begin{align}
\vec{I}_{\mathit{o{_{_4,_5,_6}}}} = -\frac{o_{_4,_5,_6} e{_s} c \vec{e}{_{So}}}{4 \pi r_{\mathit{o{_{_4,_5,_6}}}}}
\quad \vec{I}_{\mathit{o{_{_7}}}} = -\frac{o_{7} e{_e} c \vec{e}{_{So}}}{12 \pi r_{\mathit{o{_{_7}}}}} 
\quad \vec{I}_{\mathit{o{_{_8,_9}}}} = -\frac{o_{_8,_9} e{_w} c \vec{e}{_{So}}}{4 \pi r_{\mathit{o{_{_8,_9}}}}}
\end{align}

of the charges (i=4,5,6,7,8,9) divided by the time the excitation needs for a full turn around on a cylinder is the source for the induced fields of the strong, electric, and weak interaction. Here the unit vector $\vec{e}{_{So}}$ indicates the spin-direction of a quark and points either in the positive or negative $z$-direction.

\begin{align}
\vec{B}{_o{_{_i}}}=\frac{\mu_0 \vec{\mu}{_o{_{_i}}}}{2 \pi {r_i}^3} = \frac{2\mu_0 \vec{I}_{\mathit{o{_{_i}}}} A_{\mathit{o{_{_i}}}}  }{2 \pi {r_i}^3}
\end{align}

The cross section of the compactified dimensions are $A_{\mathit{o{_{_i}}}} = \pi \cdot r_{\mathit{o{_{_i}}}}^2$. The $2$ at the beginning of the right term is caused by the property of the cylinders having circumferences of $4 \pi \cdot r_{\mathit{o{_{_i}}}}$. The magnetic momentum creating the magnetic fields of the compactified dimensions is $\vec{\mu}{_o{_{_i}}} = 2 \vec{I}_{o_{i}} A_{o_{i}} = -\frac{o_{i} e_{i} r_{\mathit{oi}} c \vec{e}{_o{_{_i}}}}{k}$ with $k=6$ for the electric and $k=2$ or the other interactions. In contrast, the magnetic moment of a quark effected by magnetic fields is\\
\begin{align}
\vec{\mu}_{s,e,w} = -\frac{g_{{s,e,w}} \mu_{S,E,W} \vec{S}{_q}}{\hslash}
\end{align}\\
With the three analogue of the Bohr magneton $\mu_{S}=\frac{e_{s} \hslash_{s}}{2 m_{q}}$, $\mu_{E}=\frac{\left|o_{7}\right| e_{e} \hbar_{1}}{6 m_{q}}$,  $\mu_{W}=\frac{\left|o_{8}\right| e_{w} \hbar}{2 m_{q}}$ and $A$ the flavour constant. The g-factors are

\begin{align}
g_{{s,e,w}}=\frac{2 m_{q} c r_{\mathit{s,e,w}}}{(1+A) \ o_{s,e,w} \hbar}
\end{align}\\
and $\vec{S}{_q}=\frac{ \hslash_{s} \vec{e}{_o{_{_i}}}}{2}$ the quark's spin. The mass $m_{q}$ is of equation (52), the radii of (18), (31), and (44). In contrast to gyromagnetic ratio $g_{electron}$ precisely known, $g_{s,e,w}$ for quarks are not experimentally determined, but are constructed in analogy to the electron case.\\
Table 14: Particles' g-factors rounded

\begin{center}
\begin{tabular}{|l|c|c|c|}
\hline
Particle & $g_{s}$ & $g_{e}$ & $g_w$ \\
\hline
$\mathrm{d}, \mathrm{s}, \mathrm{b}$ & 36 & 2536 & 729164 \\
\hline
$\mathrm{u}, \mathrm{c}, \mathrm{t}$ & 36 & 1269 & $1.26 \cdot 10^6$ \\
\hline
$\mathrm{e}, \mu, \tau$ & --- & 2.0023 & 1726 \\
\hline
\end{tabular}
\end{center}

The energy of the magnetic fields are proportional to $g_{s,e,w}^{2}$. But, weak fields always cancel (see equations (101), (102), (103), (104)) and strong fields repeal within mesons (equations (101), (102)). As a consequence of $g_{e}$ being much bigger within quarks compared to leptons the magnetic field energy of a u-quark is about a factor $4 \cdot 10^{5}$ bigger than the electron's field energy. The quark radii of the second and third generation having the same values as those of generation one for non-relativistic quark velocities relative to the hadron's center of gravity. The field energy of a meson - a quark/anti-quark pair - or of a baryon's interaction of two quarks with the charges ${o_{4,5,6} e{_s}},\frac{o_{7} e{_e}}{3},o_{8,9} e{_w}$ and ${n_{4,5,6} e{_s}},\frac{n_{7} e{_e}}{3},n_{8,9} e{_w}$ is:

\begin{align}
U_{Bi} = 
\int \frac{\left(B_{\mathit{oi}} +B_{\mathit{ni}} \right)^{2}}{2 \mu_{0}}d V \\ = 
\int \frac{\left(\frac{3 \mu_{0} |o_{i}| e_{i} c e_{\mathit{So}} \rho  }{4 \alpha_{i} \left(1-f_{\mathit{no}} n_{i} \mathrm{signum}\left(o_{i} \right) h_{i} \left(z \right)\right) \pi  \,r_{o}^{3}}+\frac{3 \mu_{0} |n_{i}| e_{i} c e_{\mathit{Sn}} e_{q} \rho }{4 \alpha_{i} \left(1-f_{\mathit{no}} o_{i} \mathrm{signum}\left(n_{i} \right) h_{i} \left(z \right)\right) \pi  \,r_{n}^{3}}\right)^{2}}{2 \mu_{0}}
d V \nonumber
\end{align}

Here $h_{s} \! \left(z \right) = z$, $h_{e} \! \left(z \right) = \frac{1}{z}$, and $h_{w} \! \left(z \right) = \ln \! \left(z \right)$, $\mu_{0}$, the permeability of vacuum. Following, the energy of the quarks magnetic dipole moments within the magnetic fields of the associated quark can be stated as

\begin{align}
U_{Mi} = -{\mu_{\mathit{o_{i}}} B_{\mathit{n_{i}}} - {\mu_{\mathit{n_{i}}}} B_{\mathit{o_{i}}}} \\ =
-\frac{\left(\frac{e_{\mathit{So}} {| n_{i} |} e_{\mathit{Sn}} e_{q}}{\mathit{A1} +1}+\frac{e_{\mathit{Sn}} e_{q} {| o_{i} |} e_{\mathit{So}}}{\mathit{A2} +1}\right) l c \hslash}{2 \alpha_{i} \left(1-f_{\mathit{no}} n_{i} \mathrm{signum}\! \left(o_{i} \right) h_{i} \! \left(z \right)\right) \left(1-f_{\mathit{no}} o_{i} \mathrm{signum}\! \left(n_{i} \right) h_{i} \! \left(z \right)\right) \rho  \,z^{3}} \nonumber
 \end{align}

with $l=9$ for the electric interaction and $l=1$ for the others. $e_{\mathit{So}}$ and $e_{\mathit{Sn}}$ stand for the quark spin up if equal to 1 and -1 for spin down. $e_{q}$ is 1 for a quark and -1 for an anti-quark and the substitution $\alpha_{i} = \frac{\mu_{0} e{_i}^{2} c}{4 \pi  \hslash}$ is used. The expression of (97) could not be integrated fully. However, an approximated version exhibits a $1/z^3$ dependency as (98) does as well. It is assumed that $U_{Bi}+U_{Mi} = constant$ and not dependent on $z$. By choosing  $h_{s,e} \approx 0$ and $h_{w} \approx 1$ the additional mass contributions caused by the magnetic fields and the dipoles within fields are approximated as\begin{align}
m_{q \bar{q}}^{Bi} = \frac{U_{Bi}}{c^2}=
\frac{l \left(|o_{i}| e_{\mathit{So}} +|n_{i}| e_{\mathit{Sn}} e_{q} \right)^{2}}{\alpha_{i}} \cdot N 
\end{align} and \begin{align}
m_{\mathit{oiM}} = -\frac{U_{Mi}}{c^2}= \left(\frac{\  e_{\mathit{So}} |n_{i}| e_{\mathit{Sn}} e_{q} }{\left(1+\mathit{A1} \right)}  +\frac{\  e_{\mathit{Sn}} e_{q} |o_{i}| e_{\mathit{So}} }{ \left(1+\mathit{A2} \right)}\right) \cdot \frac{l \cdot M}{\alpha_{i}}
\end{align}

$N$ and $M$ are two constants, which are calculated simultaneously with $\alpha_{s}$ by comparing the calculated and the measured masses of the proton, neutron, and charged pion.

\subsubsection{Scalar and vector meson contribution}
The full mass contributions of the mesons' field energies are with the spin setting $e_{\mathit{So}}, e_{\mathit{Sn}} = \pm1$ with 1: not flipped, -1: flipped,  $e_{\mathit{q}} = 1$ for quarks, -1 for anti-quarks:\\
\begin{align}
m_{q \bar{q}}^{B} =\frac{9 \left(|o_{7}| e_{\mathit{So}} \! + |n_{7}| e_{\mathit{Sn}} e_{q} \! \right)^{2} \cdot N}{\alpha_{e} }  
\end{align}

, while the dipole energies are

\begin{align}
m_{q \bar{q}}^{dipole} = \left(\frac{9  e_{\mathit{So}} |n_{7}| e_{\mathit{Sn}} e_{q} }{\alpha_{e} \left(1+\mathit{A1} \right)}+\frac{9  e_{\mathit{Sn}} e_{q} |o_{7}| e_{\mathit{So}} }{\alpha_{e} \left(1+\mathit{A2} \right)}\right) \cdot M 
\end{align}

$A1$ and $A2$ stand for the flavor constants.

\subsubsection{Spin $\frac{ \hslash}{2}$ baryon contribution}
The additional masses for baryons with the $p$-quark being flipped are

\begin{align}
m_{\mathit{qqq}}^{B} = 
\left[\frac{\left({|n_{4}}|  +|o_{4}|  \right)^{2}+p_{4}^{2} }{\alpha_{s}}
-\frac{\left(|n_{5}| +|o_{5}| \right)^{2}+p_{5}^{2} }{\alpha_{s}}\right. 
\\
\left.+\frac{\left(|n_{6}| +|o_{6}| e_{\mathit{So}}  \right)^{2}+p_{6}^{2} }{\alpha_{s}} \right. 
\left.-\frac{9 \left(|o_{7}| +|n_{7}| e_{\mathit{Sn}} \right)^{2}+p_{7}^{2} }{\alpha_{e}} \right.\nonumber 
\\ \left.+\frac{\left(|o_{8}| +|n_{8}|  \right)^{2}+p_{8}^{2} }{2\alpha_{w}}-\frac{\left(|o_{9}| +|n_{9}| \right)^{2}+p_{9}^{2} }{2\alpha_{w}}\right] \cdot N \nonumber
\end{align}

In equation (104) the following setting takes place: $e_{\mathit{So}} = 1$, $e_{\mathit{Sn}} = 1$, $e_{\mathit{Sp}} = 0$. The definition $\mathrm{signum} \left(0 \right) = 0$ is used.

\begin{align}
m_{\mathit{qqq}}^{B} = \sum_{i=4}^{9}(-1)^{i} \left[ \frac{|\mathrm{signum}\! \left(o_{i} \right)| e_{\mathit{So}} \left(|n_{i}| e_{\mathit{Sn}} +|p_{i}| e_{\mathit{Sp}} \right)}{\alpha_{i} \left(1+\mathit{A1} \right)} \right.
\\
+\left. \frac{|\mathrm{signum}\! \left(n_{i} \right)| e_{\mathit{Sn}} \left(|o_{i}| e_{\mathit{So}} +|p_{i}| e_{\mathit{Sp}} \right)}{\alpha_{i} \left(1+\mathit{A2} \right)} \right. \nonumber 
\\
+\left. \frac{|\mathrm{signum}\! \left(p_{i} \right)| e_{\mathit{Sp}} \left(|n_{i}| e_{\mathit{Sn}} +|o_{i}|e_{\mathit{So}} \right)}{\alpha_{i} \left(1+\mathit{A3} \right)} \right] \nonumber
\\ 
\left(1+8 \cdot \delta \left(i-7 \right) - \frac{\delta \left(i-8 \right) + \delta \left(i-9 \right)}{2} \right) \cdot M \nonumber
\end{align}

With $\alpha_{4,5,6}=\alpha_{s}$, $\alpha_{7}=\alpha_{e}$, $\alpha_{8,9}=\alpha_{w}$.

In addition the following selection rules for $\frac{\hslash}{2}$-baryons take place:

\begin{enumerate}
  \item For baryons consisting of a mixture of u-like (u, c, t) and d-like (d, s, b) quarks, the non-flipped quarks always represent a u-d-like combination.
  \item In case of a baryon consisting of a u-d-mixture and build of three different quark species, there always exist two combinations with a different quark flipped. These are the spin $\frac{\hslash}{2}$ baryons with identical excitation numbers but dissimilar mass.
  \item For those $\frac{\hslash}{2}$-baryons consisting of purely u- or d-like quarks the heaviest quark is flipped.
\end{enumerate}

\subsubsection{Spin $\frac{3 \hslash}{2}$ baryon contribution}
For spin $\frac{3 \hslash}{2}$ baryons the mass created by the magnetic field energy is

\begin{align}
m_{\mathit{qqq}}^{B} = \sum_{i=4}^{9}(-1)^{i} 
\left[\frac{\left(|n_{i}| + |o_{i}| + |p_{i}| \right)^{2}}{\alpha_{i}} \right] \\
\left(1+8 \cdot \delta \left(i-7 \right) - \frac{\delta \left(i-8 \right) + \delta \left(i-9 \right)}{2} \right) \cdot N
\nonumber
\end{align}

The additional mass $m_{\mathit{qqq}}^{dipole}$ caused by the dipoles within the fields of the other two quarks is identical to (104) but with the different setting of $e_{\mathit{So}} = 1$, $e_{\mathit{Sn}} = 1$, $e_{\mathit{Sp}} = 1$.

\subsection{Coordinates and velocities within the ten dimensions}
The coordinate origin is chosen to be at the entangled partner-particle location. So, $z^{1}=z$ is the normalized distance between two entangled particles. The normalized distance between the two particles is of interest only. Therefore, the coordinate systems $z^{1}$-axis is selected to point in the direction of the particle and the two other spatial coordinates are zero. Within the compact dimensions a stable excitation is spread out onto the complete cycle $4 \pi  r_{i}$. A specific point on a compactified coordinate is described below with $0\le K <1$.

\begin{align}
z^{1} = z\:\:\:\:z^{2} = z^{3}=0\:\:\:\:z^{i} = \frac{K \cdot 4 \pi  r_{i}}{\rho}\:\:\:\:i = 4, ..., 9 
\end{align}

The velocities in 3-space is smaller $c$ for massive particle (induced by the higher dimensions) and equal to the speed of light for the photon and excitations of the 4D-spacetime (gravitational waves). Within the higher dimensions velocity is always the speed of light.

\begin{align}
u^{0} = \gamma  c\:\:\:\:u^{1} = \gamma  \beta  c\:\:\:\:u^{2} = u^{3}=0\\ u^{i} = |signum(o_{i})| \ c\:\:\:\: i = 4, ..., 9 \nonumber
\end{align}\\
Also, see comment below equation (11).

\subsection{The metric tensor}
The components of the metric tensor are influenced by the distance to the next particle and marginally, but omitted below (112), by the average distances to the next neutrino and anti-neutrino. There exist three metric tensors, one for each interaction with the general form

\begin{align}
g_{\mu \nu}^{s,e,w} \! \left(\mathit{z} \right)= \nonumber
\\ \left[\begin{array}{cccccccccc}
1 & 0 & 0 & 0 & 0 & 0 & 0 & 0 & 0 & 0 
\\
 0 & g_{11}^{s,e,w} \! \left(\mathit{z} \right) & 0 & 0 & g_{14}^{s} \! \left(\mathit{z} \right) & g_{15}^{s} \! \left(\mathit{z} \right) & g_{16}^{s} \! \left(\mathit{z} \right) & g_{17}^{e} \! \left(\mathit{z} \right) & \frac{g_{18}^{w} \! \left(\mathit{z} \right)}{2} & \frac{g_{19}^{w} \! \left(\mathit{z} \right)}{2} 
 0  
\\
 0 & 0 & -1 & 0 & 0 & 0 & 0 & 0 & 0 & 0 
\\
 0 & 0 & 0 & -1 & 0 & 0 & 0 & 0 & 0 & 0 
\\
 0 & g_{41}^{s} \! \left(\mathit{z} \right) & 0 & 0 & -1 & 0 & 0 & 0 & 0 & 0 
\\
 0 & g_{51}^{s} \! \left(\mathit{z} \right) & 0 & 0 & 0 & -1 & 0 & 0 & 0 & 0 
\\
 0 & g_{61}^{s} \! \left(\mathit{z} \right) & 0 & 0 & 0 & 0 & -1 & 0 & 0 & 0 
\\
 0 & g_{71}^{e} \! \left(\mathit{z} \right) & 0 & 0 & 0 & 0 & 0 & -1 & 0 & 0 
\\
 0 & \frac{g_{81}^{w} \! \left(\mathit{z} \right)}{2} & 0 & 0 & 0 & 0 & 0 & 0 & -1 & 0 
\\
 0 & \frac{g_{91}^{w} \! \left(\mathit{z} \right)}{2} & 0 & 0 & 0 & 0 & 0 & 0 & 0 & -1 
\end{array}\right] 
\end{align}

For $1\ll z$ all off-diagonal elements approaching zero and $g_{11} \! \left(\mathit{1\ll z} \right) = -1$. The components can be calculated directly using equation (3). The specific masses related to the strong, electric, and weak interactions couple to their interaction specific metric tensor only. E.g. for the electric interaction the non-vanishing two off-diagonal elements are $g_{71}\! \left(\mathit{z} \right)=g_{17}\! \left(\mathit{z} \right)$. Also, there exist different diagonal elements for each interaction (108).

\begin{align}
g_{11}^{s} \! \left(z \right)\:\:\:\:\:\:\:\:g_{11}^{e} \! \left(z \right)\:\:\:\:\:\:\:\:g_{11}^{w} \! \left(z \right)
\end{align}

The off-diagonal elements are of the form:\\
\begin{align}g_{14}^{s} \! \left(z , q q (q)\right) = \nonumber
\\
-\frac{z^{4} \left(\left(-1-f_{\mathit{no}} \! \left(z \right) z \right) \left(\frac{d}{d z}\gamma_{o} \! \left(z \right)\right)+\gamma_{o} \! \left(z \right) \left(z \left(\frac{d}{d z}f_{\mathit{no}} \! \left(z \right)\right)+f_{\mathit{no}} \! \left(z \right)\right)\right)}{\gamma_{o} \! \left(z \right) \left(1+f_{\mathit{no}} \! \left(z \right) z \right)}
\end{align}

\begin{align}g_{17}^{e} \! \left(z , d d(q)\right) = \nonumber 
\\
\frac{\left(z \left(z -f_{\mathit{no}} \! \left(z \right)\right) \left(\frac{d}{d z}\gamma_{o} \! \left(z \right)\right)+\gamma_{o} \! \left(z \right) \left(z \left(\frac{d}{d z}f_{\mathit{no}} \! \left(z \right)\right)-f_{\mathit{no}} \! \left(z \right)\right)\right) \mathit{z}^{7}}{\gamma_{o} \! \left(z \right) z \left(z -f_{\mathit{no}} \! \left(z \right)\right)}
\end{align}

\begin{align}
g_{18}^{w} \! \left(z , d d(q)\right) = \frac{1}{\gamma_{o} \! \left(z \right) \left(2 f_{\mathit{no}} \! \left(z \right) \ln \! \left(z \right)-1\right)}
\\
\cdot z^{8} \left[2 f_{\mathit{no}} \! \left(z \right) \ln \! \left(z \right) \left(\frac{d}{d z}\gamma_{o} \! \left(z \right)\right) z -2 \ln \! \left(z \right) \gamma_{o} \! \left(z \right) \left(\frac{d}{d z}f_{\mathit{no}} \! \left(z \right)\right) z \right. \nonumber
\\
\left. -2 \gamma_{o} \! \left(z \right) f_{\mathit{no}} \! \left(z \right)-\left(\frac{d}{d z}\gamma_{o} \! \left(z \right)\right) z \right] \nonumber
\end{align}\\
$4$,$7$,$8$ represent compactified dimensions' contravariant coordinate indices.

\subsection{The invariant expression $f_{\mathit{no}} \! \left(z \right) = \frac{w^{\mu} v_{\mu}}{c^{2}}$  }
The numerical value of the expression $f_{\text {no }}(z)$ for charged leptons at rest is\\
\begin{align}
f_{\text {no }}(z,\beta_{o}^{1} =  0,\beta_{n}^{1} =  0) = -1 \\ = g_{00} + g_{77}  +  g_{kk} \nonumber
\end{align}

With $k=8,9$ stands for superposition of the weak dimensions.\\
For quarks within a baryon (at rest) in respect of one of the two other quarks

\begin{align}
f_{\text {no }}(z,\beta_{o}^{1} =  0,\beta_{n}^{1} =  0) = -2 \\ = g_{00} + g_{ii}  + g_{77}  + g_{kk} \nonumber
\end{align}

and for quarks within a meson (at rest)\\
\begin{align}
f_{\text {no }}(z,\beta_{o}^{1} =  0,\beta_{n}^{1} =  0) = -3 \\ = g_{00} + g_{ii}  + g_{jj}  + g_{77}  + g_{kk} \nonumber
\end{align}

, depends on the active color - $i,j=4,5,6$ with $i \neq j$ - and the contribution of the weak interaction. For non-zero velocities $f_{\text {no }}(z)$ must result in one single value for a quark valid for each kind of interaction. Following $f_{\text {no }}(z)$ is calculated for each force separately. The calculation of $f_{\text {no }}(z)$ follows a common procedure. It starts with the radii (18), (31), or (44) as input for (1). With the help of (3) the metric tensors are calculated. Next step is using (60) to compute the normalized velocities, which are dependent on $f_{\text {no }}(z)$. The normalized velocities $\beta_{o}$ and $\beta_{n}$ separately calculated for the strong, electric, and weak interactions are inserted into $f_{\mathit{no}} \! \left(z \right) = \frac{w^{\mu} v_{\mu}}{c^{2}}$. This results in differential equations of $f_{\mathit{no}} \! \left(z \right)$ for lepton combinations, quark/quark combination of baryons and quark/anti-quark combinations of mesons. The solutions depend on the coordinate of the higher dimensions. With the setting $z^{i} = \frac{K 4 \pi  \,r^{i}(z)}{\rho}$ and $0\le K < 1$ always two solutions exist $f_{\mathit{no}}^{leptons} \! \left(z \right) = -1$, $f_{\mathit{no}}^{leptons} \! \left(\mathit{z} \right) = -\frac{N_{1} K^{2} \pi^{2}+1}{N_{2} K^{2} \pi^{2}+1}$, $f_{\mathit{no}}^{baryons} \! \left(z \right) = -2$, $f_{\mathit{no}}^{baryons} \! \left(\mathit{z} \right) = -\frac{N_{1} K^{2} \pi^{2}+2}{N_{2} K^{2} \pi^{2}+1}$ and $f_{\mathit{no}}^{mesons} \! \left(z \right) = -3$, $f_{\mathit{no}}^{mesons} \! \left(\mathit{z} \right) = -\frac{N_{1} K^{2} \pi^{2}+3}{N_{2} K^{2} \pi^{2}+1}$ with $N_{1},N_{2} \in \mathbb{N}$. For $\beta_{o}$ approaching zero $\lim _{\beta_{o} \rightarrow 0}\left(-\frac{N_{1} \pi^{2} K^{2}+n}{N_{2} \pi^{2} K^{2}+1}\right)=-n$ the constant $K$ must be zero for consistency. This results in\\
\begin{align}
f_{\mathit{no}}^{leptons} \! \left(z \right) = -1\:\:\:\:\:\:\:\:f_{\mathit{no}}^{baryons} \! \left(z \right) = -2 \:\:\:\:\:\:\:\: f_{\mathit{no}}^{mesons} \! \left(z \right) = -3\end{align}

\subsection{The strong, electric, and weak forces within baryons and mesons and between leptons }
The effect of the curved spacetime onto the specific mass induced by the strong, electric, or weak dimensions result in the forces (63) between quark/quark respective quark/anti-quark pairs and fermions.

\subsubsection{The strong, electric, and weak forces of baryons }
\begin{align}
F_{s}\! \left(z , \mathit{qq} \ q\right) = 
\frac{\alpha_{s} \hslash  \left(\left(\beta_{o}+1\right)^{2} \left(z -\frac{1}{2}\right)^{4} \left(\beta_{o}-1\right)^{2} \alpha_{s}^{2}-\frac{\beta_{o}^{2}}{2}\right) c}{\left(\beta_{o}+1\right) \rho^{2} \left(-\frac{1}{4}+\left(\beta_{o}+1\right) \left(z -\frac{1}{2}\right)^{4} \left(\beta_{o}-1\right) \alpha_{s}^{2}\right) \left(z -\frac{1}{2}\right) \left(1-\beta_{o}\right)}
 \nonumber
\end{align}\includegraphics[max width=\textwidth]{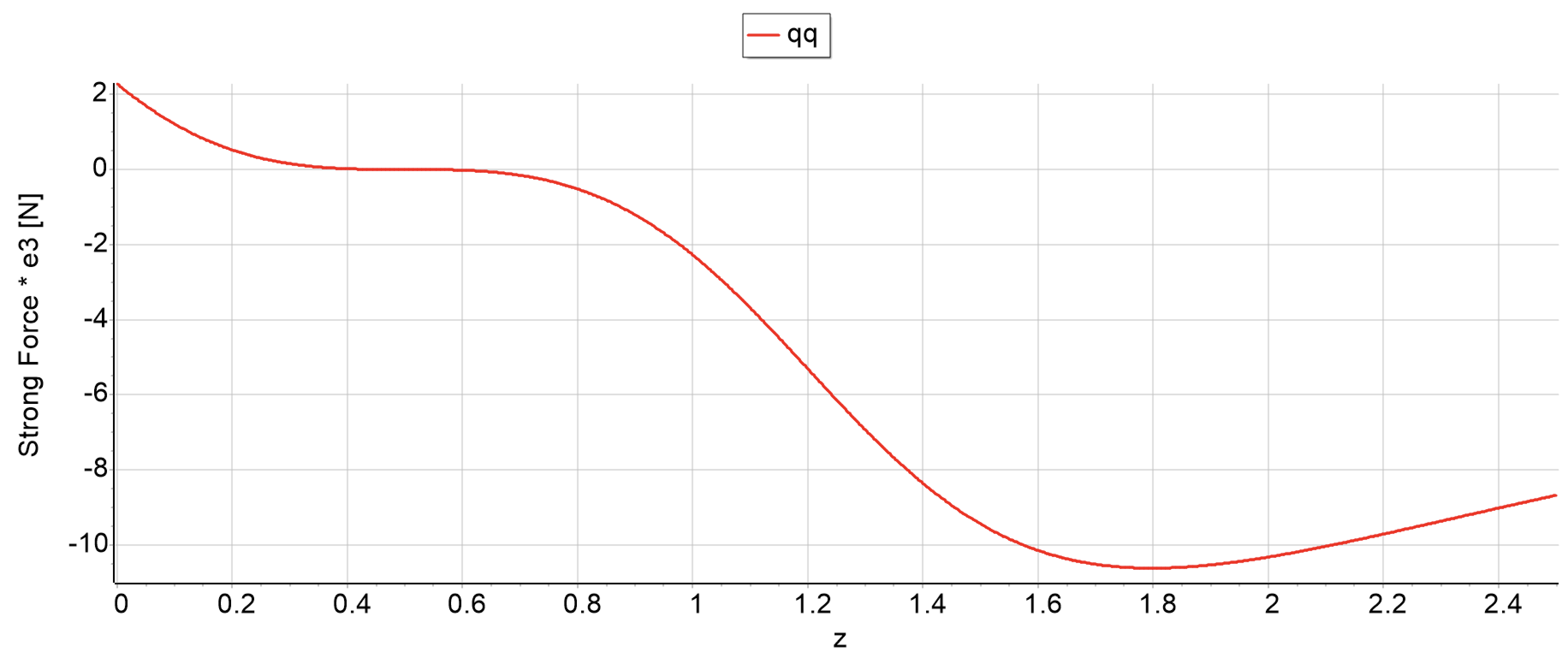}\\
Figure 4: Strong force between $qq$ within a baryon ($\beta_{o}=0$).

For $\beta_{o} \neq 0$ a transition appears within the force curve of the form as shown in Figure 5.\\
\includegraphics[max width=\textwidth]{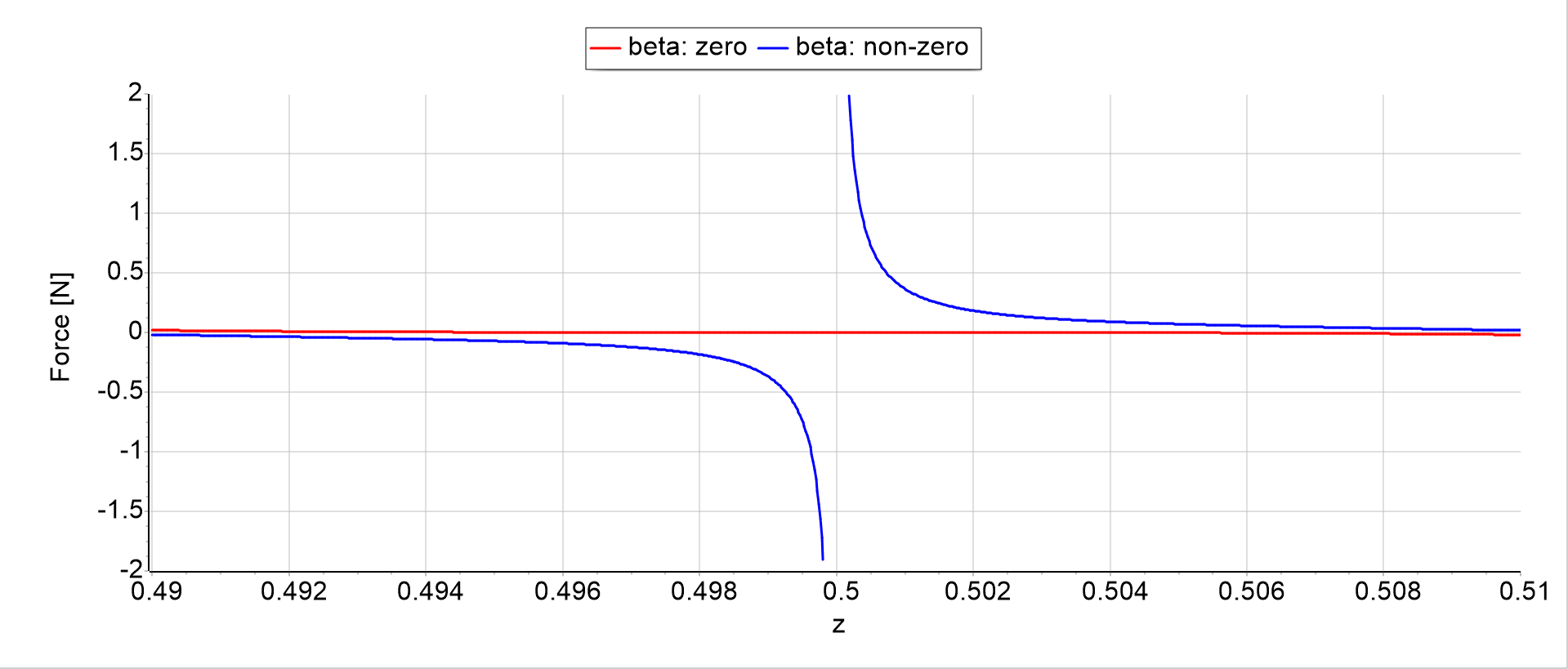} \\ Figure 5: Strong force transit from negative to positive force at $z=0.5$.

As for the strong force within a baryon at $z=1/2$ for other forces such transitions arise as well.

\begin{align}F_{e} \! \left(z , \mathit{uuq} \right) = 
\frac{8 \alpha_{e} \hslash  c \left(\left(\beta_{o} -1\right)^{2} \left(\beta_{o} +1\right)^{2} \left(z +4\right)^{4} \alpha_{e}^{2}+648 \beta_{o}^{2} z \right)}{9 \left(\beta_{o} +1\right) \rho^{2} \left(-1296+\left(\beta_{o} -1\right) \left(\beta_{o} +1\right) \left(z +4\right)^{4} \alpha_{e}^{2}\right) \left(z +4\right) z \left(\beta_{o} -1\right)}
 \nonumber
\end{align}

\begin{align}F_{e} \! \left(z , \mathit{udq} \right) = 
\frac{4 \alpha_{e} \hslash  c \left(\left(\beta_{o} -1\right)^{2} \left(\beta_{o} +1\right)^{2} \left(z -2\right)^{4} \alpha_{e}^{2}-324 \beta_{o}^{2} z \right)}{9 \left(z -2\right) \left(\beta_{o} +1\right) \left(1-\beta_{o} \right) z \left(-324+\left(\beta_{o} -1\right) \left(\beta_{o} +1\right) \left(z -2\right)^{4} \alpha_{e}^{2}\right) \rho^{2}}
 \nonumber
\end{align}

\begin{align}
F_{e} \! \left(z , \mathit{duq} \right) = 
\frac{4 \alpha_{e} \hslash  c \left(\left(\beta_{o} -1\right)^{2} \left(\beta_{o} +1\right)^{2} \left(z -4\right)^{4} \alpha_{e}^{2}-648 \beta_{o}^{2} z \right)}{9 \left(-1296+\left(\beta_{o} -1\right) \left(\beta_{o} +1\right) \left(z -4\right)^{4} \alpha_{e}^{2}\right) \left(\beta_{o} +1\right) \left(1-\beta_{o} \right) z \left(z -4\right) \rho^{2}}
 \nonumber
\end{align}

\begin{align}F_{e} \! \left(z , \mathit{ddq} \right) = 
\frac{2 \alpha_{e} \hslash  c \left(\left(\beta_{o} -1\right)^{2} \left(\beta_{o} +1\right)^{2} \left(z +2\right)^{4} \alpha_{e}^{2}+324 \beta_{o}^{2} z \right)}{9 \left(\beta_{o} +1\right) \rho^{2} \left(-324+\left(\beta_{o} -1\right) \left(\beta_{o} +1\right) \left(z +2\right)^{4} \alpha_{e}^{2}\right) \left(z +2\right) z \left(\beta_{o} -1\right)}
 \nonumber
\end{align} \includegraphics[max width=\textwidth]{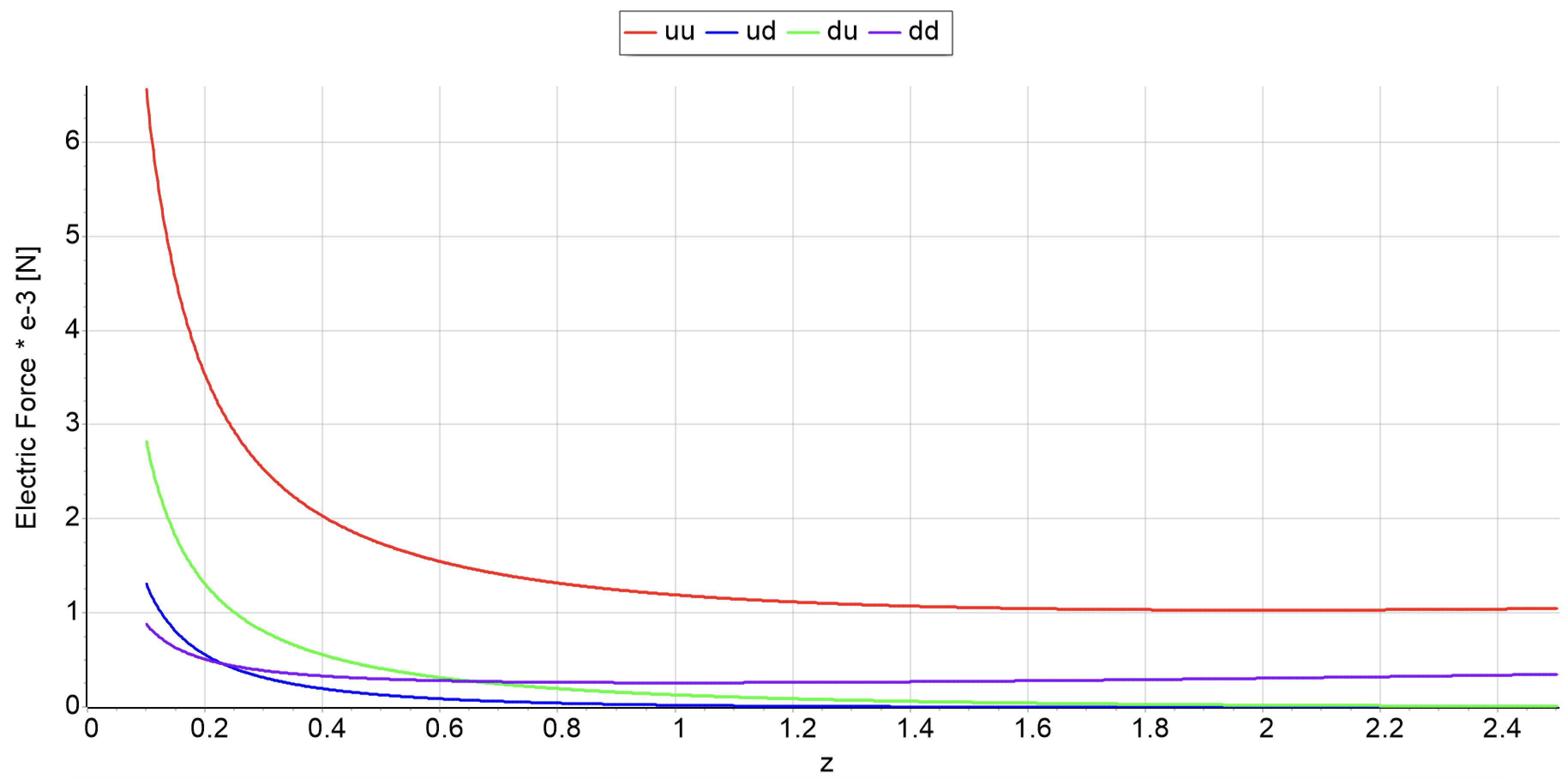}\\
Figure 6: Electric force between $uu$, $ud$,$du$, and $dd$ within a baryon ($\beta_{o}=0$).

Table 15: Transition points of the electric force within baryons for $\beta_{o}\neq0$

\begin{center}
\begin{tabular}{|l|c|c|}
\hline
Electric Force & Transition Point $z$ & Transit \\
\hline
$F_{e} \! \left(z , \mathit{uuq} \right)$ & non & --- \\
\hline
$F_{e} \! \left(z , \mathit{udq} \right)$ & 2 & negative/positive \\
\hline
$F_{e} \! \left(z , \mathit{duq} \right)$ & 4 & negative/positive \\
\hline
$F_{e} \! \left(z , \mathit{ddq} \right)$ & non & --- \\
\hline
\end{tabular}
\end{center}

The forces above are the forces felt by the first quark denoted within the bracket of the left side of the equal sign. The strong force's strength is independent of the quark flavour. At $z=1.72 \cdot 10^{-20}$, the equivalence of the Planck-length, the electric forces increase to $>10^{15} N$ and are repulsive for all quark/quark combinations independent of their load signs! For $\beta_{o}\neq0$ the weak force approaches infinity for $z=0$.

\begin{center}
\includegraphics[max width=\textwidth]{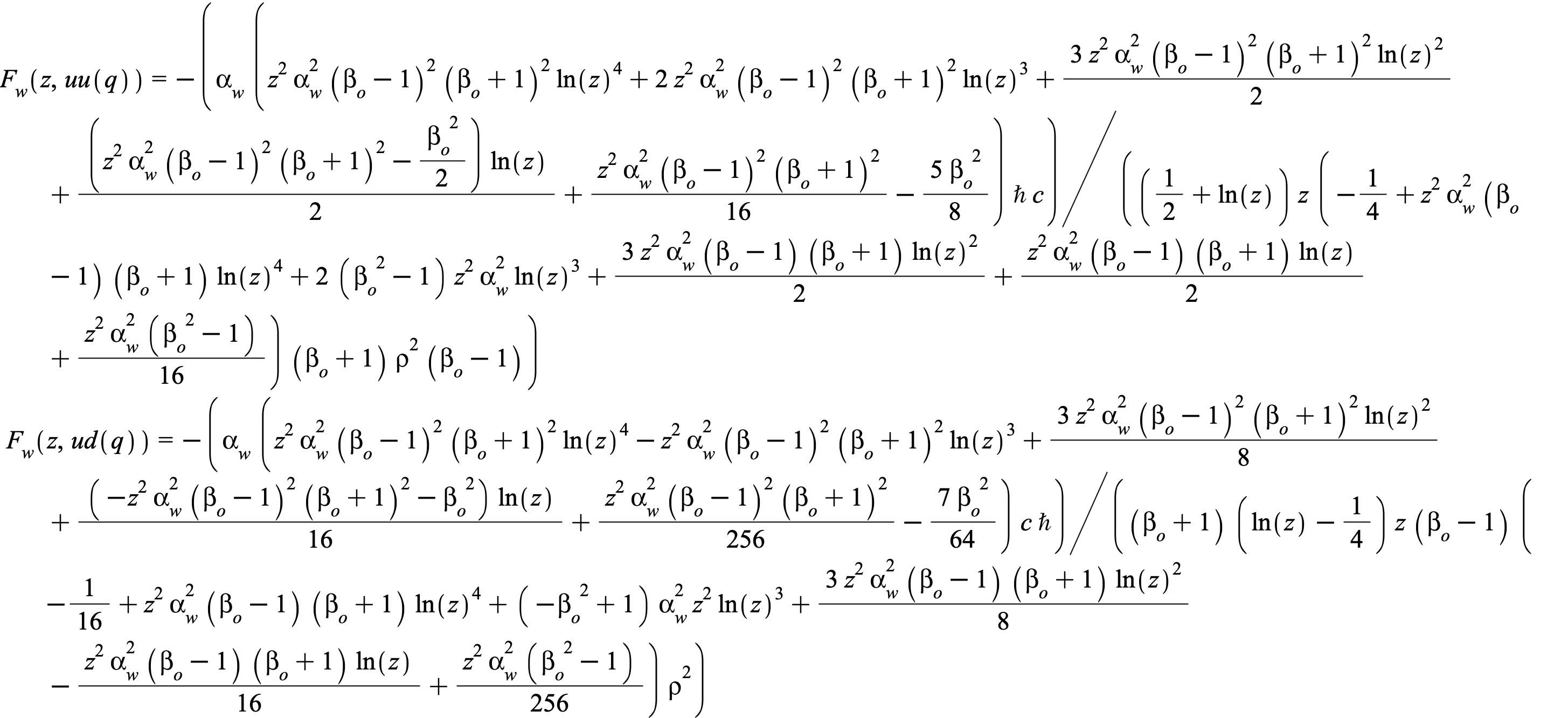}
\end{center}

\begin{center}
\includegraphics[max width=\textwidth]{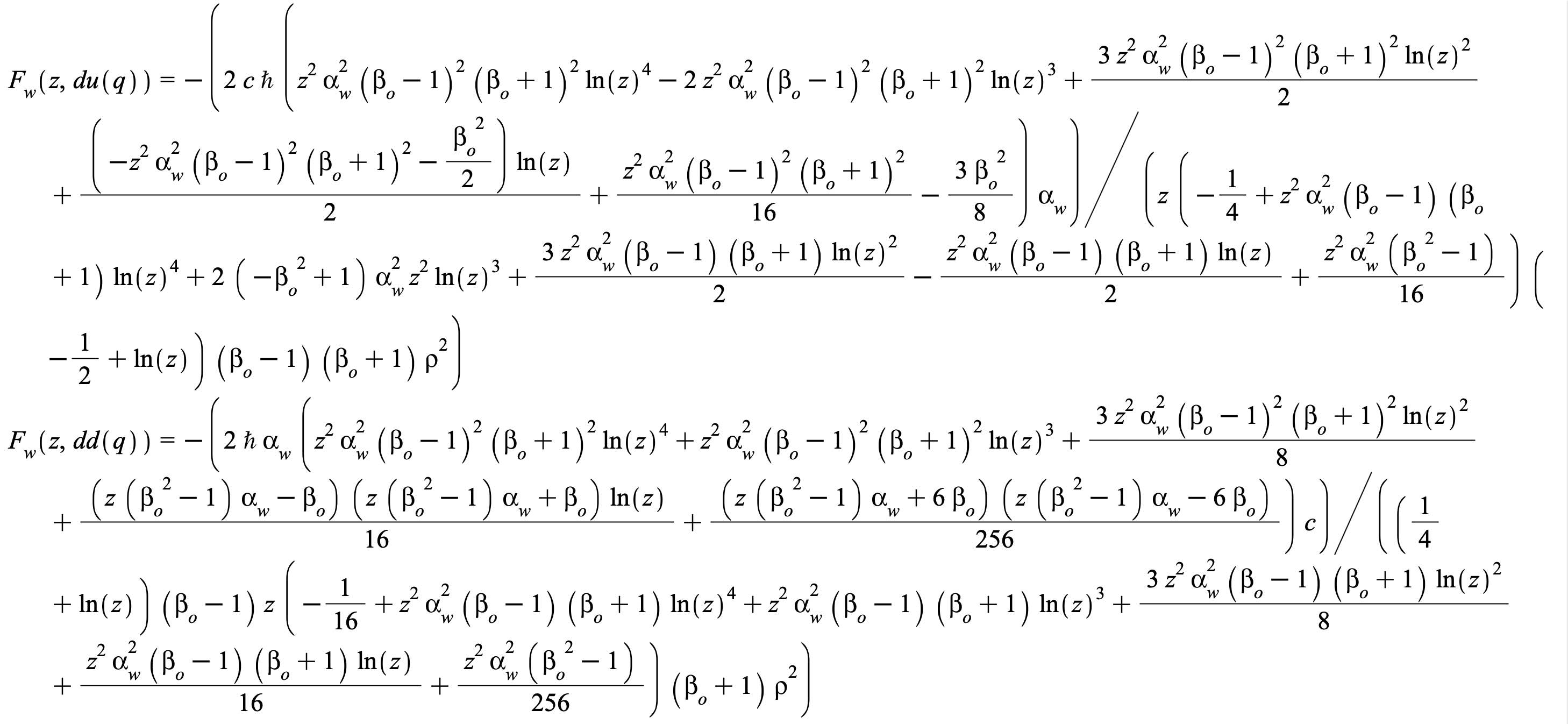}
\end{center}

\includegraphics[max width=\textwidth]{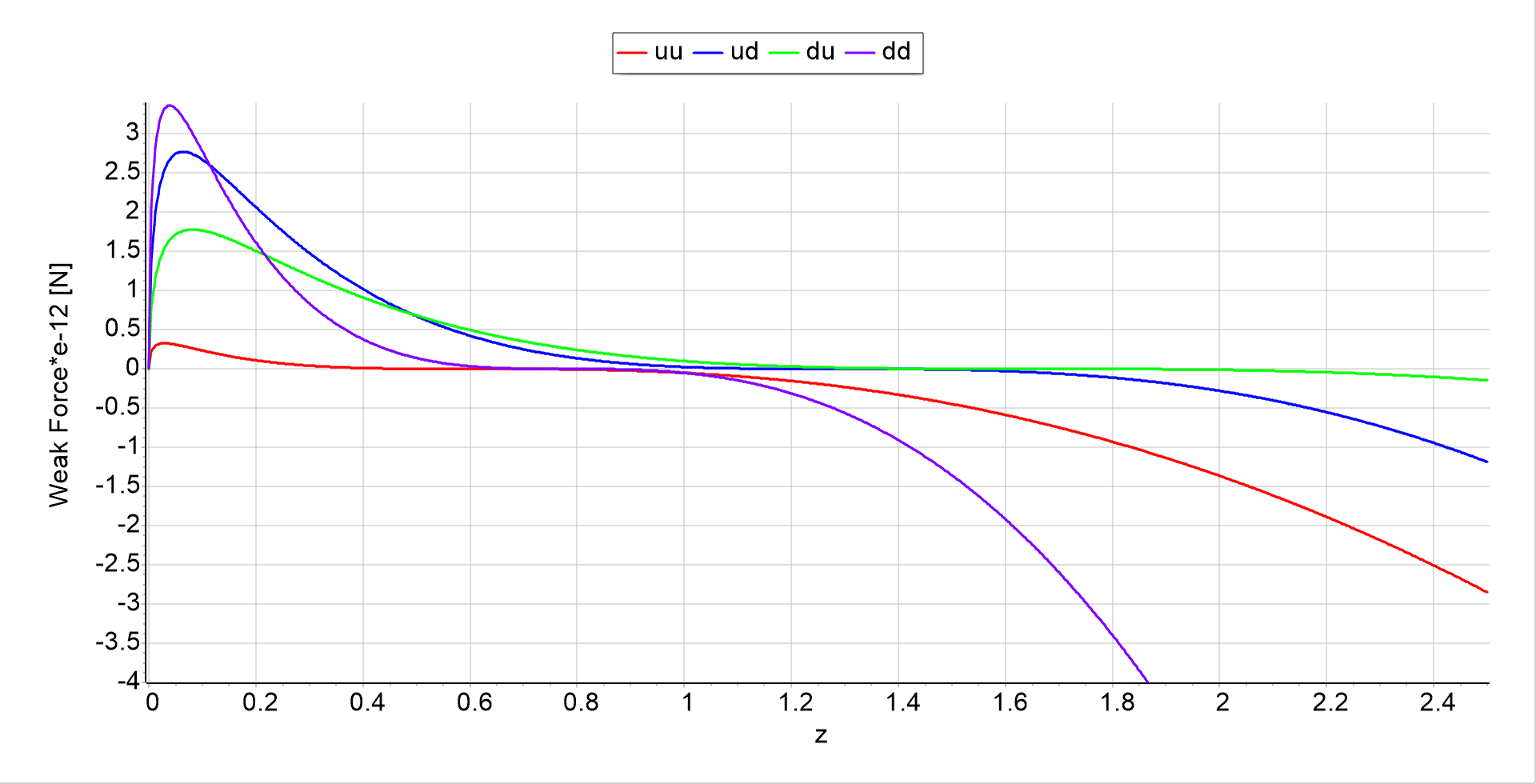} Figure 7: Weak force between $uu$, $ud$,$du$, and $dd$ within a baryon ($\beta_{o}=0$).

Table 16: Transition points of the weak force within baryons for $\beta_{o}\neq0$

\begin{center}
\begin{tabular}{|l|c|c|}
\hline
Weak Force & Transition Point $z$ & Transit \\
\hline
$F_{w} \! \left(z , \mathit{uuq} \right)$ & ${\mathrm e}^{-\frac{1}{2}} \approx 0.60653$ & negative/positive \\
\hline
$F_{w} \! \left(z , \mathit{udq} \right)$ & ${\mathrm e}^{\frac{1}{4}} \approx 1.2840$ & negative/positive \\
\hline
$F_{w} \! \left(z , \mathit{duq} \right)$ & ${\mathrm e}^{\frac{1}{2}} \approx 1.6487$ & negative/positive \\
\hline
$F_{w} \! \left(z , \mathit{ddq} \right)$ & ${\mathrm e}^{-\frac{1}{4}} \approx 0.77880$ & negative/positive \\
\hline
\end{tabular}
\end{center}

Also, for $\beta_{o}\neq0$ the weak force approaches infinity for $z=0$.

\subsubsection{The strong, electric, and weak forces of mesons }
\begin{align}
F_{s} \! \left(z , q \overline{q}\right) = 
\frac{4 \left(\left(-\frac{1}{3}+z \right)^{4} \left(\beta_{o} -1\right)^{2} \left(\beta_{o} +1\right)^{2} \alpha_{s}^{2}-\frac{2 \beta_{o}^{2}}{9}\right) \hslash  \alpha_{s} c}{\left(-\frac{1}{3}+z \right) \left(1-\beta_{o} \right) \left(\beta_{o} +1\right) \left(-\frac{2}{9}+\left(-\frac{1}{3}+z \right)^{4} \left(\beta_{o} -1\right) \left(\beta_{o} +1\right) \alpha_{s}^{2}\right) \rho^{2}}
 \nonumber
\end{align} \includegraphics[max width=\textwidth]{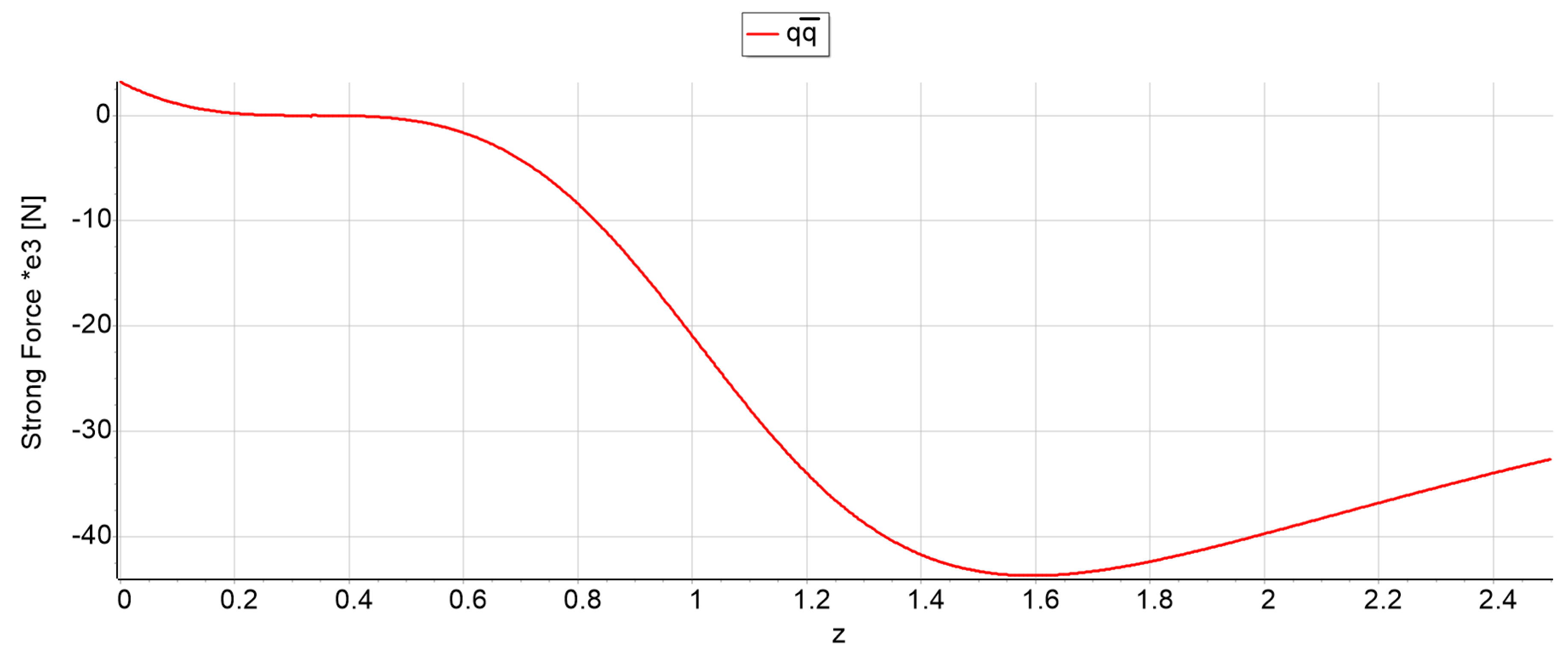}\\
Figure 8: Strong force between $q\bar{q}$ within a meson ($\beta_{o}=0$) with the transit at $z=1/3$ from negative to positive for $\beta_{o} \neq 0$.

\begin{align}F_{e} \! \left(z , u \overline{u}\right) = 
\frac{4 \hslash  \alpha_{e} \left(\left(\beta_{o} -1\right)^{2} \left(\beta_{o} +1\right)^{2} \left(z -6\right)^{4} \alpha_{e}^{2}-972 \beta_{o}^{2} z \right) c}{3 z \left(z -6\right) \left(1-\beta_{o} \right) \left(\beta_{o} +1\right) \rho^{2} \left(-2916+\left(\beta_{o} -1\right) \left(\beta_{o} +1\right) \left(z -6\right)^{4} \alpha_{e}^{2}\right)}
 \nonumber
\end{align}

\begin{align}
F_{e} \! \left(z , u \overline{d}\right) = 
\frac{1944 \left(\frac{\left(\beta_{o} -1\right)^{2} \left(\beta_{o} +1\right)^{2} \left(z +3\right)^{4} \alpha_{e}^{2}}{2916}+\frac{\pi  \beta_{o} \left(\beta_{o} -1\right) \left(\beta_{o} +1\right) \left(z +3\right)^{2} \alpha_{e}}{27}+\beta_{o}^{2} \left(\pi^{2}+\frac{z}{6}\right)\right) c \hslash  \alpha_{e}}{\left(z +3\right) \rho^{2} \left(\beta_{o} +1\right) \left(\beta_{o} -1\right) z \left(-729+\left(\beta_{o} -1\right) \left(\beta_{o} +1\right) \left(z +3\right)^{4} \alpha_{e}^{2}\right)}
 \nonumber
\end{align}

\begin{align}
F_{e} \! \left(z , \overline{d} u \right) = 
\frac{7776 \left(\frac{\left(\beta_{o} -1\right)^{2} \left(\beta_{o} +1\right)^{2} \left(z +6\right)^{4} \alpha_{e}^{2}}{11664}+\frac{\pi  \beta_{o} \left(\beta_{o} -1\right) \left(\beta_{o} +1\right) \left(z +6\right)^{2} \alpha_{e}}{54}+\beta_{o}^{2} \left(\pi^{2}+\frac{z}{12}\right)\right) \alpha_{e} \hslash  c}{\left(\beta_{o} -1\right) \left(-2916+\left(\beta_{o} -1\right) \left(\beta_{o} +1\right) \left(z +6\right)^{4} \alpha_{e}^{2}\right) z \left(\beta_{o} +1\right) \left(z +6\right) \rho^{2}}
 \nonumber
\end{align}

\begin{align}
F_{e} \! \left(z , d \overline{u}\right) = 
\frac{7776 \alpha_{e} \hslash  \left(\frac{\left(\beta_{o} -1\right)^{2} \left(\beta_{o} +1\right)^{2} \left(z +6\right)^{4} \alpha_{e}^{2}}{11664}-\frac{\pi  \beta_{o} \left(\beta_{o} -1\right) \left(\beta_{o} +1\right) \left(z +6\right)^{2} \alpha_{e}}{54}+\beta_{o}^{2} \left(\pi^{2}+\frac{z}{12}\right)\right) c}{\left(\beta_{o} -1\right) \left(-2916+\left(\beta_{o} -1\right) \left(\beta_{o} +1\right) \left(z +6\right)^{4} \alpha_{e}^{2}\right) z \left(\beta_{o} +1\right) \left(z +6\right) \rho^{2}}
\nonumber
\end{align}

\begin{align}F_{e} \! \left(z , \overline{u} d\right) = 
\frac{2 \left(\left(\beta_{o} -1\right)^{2} \left(\beta_{o} +1\right)^{2} \left(z +3\right)^{4} \alpha_{e}^{2}+486 \beta_{o}^{2} z \right) \hslash  \alpha_{e} c}{3 z \left(\beta_{o} -1\right) \left(\beta_{o} +1\right) \left(z +3\right) \rho^{2} \left(-729+\left(\beta_{o} -1\right) \left(\beta_{o} +1\right) \left(z +3\right)^{4} \alpha_{e}^{2}\right)}
 \nonumber
\end{align}

\begin{align}
F_{e} \! \left(z , d \overline{d}\right) = 
\frac{\hslash  \alpha_{e} \left(\left(\beta_{o} -1\right)^{2} \left(\beta_{o} +1\right)^{2} \left(z -3\right)^{4} \alpha_{e}^{2}-486 \beta_{o}^{2} z \right) c}{3 z \left(z -3\right) \left(-729+\left(\beta_{o} -1\right) \left(\beta_{o} +1\right) \left(z -3\right)^{4} \alpha_{e}^{2}\right) \left(1-\beta_{o} \right) \left(\beta_{o} +1\right) \rho^{2}}
 \nonumber
\end{align} \includegraphics[max width=\textwidth]{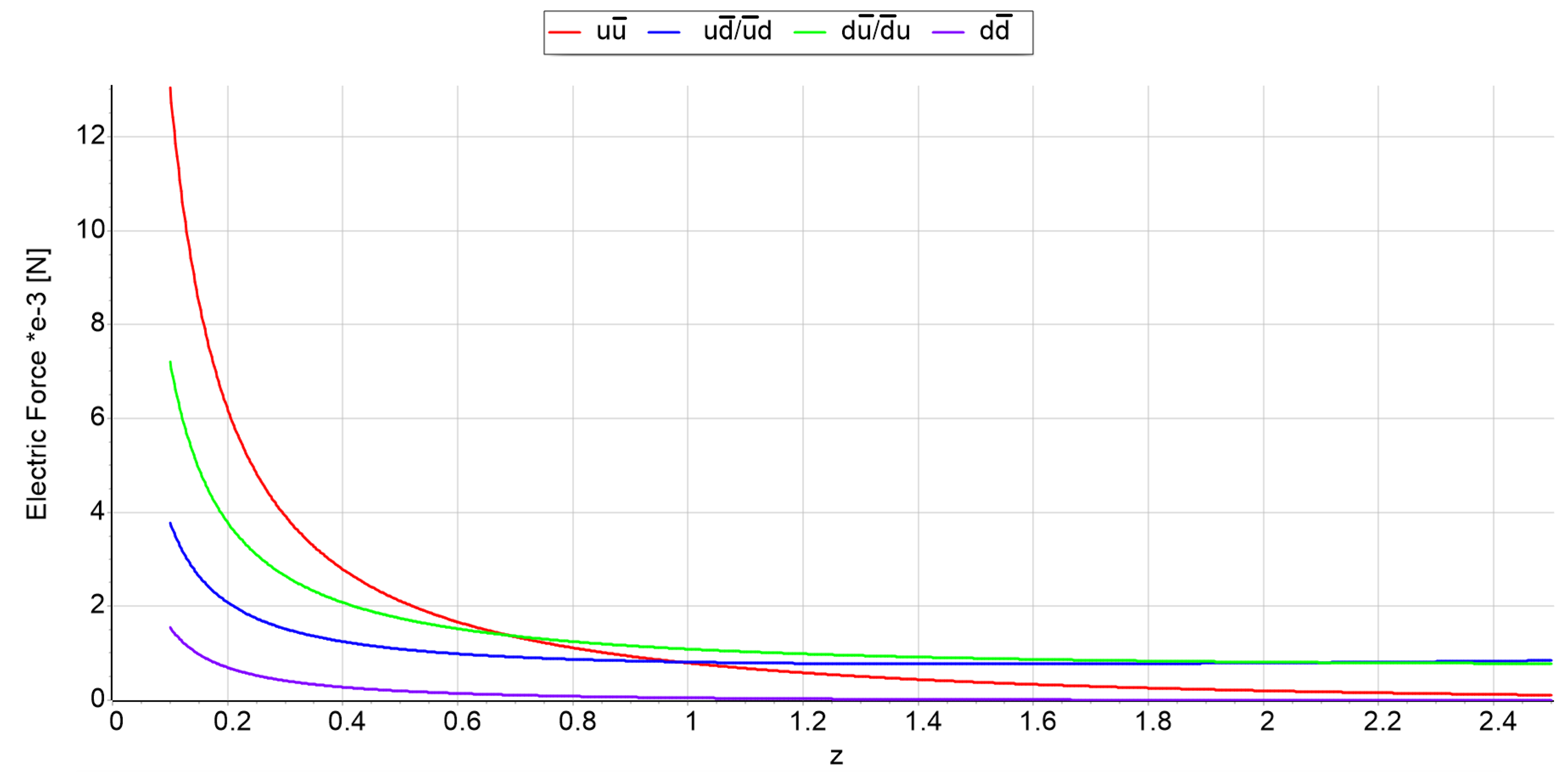}\\
Figure 9: Electric force between $u \bar{u}$, $u\bar{d}/\bar{u}d$,$d\bar{u}/\bar{d}u$, and $d\bar{d}$ within a meson ($\beta_{o}=0$).

Table 17: Transition points of the electric force within mesons for $\beta_{o}\neq0$

\begin{center}
\begin{tabular}{|l|c|c|}
\hline
Electric Force & Transition Point $z$ & Transit \\
\hline
$F_{e} \! \left(z , u \overline{u}\right)$ & 6 & negative/positive \\
\hline
$F_{e} \! \left(z , u \overline{d}\right)$ & non & --- \\
\hline
$F_{e} \! \left(z , \overline{d} u \right)$ & non & --- \\
\hline
$F_{e} \! \left(z , d \overline{u}\right)$ & non & --- \\
\hline
$F_{e} \! \left(z , \overline{u} d\right)$ & non & --- \\
\hline
$F_{e} \! \left(z , d \overline{d}\right)$ & 3 & negative/positive \\
\hline
\end{tabular}
\end{center}

\includegraphics[max width=\textwidth]{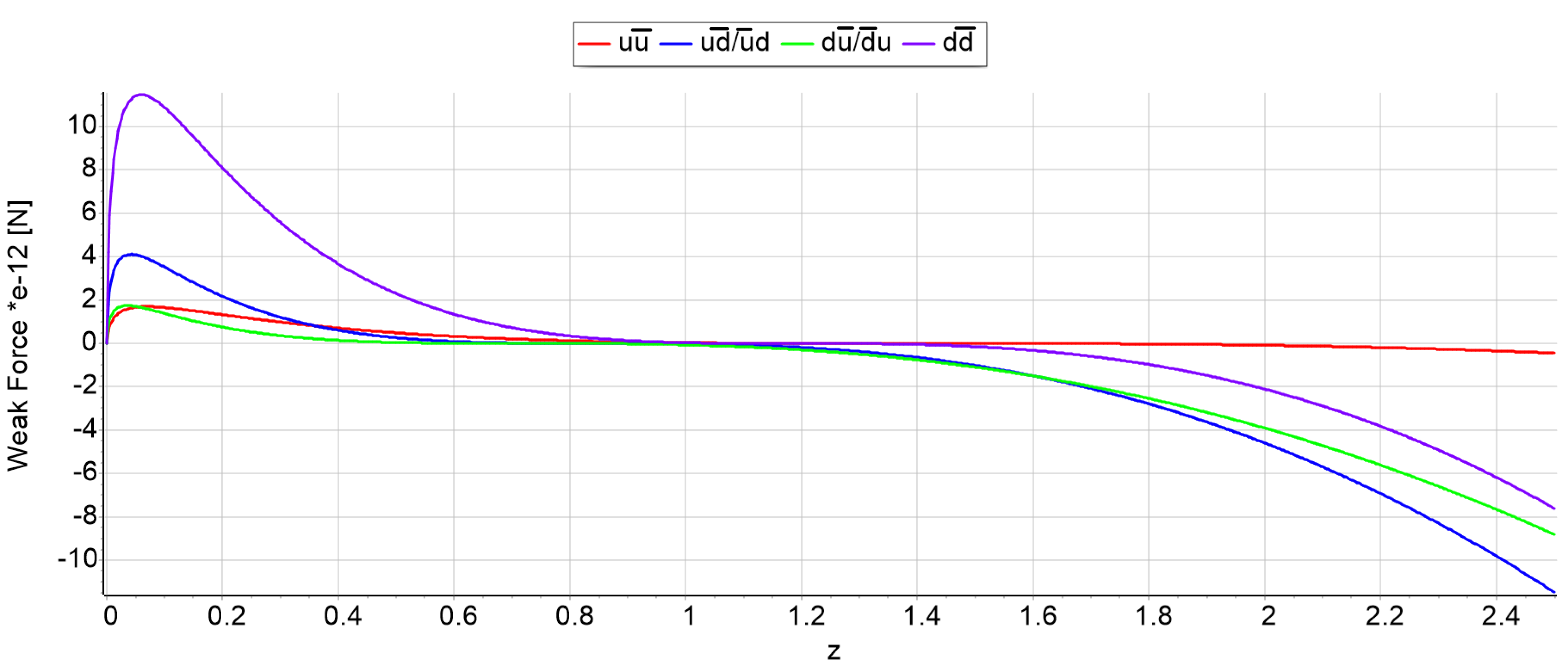}\\
Figure 10: Weak force between $u \bar{u}$, $u\bar{d}/\bar{u}d$,$d\bar{u}/\bar{d}u$, and $d\bar{d}$ within a meson ($\beta_{o}=0$).

Table 18: Transition points of the weak force within mesons for $\beta_{o}\neq0$

\begin{center}
\begin{tabular}{|l|c|c|}
\hline
Weak Force & Transition Point $z$ & Transit \\
\hline
$F_{w} \! \left(z , u \overline{u}\right)$ & ${\mathrm e}^{\frac{1}{3}} \approx$  1.3956 & negative/positive \\
\hline
$F_{w} \! \left(z , u \overline{d}\right)$ & ${\mathrm e}^{-\frac{1}{6}} \approx 0.84648$ & negative/positive \\
\hline
$F_{w} \! \left(z , \overline{d} u \right)$ & ${\mathrm e}^{-\frac{1}{3}} \approx 0.71653$ & positive/negative \\
\hline
$F_{w} \! \left(z , d \overline{u}\right)$ & ${\mathrm e}^{-\frac{1}{3}} \approx 0.71653$ & positive/negative \\
\hline
$F_{w} \! \left(z , \overline{u} d\right)$ & ${\mathrm e}^{-\frac{1}{6}} \approx 0.84648$ & negative/positive \\
\hline
$F_{w} \! \left(z , d \overline{d}\right)$ & ${\mathrm e}^{\frac{1}{6}} \approx 1.1814$ & negative/positive \\
\hline
\end{tabular}
\end{center}

For quark/anti-quark combinations strong forces are flavour independent. Electric forces within mesons display the same characteristic as in the case of baryons. At $z=1.72 \cdot 10^{-20}$, the equivalence of the Planck-length, the electric forces increase to $>10^{15} N$ and are repulsive for all quark/anti-quark combinations independent of their load signs! Weak forces are repulsive for $z<1$ and approaching infinity for $z=0$ if $\beta_{o}\neq0$.

\subsubsection{The electric and weak forces of the charged leptons }
\includegraphics[max width=\textwidth, center]{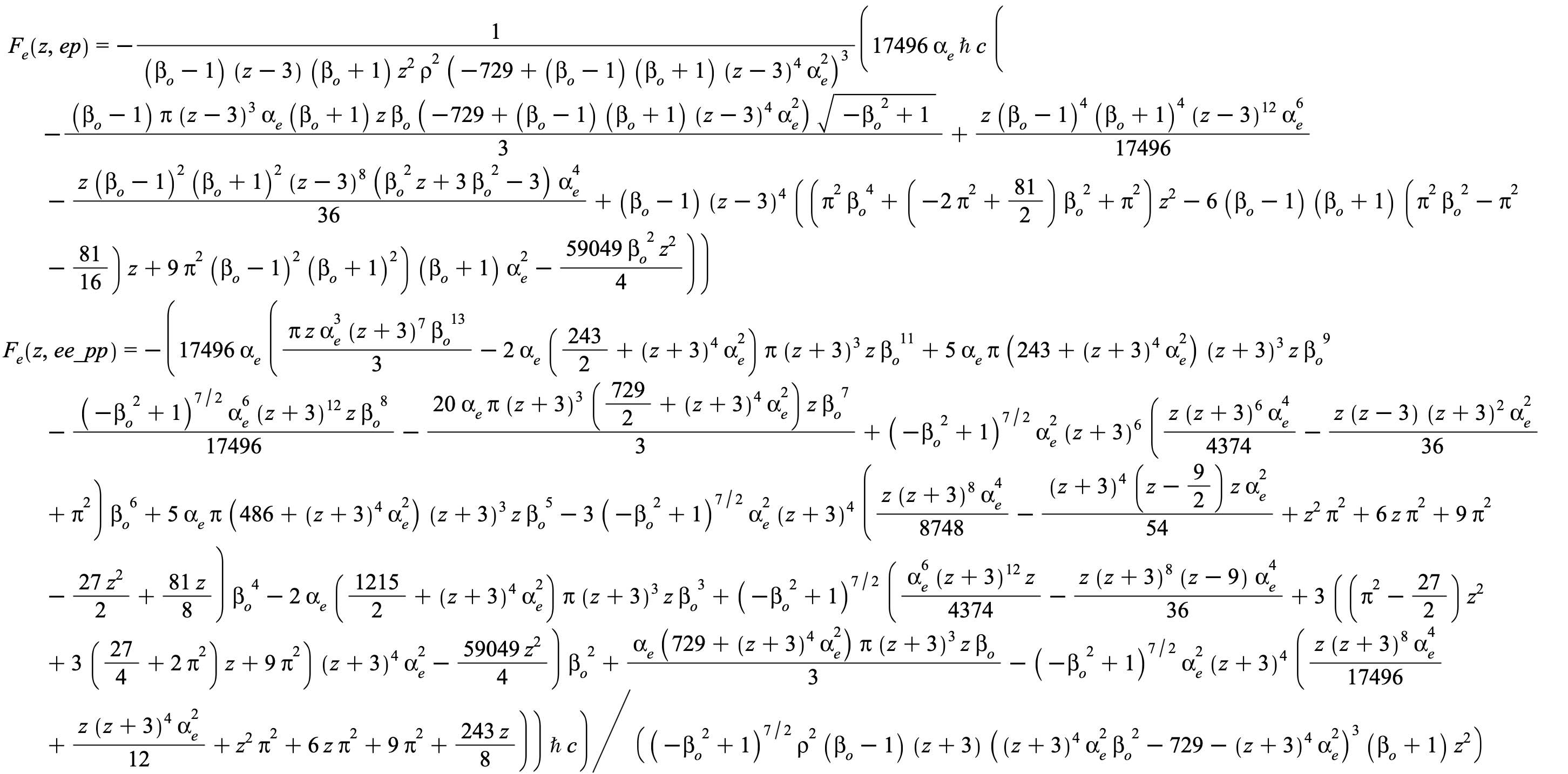}\\
\includegraphics[max width=\textwidth, center]{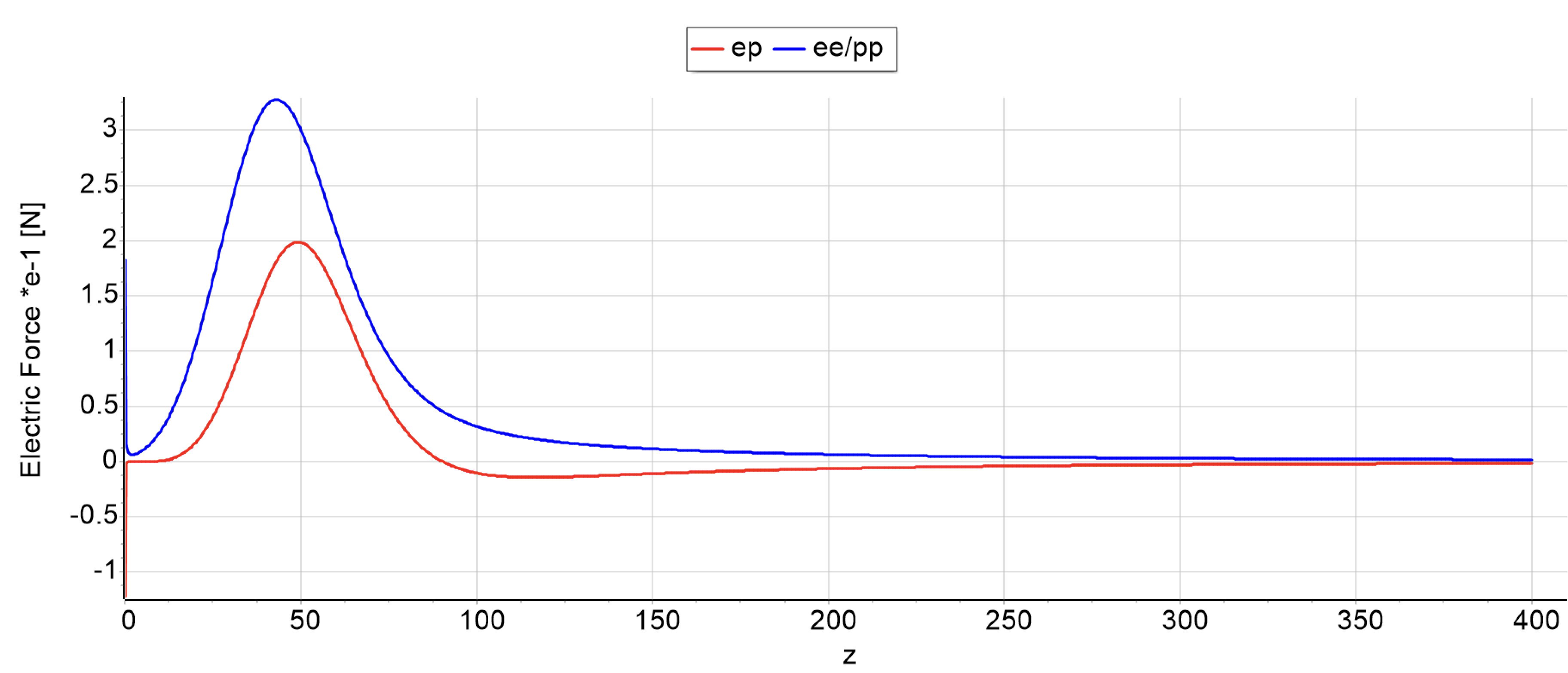}

Figure 11: Electric force between $e^{-}e^{+}$ and $e^{-}e^{-}/e^{+}e^{+}$ ($\beta_{o}=0$).\\
Leptons act in the expected way of $\lim _{z \rightarrow 0} F_{c}\left(z, e^{-} e^{+}\right)=-\infty$ and \\ 
$\lim _{z \rightarrow 0} F_{c}\left(z, e^{-}e^{-}/e^{+}e^{+}\right)=+\infty$. They do, however, show repulsion independent of the load sign combinations in the range $8 < z < 90$. For $200 < z$ the force follows the Coulomb law.\\
\includegraphics[max width=\textwidth, center]{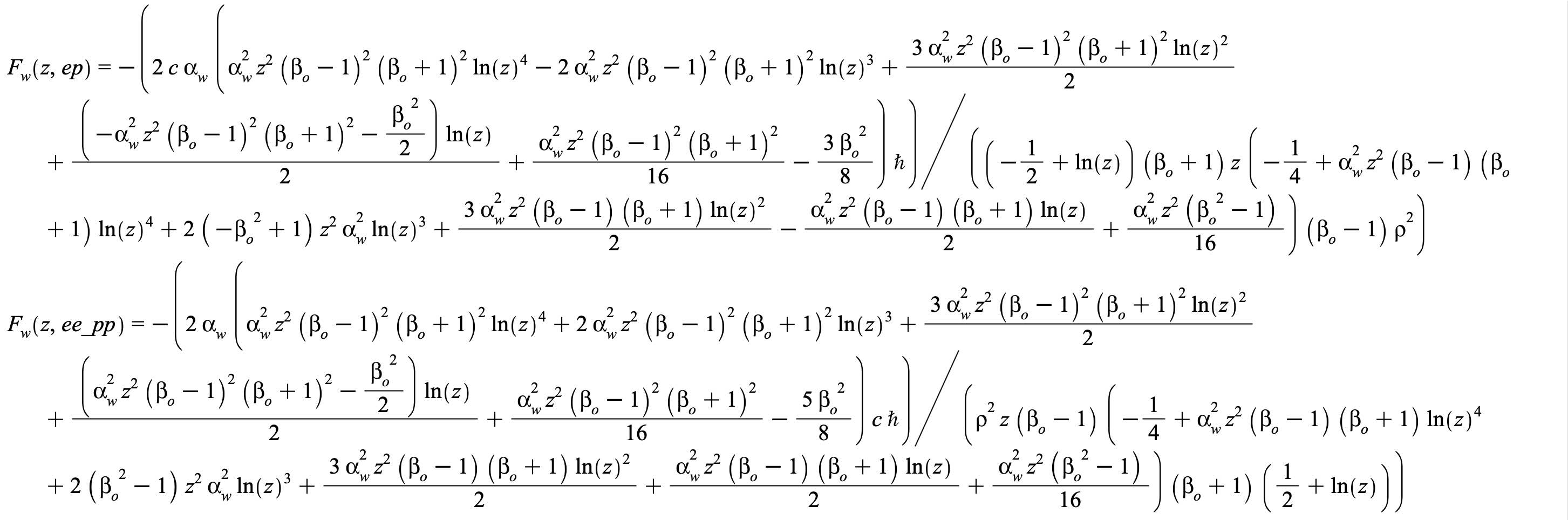}\\
\includegraphics[max width=\textwidth]{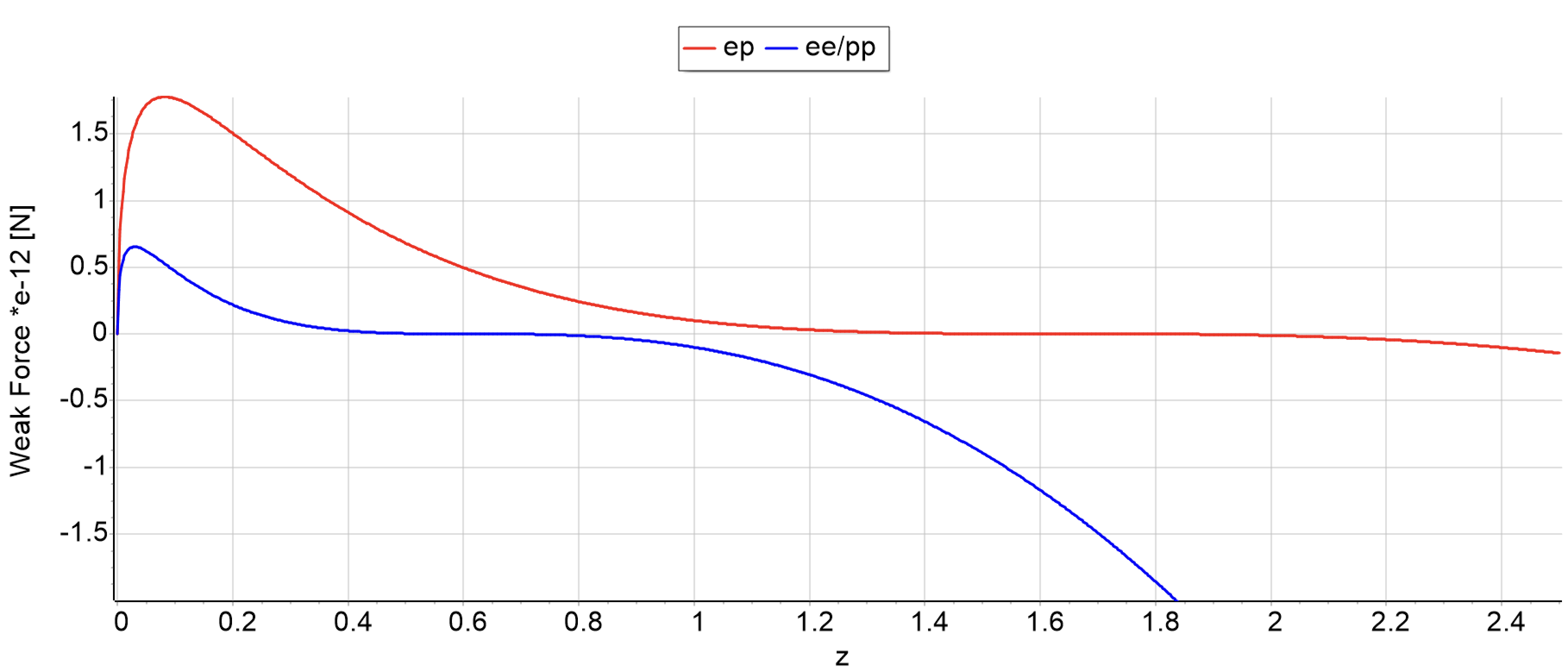} \\
Figure 12: Weak force between $e^{-}e^{+}$ and $e^{-}e^{-}/e^{+}e^{+}$ ($\beta_{o}=0$).

Table 19: Transition points of the electric and weak forces of leptons for $\beta_{o}\neq0$

\begin{center}
\begin{tabular}{|l|c|c|}
\hline
Electric/Weak Force & Transition Point $z$ & Transit \\
\hline
$F_{e} \! \left(z , ep\right)$ & 3 & negative/positive \\
\hline
$F_{e} \! \left(z , ee/pp\right)$ & non & --- \\
\hline
$F_{w} \! \left(z , ep\right)$ & ${\mathrm e}^{\frac{1}{2}} \approx 1.6487$ & negative/positive \\
\hline
$F_{w} \! \left(z , ee/pp\right)$ & ${\mathrm e}^{-\frac{1}{2}} \approx 0.60653$ & negative/positive \\
\hline
\end{tabular}
\end{center}

The weak force at $\beta_{o}=0$ has a strength of about $10^{-12} N$ and the force is zero for $z=0$ if $\beta_{o}=0$, but infinit for $\beta_{o}\neq0$.

\subsection{Lorentz factors of strong, electric, and weak interactions at their fix points}
The calculation of the velocities at the fix points for the three forces are calculated separately. It starts with equation (60) were the metric tensors (3) is used with the input of (1).

\begin{enumerate}
  \item For the strong interaction (60) becomes a differential equation that allows to calculate  $\beta(z=0)$ with the result $\beta  = 0$ and following\\
\begin{align}
\gamma_{os}(0) = 1
\end{align}

  \item Solving equation (60) of the electric interaction results in $\beta=0$ as well. $=>$\\
\begin{align}
\gamma_{oe}(\infty) = 1
\end{align}

  \item Finally the weak interaction, solving (60) equates to $\beta=0$ and\\
\begin{align}
\gamma_{ow}(1) = 1
\end{align}

\end{enumerate}

\section{Literature}
\begin{enumerate}
  \item Goldberg D. The standard model in a nutshell. Princeton ; Oxford: Princeton University Press; 2017. xvii, 295 pages $p$.

  \item Langacker P. The standard model and beyond. Boca Raton, FL: CRC Press; 2010. xii, 663 p. p.

  \item Thomson M. Modern particle physics. Cambridge, United Kingdom ; New York: Cambridge University Press; 2013. xvi, 554 pages $p$.

  \item Vergados JD. The standard model and beyond. New Jersey: World Scientific; 2018. xvi, 435 pages $p$.

  \item Bettini A. Introduction to elementary particle physics. Second edition. ed. New York: Cambridge University Press; 2014. xvii, 474 pages $p$.

  \item Griffiths DJ. Introduction to elementary particles. New York: Harper \& Row; 1987. vii, 392 p. p.

  \item Cottingham WN, Greenwood DA. An introduction to the standard model of particle physics. 2nd ed. Cambridge ; New York: Cambridge University Press; 2007. xix, 272 p. p.

  \item Jaeger G. Exchange Forces in Particle Physics. Foundations of Physics. 2021;51(1):13.

  \item Schwinger J. On Quantum-Electrodynamics and the Magnetic Moment of the Electron. Physical Review. 1948;73(4):416-7.

  \item Aoyama T, Hayakawa M, Kinoshita T, Nio M. Tenth-order electron anomalous magnetic moment: Contribution of diagrams without closed lepton loops. Physical Review D. 2015;91(3):033006.

  \item Wang B, Abdalla E, Atrio-Barandela F, Pavón D. Dark matter and dark energy interactions: theoretical challenges, cosmological implications and observational signatures. Reports on Progress in Physics. 2016;79(9):096901.

  \item Rovelli C. Quantum gravity. 1st pbk. ed. Cambridge ; New York: Cambridge University Press; 2008. 458 p. p.

  \item Einstein A. Die Grundlage der allgemeinen Relativitätstheorie. Leipzig: Verlag von Johann Ambrosious Barth; 1916. 64 p. p.

  \item Weinberg S. Cosmology. Oxford ; New York: Oxford University Press; 2008. xvii, 593 p. $p$

  \item Kaluza T. Zum Unitätsproblem der Physik. Sitzungsberichte der Königlich Preußischen Akademie der Wissenschaften (Berlin. 1921:966.

  \item Kaluza T, editor On the Problem of Unity in Physics1983 January 01, $1983 .$

  \item Klein $O$. Quantentheorie und fünfdimensionale Relativitätstheorie. Zeitschrift fur Physik. 1926;37:895.

  \item Klein O. The Atomicity of Electricity as a Quantum Theory Law. Nature. 1926;118(2971):516-

  \item Appelquist T, Chodos A, Freund PGO. Modern Kaluza-Klein theories. Menlo Park, Calif.: Addison-Wesley Pub. Co.; 1987. xvii, 619 p.

  \item Wesson PS. Space-Time-Matter: Modern Kaluza-Klain Theory. Singapore: World Sciency Publishing $1999 .$

  \item Wesson PS, Overduin JM. Principles of space-time-matter : cosmology, particles and waves in five dimensions. Singapore ; Hackensack, NJ: World Scientific Publishing Co. Pte. Ltd.; 2019. pages $\mathrm{cm} p$.

  \item Witten E. String theory dynamics in various dimensions. Nuclear Physics B. 1995;443(1):85-126.

  \item Zwiebach B. A first course in string theory. New York: Cambridge University Press; 2004. xx, 558 p. p.

  \item Polchinski JG. String theory. Cambridge, UK ; New York: Cambridge University Press; $1998 .$

  \item Urbanowski K. Decay law of relativistic particles: Quantum theory meets special relativity. Physics Letters B. 2014;737:346-51.

  \item DeGrand $\mathrm{T}$, Jaffe RL, Johnson $\mathrm{K}$, Kiskis J. Masses and other parameters of the light hadrons. Physical Review D. 1975;12(7):2060-76.

  \item Palazzi P. Particles and shells. arXiv preprint physics/0301074. $2003 .$

  \item Carnal O, Mlynek J. Young's double-slit experiment with atoms: A simple atom interferometer. Physical Review Letters. 1991;66(21):2689-92.

  \item Group PD, Zyla PA, Barnett RM, Beringer J, Dahl O, Dwyer DA, et al. Review of Particle Physics. Progress of Theoretical and Experimental Physics. 2020;2020(8).

  \item Wiechert E. Elektrodynamische Elementargesetze. Annalen der Physik. 1901;309:667.

  \item Ishii N, Aoki S, Hatsuda T. Nuclear Force from Lattice QCD. Physical Review Letters. 2007;99(2):022001.

  \item Aoki S, Hatsuda T, Ishii N. Theoretical foundation of the nuclear force in QCD and its applications to central and tensor forces in quenched lattice QCD simulations. Progress of theoretical physics. 2010;123(1):89-128.

  \item Schröder Y. The static potential in QCD to two loops. Physics Letters B. 1999;447(3):321-6.

  \item Anzai C, Kiyo $Y$, Sumino $Y$. Static QCD potential at three-loop order. Physical review letters. 2010;104(11):112003.

  \item Del Debbio L, Ramos A. Lattice determinations of the strong coupling. Physics Reports. 2021;920:1-71.

  \item Nave R. Coupling Constants for the Fundamental Forces 2021 [Available from: \href{http://hyperphysics.phy-astr.gsu.edu/hbase/Forces/couple.html}{http://hyperphysics.phy-astr.gsu.edu/hbase/Forces/couple.html}.

  \item Mohr PJ, Newell DB, Taylor BN. CODATA recommended values of the fundamental physical constants: 2014. Journal of Physical and Chemical Reference Data. 2016;45(4):043102.

  \item Berends FA, Kleiss R. Distributions for electron-positron annihilation into two and three photons. Nuclear Physics B. 1981;186(1):22-34.

  \item Wikipedia. List of mesons: Wikipedia; 2021 [\href{https://en.wikipedia.org/wiki/List_of_mesons}{https://en.wikipedia.org/wiki/List\_of\_mesons}

  \item Wikipedia. List of baryons: Wikipedia; 2021 [\href{https://en.wikipedia.org/wiki/List_of_baryons}{https://en.wikipedia.org/wiki/List\_of\_baryons}

  \item Collaboration K, Aker M, Altenmüller K, Arenz M, Babutzka M, Barrett J, et al. Improved Upper Limit on the Neutrino Mass from a Direct Kinematic Method by KATRIN. Physical Review Letters. 2019;123(22):221802.

  \item Vikhlinin A, Kravtsov AV, Burenin RA, Ebeling H, Forman WR, Hornstrup A, et al. CHANDRA CLUSTER COSMOLOGY PROJECT III: COSMOLOGICAL PARAMETER CONSTRAINTS. The Astrophysical Journal. 2009;692(2):1060-74.

  \item de Broglie L. The reinterpretation of wave mechanics. Foundations of Physics. 1970;1(1):5-15.

  \item Gamow G. Physics and Feynman's Diagrams. $2005$.

\end{enumerate}

\end{document}